\documentclass[10pt,english]{article}
\usepackage[T1]{fontenc}
\usepackage[latin9]{inputenc}
\usepackage{geometry}
\usepackage{color}
\usepackage{float}
\usepackage{mathtools}
\usepackage{amsmath}
\usepackage{amssymb}
\usepackage{stmaryrd}
\usepackage{graphicx}
\usepackage{esint}
\usepackage{subscript}

\makeatletter

\providecommand{\tabularnewline}{\\}

\usepackage[english]{babel}
\usepackage{dsfont}

\let\@fnsymbol\@arabic

\usepackage{babel}
\newcommand{\PP}{P}
\newcommand{\uu}{u}
\newcommand{\HH}{H}
\newcommand{\sis}{\sigma}
\newcommand{\XX}{X}
\newcommand{\mm}{m}
\newcommand{\MM}{M}
\newcommand{\nn}{n}
\newcommand{\ff}{f}
\newcommand{\gs}{s}
\newcommand{\gt}{t}
\newcommand{\gtau}{\tau}
\newcommand{\geps}{\epsilon}
\newcommand{\gA}{A}
\newcommand{\BB}{B}
\newcommand{\CC}{C}
\newcommand{\vv}{v}
\newcommand{\id}{\mathds1}
\newcommand{\Div}{\mathrm{Div}}
\newcommand{\Curl}{\mathrm{Curl}}
\newcommand{\tr}{\mathrm{tr}}
\newcommand{\R}{\mathbb{R}}
\newcommand{\jump}[1]{ \llbracket #1 \rrbracket }

\makeatother

\usepackage{babel}
\begin{document}

\title{Reflection and transmission of elastic waves in non-local band-gap
metamaterials: a comprehensive study via the relaxed micromorphic
model}

\author{Angela Madeo\thanks{Angela Madeo, \ \ Laboratoire de G\'{e}nie Civil et Ing\'{e}nierie Environnementale, SMS-ID,
Universit\'{e} de Lyon-INSA, B\^{a}timent Coulomb, 69621 Villeurbanne
Cedex, France; and member of ``Institut Universitaire de France'', 
email: angela.madeo@insa-lyon.fr} \quad{}and \quad{}Patrizio Neff\thanks{Patrizio Neff, \ \  Head of Lehrstuhl f\"{u}r Nichtlineare Analysis und Modellierung, Fakult\"{a}t f\"{u}r
Mathematik, Universit\"{a}t Duisburg-Essen,  Thea-Leymann Str. 9, 45127 Essen, Germany, email: patrizio.neff@uni-due.de} \quad{}and \quad{}Ionel-Dumitrel Ghiba\thanks{Ionel-Dumitrel Ghiba, \ \ \ \  Lehrstuhl f\"{u}r Nichtlineare Analysis und Modellierung, Fakult\"{a}t f\"{u}r Mathematik,
Universit\"{a}t Duisburg-Essen, Thea-Leymann Str. 9, 45127 Essen, Germany; and  Alexandru Ioan Cuza University of Ia\c si, Department of Mathematics,  Blvd. email: dumitrel.ghiba@uaic.ro} \quad{}and \quad{}Giuseppe Rosi\thanks{Giussepe Rosi,\ \ Laboratoire Mod\'elisation Multi-Echelle, MSME UMR
8208 CNRS, Universit\'e Paris-Est, 61 Avenue du General de Gaulle, Creteil
Cedex 94010, France; email: giuseppe.rosi@u-pec.fr } }
\maketitle
\begin{abstract}
\textcolor{black}{In this paper we propose to study wave propagation,
transmission and reflection in band-gap mechanical metamaterials via
the relaxed micromorphic model. To do so, guided by a suitable variational
procedure, we start deriving the jump duality conditions to be imposed
at surfaces of discontinuity of the material properties in non-dissipative,
linear-elastic, isotropic, relaxed micromorphic media. Jump conditions
to be imposed at surfaces of discontinuity embedded in Cauchy and
Mindlin continua are also presented as a result of the application
of a similar variational procedure. The introduced theoretical framework
subsequently allows the transparent set-up of different types of micro-macro
connections granting the description of both i) internal connexions
at material discontinuity surfaces embedded in the considered continua
and, as a particular case, ii) possible connections between different
(Cauchy, Mindlin or relaxed micromorphic) continua. The established
theoretical framework is general enough to be used for the description
of a wealth of different physical situations and can be used as reference
for further studies involving the need of suitably connecting different
continua in view of (meta-)structural design. In the second part of
the paper, we focus our attention on the case of an interface between
a classical Cauchy continuum on one side and a relaxed micromorphic
one on the other side in order to perform explicit numerical simulations
of wave reflection and transmission. This particular choice is descriptive
of a specific physical situation in which a classical material is
connected to a phononic crystal. The reflective properties of this
particular interface are numerically investigated for different types
of possible micro-macro connections, so explicitly showing the effect
of different boundary conditions on the phenomena of reflection and
transmission. Finally, the case of the connection between a Cauchy
continuum and a Mindlin one is also presented as a numerical study,
so showing that band-gap description is not possible for such continua,
in strong contrast with the relaxed micromorphic case.}

\bigskip{}

\noindent \textbf{{Key words:}} micromorphic elasticity, dynamic
problem, wave propagation, band-gap phenomena, interface, reflection,
transmission.

\bigskip{}

\noindent {\textbf{AMS 2010 subject classification}: } 74A10 (stress),
74A30 (nonsimple materials), 74A35 (polar materials), 74A60 (micromechanical
theories), 74B05 (classical linear elasticity), 74M25 (micromechanics),
74Q15 (effective constitutive equations), 74J05 (Linear waves), 74A50
(Structured surfaces and interfaces, coexistent phases) 
\end{abstract}
\newpage{}

\title{\tableofcontents{}}

\newpage{}

\section{Introduction\protect\protect\textsubscript{}}

\subsection{Band gap metamaterials and the relaxed micromorphic model}

Engineering metamaterials showing exotic behaviors with respect to
both mechanical and electromagnetic wave propagation are recently
attracting growing attention for their numerous possible astonishing
applications \cite{phononic1,phononic,phononic3,phononic2}. Actually,
materials which are able to ``stop\textquotedblright{} or ``bend\textquotedblright{}
the propagation of waves of light or sound with no energetic cost
could suddenly disclose rapid and extremely innovative technological
advancements.

In this paper, we focus our attention on those metamaterials which
are able to ``stop\textquotedblright{} wave propagation, i.e. metamaterials
in which waves within precise frequency ranges cannot propagate. Such
frequency intervals at which wave inhibition occurs are known as frequency
band-gaps and their intrinsic characteristics (characteristic values
of the gap frequency, extension of the band-gap, etc.) strongly depend
on the metamaterial microstructure. Such unorthodox dynamical behavior
can be related to two main different phenomena occurring at the micro-level: 
\begin{itemize}
\item local resonance phenomena (Mie resonance): the micro-structural components,
excited at particular frequencies, start oscillating independently
of the matrix thus capturing the energy of the propagating wave which
remains confined at the level of the microstructure. Macroscopic wave
propagation thus results to be inhibited. 
\item micro-diffusion phenomena (Bragg scattering): when the propagating
wave has wavelengths which are small enough to start interacting with
the microstructure of the material, reflection and transmission phenomena
occur at the micro-level that globally result in an inhibited macroscopic
wave propagation. 
\end{itemize}
Such resonance and micro-diffusion mechanisms (usually a mix of the
two) are at the basis of both electromagnetic and elastic band-gaps
(see e.g. \cite{phononic1}) and they are manifestly related to the
particular microstructural topologies of the considered metamaterials.
Indeed, it is well known (see e.g. \cite{phononic1,phononic2,phononicTorq})
that the characteristics of the microstructures strongly influence
the macroscopic band gap behavior. In this paper, we will be concerned
with mechanical waves, even if some of the used theoretical tools
can be thought to be suitably generalized for the modeling of electromagnetic
waves as well. Such generalizations could open new long-term research
directions, for example in view of the modeling of so called ``phoxonic
crystals\textquotedblright{} which are simultaneously able to stop
both electromagnetic and elastic wave propagation \cite{phononic1,phononic2}.

\smallskip{}

In recent contributions \cite{BandGaps1,BandGaps2} we proposed a
new generalized continuum model, which we called \textit{relaxed micromorphic}
which is able to account for the onset of microstructure-related frequency
band-gaps while remaining in the macroscopic framework of continuum
mechanics. Well posedness results have already been proved for this
model \cite{Ghiba,Ghiba1}. It turns out that the relaxed micromorphic
model is the only macroscopic continuum model known to date which
is simultaneously able to account for 
\begin{itemize}
\item the onset and prediction of complete band-gaps in metamaterials 
\item the possibility of non-local effects via the introduction of higher
order derivatives of the micro distortion tensor in the strain energy
density. 
\end{itemize}
Effective numerical homogenization methods \cite{Kutznetzova} have
been recently introduced which allow to account for frequency band
gap at the homogenized level. Such methods make use of a ``separation
of scales hypothesis'' which basically implies that the vibrations
of the microstructural elements remain confined in the considered
unit cells. Such hypothesis intrinsically leads, at the homogenized
level, to generalized models, sometimes called internal variable models,
in which no space derivatives of the introduced internal variable
appear. In other words, non-local effects cannot be accounted for
at the homogenized level.

Our relaxed micromorphic model allows to account for the possibility
of non-local effects in band-gap metamaterials and contains the internal
variable model as a degenerate limit case when suitably setting the
characteristic length to be zero. Even if the internal variable model
can be considered to be an acceptable tool for studying the behavior
of a certain sub-class of band gap metamaterials, the fact of neglecting
\textit{a priori} any non-locality might be hazardous since microstructered
materials are intrinsically expected to exhibit non-local behaviors
when subjected to particular loading and/or boundary conditions \cite{NonLoc}.
\smallskip{}

\textcolor{black}{The main interest of using macroscopic theories
for the description of the behavior of materials with microstructures
can be found in the fact that they feature the introduction of few
parameters which are, in an averaged sense, reminiscent of the presence
of an underlying microstructure. If, on the one hand, this fact provides
a drastic modeling simplification which is optimal to proceed towards
(meta-)structural design, some drawbacks can be reported which are
mainly related to the difficulty of directly relating the introduced
macroscopic parameters to the specific characteristics of the microstructure
(topology, microstructural mechanical properties, etc.). In order
to account in detail for the effect of the underlying microstructures
on the overall mechanical properties of the material at the homogenized
level, enhanced homogenization techniques, including higher order
terms in the performed expansions of the micro-fields, may be used
\cite{Hui1,Hui2,Dontsov1,guzina2}.}

\textcolor{black}{The aforementioned difficulty of explicitly relating
macro-parameters to micro-properties is often seen as a limit for
the effective application of enriched continuum models. As a matter
of fact, it is the authors' belief that such models are a necessary
step if one wants to proceed towards the engineering design of metastructures,
i.e. structures which are made up of metamaterials as building blocks.
Of course, the proposed model will introduce a certain degree of simplification,
but it is exactly this simplicity that makes possible to envision
the next step which is that of proceeding towards the design of complex
structures made of metamaterials.}

\textcolor{black}{To be more precise, the relaxed micromorphic model
proposed here, is able to describe the onset of the first (and sometimes
the second) band-gap which occurs at lower frequencies. In order to
catch more complex behaviors the kinematics and the constitutive relations
of the proposed model should be further enriched in a way that is
not yet completely clear. Nevertheless, we do not see this fact as
a limitation since we intend to use the unorthodox dynamical behavior
of some metamaterials exhibiting band-gaps to fit, by inverse approach,
the parameters of the relaxed micromorphic model following what has
been done e.g. in \cite{NonLoc}. This fitting, when successfully
concluded for some specific metamaterials will allow the setting up
of the design of metastructures by means of tools which are familiar
to engineers, such as Finite Element codes. Of course, as classical
Cauchy models show their limits for the description of the dynamical
behavior of metamaterials, even at low frequencies, the relaxed micromorphic
model will show its limit for higher frequencies, yet remaining accurate
enough for accounting for some macroscopic manifestations of microstructure.
To proceed in this direction, we intended to use the simplest possible
model (linear, elastic) which is able to account for the wanted phenomena
(band-gap onset and description). This allows for the introduction
of few extra parameters that may be calibrated on the basis of suitable
experimental or numerical ``discrete'' simulations. This is sufficient
when remaining in the linear-elastic framework which is the target
of the present paper. Generalized continuum models of the micromorphic
type featuring the description of band-gaps when introducing non-linearities
in the micro-inertia terms can also be found in the literature \cite{Gonella_Micro_inerzia},
but the interpretation of the introduced non-linearities would be
more complex to be undertaken.}

\subsection{\textcolor{black}{Boundary conditions and reflection and transmission
in metamaterials}}

\textcolor{black}{The problem of studying boundary conditions to be
imposed at surfaces of discontinuity of the material properties in
Cauchy continua is classically studied e.g. in \cite{Achenbach}  and
comes back to the fact of assigning jumps of forces and/or jumps of
displacements at the considered interfaces.}

\textcolor{black}{As far as generalized continua are concerned (micromorphic,
micropolar, second gradient or also porous media), the setting up
of correct boundary conditions becomes more delicate and can be successfully
achieved using suitable variational procedures \cite{BandGaps1,Guyader,FdIPlacidi,BandGaps2,damage,Sciarra1,Sciarra2,Seppecher,Gavrilyuk,EdgeIsolaSepp,FdISepp,TheseSeppecher}.}

\textcolor{black}{In this paper, basing ourselves upon an appropriate
variational procedure, we obtain the jump conditions that have to
be imposed at internal surfaces of discontinuity of the material properties
in relaxed micromorphic continua. The considered surfaces do not posses
their own material properties (mass, inertia, etc.), but such generalization
could be easily achieved using the methods presented in \cite{Placidi,SteigmannOgden}.
As a result of the use of similar variational arguments, we also present
the analogous jump conditions that have to be imposed at internal
surfaces of discontinuity in Cauchy and Mindlin continua. On the basis
of the introduced sets of jump conditions we are able to establish}
\begin{itemize}
\item \textcolor{black}{different types of (internal) micro-macro connections
which are possible between two relaxed micromorphic media, as well
as between two Cauchy or two Mindlin continua}
\item \textcolor{black}{As a particular case of the preceding point, we
are able to deduce different possible types of connections between
different continua (Cauchy, relaxed micromorphic or Mindlin). }
\end{itemize}
\textcolor{black}{The general theoretical framework introduced in
the present paper allows to deal with a wealth of different boundary
conditions which may be of use for the description of different physical
situations corresponding to specific connections between metamaterials
or between classical materials and metamaterials. In the second part
of the paper, we decide to focus our attention on the case of the
different possible connections between a Classical Cauchy medium and
a relaxed micromorphic one, since, as shown in \cite{NonLoc}, it
is of use for the simulation of experiments of reflection and transmissions
at the interface between an aluminum plate and a phononic crystal
of the type proposed e.g. by \cite{Lucklum}. Notwithstanding the
focus given to this particular case for the implementation of the
proposed numerical simulations, the present paper is intended to be
a reference for all subsequent works concerning the correct setting
up of boundary conditions at internal interfaces in Cauchy, relaxed
micromorphic and Mindlin continua, as well as for the proper description
of connections at interfaces between different media.}

\textcolor{black}{\medskip{}
}

\textcolor{black}{It has been proven \cite{Ghiba,Ghiba1} that the
relaxed micromorphic continuum is a degenerate model, in the sense
that only the tangential part of the micro-distortion tensor field
$\PP$ can be assigned in order to have a well posed problem. This
is not the case in standard Mindlin's micromorphic model in which
all the 9 components of the micro-distortion tensor $\PP$ must be
assigned at the considered boundary.}

\textcolor{black}{Due to the complexity of the kinematics of micromorphic
media (3+9=12 degrees of freedom) a great variety of connections can
be envisaged at material surfaces of discontinuity in such continua.
The fact of establishing in a clear fashion all the possible jump
conditions that can be imposed in micromorphic media is a delicate
point which is rarely treated with the due care in the literature.
Indeed, as far as boundary conditions in standard micromorphic media
are concerned, Mindlin \cite{Mindlin} and Eringen \cite{EringenBook}
propose suitable jump conditions to be imposed at surfaces of discontinuities
of the material properties. To the sake of completeness, after introducing
the jump conditions that have to be used in relaxed micromorphic media,
we also present the analogous conditions for Mindlin and Cauchy continua,
so recovering classical results. }

\textcolor{black}{After having introduced the theoretical tools which
are needed to deal with interfaces in Cauchy, Mindlin and relaxed
micromorphic media, we focus on the study of reflection and transmission
of waves at Cauchy/relaxed and Cauchy/Mindlin interfaces, respectively.
We thus present numerical simulations showing the behavior of the
reflection coefficient as function of the frequency of the traveling
waves and for different types of boundary conditions (}\textit{\textcolor{black}{internal
clamp with fixed microstructure}}\textcolor{black}{{} and }\textit{\textcolor{black}{internal
clamp with free microstructure}}\textcolor{black}{). To the author's
knowledge, such studies of reflection and transmission properties
in the spirit of micromorphic modeling are not found in the literature.
Some results of a rather simplified 1D situation can be found in \cite{berezow}
where some expressions for reflection and transmission coefficients
are presented, but finally not exploited in the proposed numerical
simulations.}

\subsection{\textcolor{black}{Organization of the paper}}

\textcolor{black}{The paper is organized as follows }
\begin{itemize}
\item \textcolor{black}{In Section 2 we start recalling the equations of
motion and jump duality conditions that can be imposed at surfaces
of discontinuity of the material properties in classical Cauchy continua.
This is useful to suitably introduce the generalizations that occur
for discontinuity surfaces in relaxed micromorphic and Mindlin continua.
In fact, the equations of motions and associated jump duality conditions
are derived for both such generalized continua. At the end of the
section we particularize our findings to the case of an interface
between a Cauchy continuum and a relaxed micromorphic (or a Mindlin)
medium. }
\item \textcolor{black}{In Section 3 we explicitly set up a series of different
connections which are possible at surfaces of discontinuity in Cauchy,
relaxed micromorphic or Mindlin's media. All the introduced connections
are conceived in order to be compatible with the jump duality conditions
presented in Section 2. If, as expected, the possible constraints
in classical Cauchy media are the }\textit{\textcolor{black}{internal
clamp}}\textcolor{black}{, the }\textit{\textcolor{black}{free boundary}}\textcolor{black}{{}
and the }\textit{\textcolor{black}{fixed boundary}}\textcolor{black}{,
a great variety of more complex connections can be envisaged in micromorphic
media, due to the richer kinematics allowing for microstructural motions.
We present the complete list of all possible micro-macro connections
both in relaxed micromorphic and Mindlin's media, but we will focus
our attention on two of them, namely the }\textit{\textcolor{black}{internal
clamp with free microstructure}}\textcolor{black}{{} and the }\textit{\textcolor{black}{internal
clamp with fixed microstructure}}\textcolor{black}{. These two constraints
impose continuity of the displacement of the macroscopic matrix and
allow particular kinematics at the level of the microstructure. Such
two constraints will be used in the sequel to study wave reflection
and transmission at interfaces between classical Cauchy and relaxed
micromorphic (or Mindlin's) media. }
\item \textcolor{black}{In section 4 we derive the principle of conservation
of total energy $E$ for Cauchy, relaxed micromorphic and Mindlin's
media in the form $E_{,t}+\mathrm{Div}\,\HH=0$, where $\HH$ is the
energy flux vector. If the definition of the total energy $E=J+W$
is straightforward once the kinetic energy $J$ and potential energy
$W$ are introduced for the considered continuum, the computation
of the energy flux for generalized media is more elaborate. The explicit
form of the energy fluxes is established in terms of micro and macro
velocities and of stresses and hyper-stresses. }
\item \textcolor{black}{In section 5 we introduce what we will call }\textit{\textcolor{black}{plane
wave ansatz}}\textcolor{black}{{} in the remainder of the paper. This
hypothesis consists of assuming that all unknown fields of displacement
$\uu$ and micro-distortion tensor $\PP$ only depend on one scalar
space variable $x_{1}$ which will also coincide with the direction
of propagation of the considered waves. This hypothesis allows us
to rewrite the governing equations of the considered continua in a
simplified form. In particular, for relaxed and standard micromophic
media, we are able to obtain systems of uncoupled partial differential
equations for longitudinal and for transverse waves, as well as for
some other waves which are only connected to purely microstructural
deformation modes and which we call ``uncoupled waves''. The jump
conditions to be imposed at surfaces of discontinuity of the material
properties and the expressions of the energy fluxes are also particularized
to the considered 1D case. }
\item \textcolor{black}{In section 6 we study plane wave propagation in
semi-infinite Cauchy, relaxed micromorphic and standard Mindlin's
media. This allows to unveil the band gap behavior of the relaxed
micromorphic media in opposition to Cauchy and Mindlin ones. To present
the dispersion curves of the considered generalized models, we follow
a procedure similar to the one presented in \cite{Chen1,Chen2} where
dispersion curves for standard micromorphic media are provided. In
addition to the previously presented results, we provide extra arguments
concerning the asymptotic behavior of the dispersion curves which
are the main feature of the relaxed micromorphic model allowing for
band-gap description. This step concerning the study of bulk wave
propagation is mandatory for the future determination of the constitutive
parameters of our relaxed micromorphic model on real band-gap metamaterials.
We identify specific }cut-off frequencies and characteristic velocities
which are related both to the macro and micro material properties
of the considered generalized continua. We finally present the case
of an internal variable model as a degenerate limit case of our relaxed
model when setting $L_{c}=0$. This model does not allow for non-local
effects, but can be still thought to describe band-gap behaviors in
particular metamaterials as the ones considered in \cite{phononic,Kutznetzova,Geers1}.
Nevertheless, the fact that a clear singularity occurs (the solution
for the case $L_{c}=0$ is different from that which is found for
very small but non-vanishing $L_{c}$) indicates that the case $L_{c}=0$
could lead to imprecise results compared to the relaxed micromorphic
model with small $L_{c}$. 
\item In section 7 we study the phenomena of reflection and transmission
at the considered interfaces, generalizing the classical definitions
of reflection and transmission coefficients. We show that the reflective
properties of the relaxed micromorphic and of Mindlin's media are
drastically different, especially in the vicinity of the frequencies
for which band-gaps are likely to occur. We repeat this study for
two different constraints (\textit{internal clamp with free microstructure}
and \textit{interna}\textit{\textcolor{black}{l clamp with fixed microstructure}}\textcolor{black}{)
and we show that the constraints which are imposed on the microstructure
of the considered generalized media have a relevant effect on the
global reflective properties of the interfaces which are investigated.
In particular, we claim that by suitably varying the parameters of
the relaxed micromorphic model, the constraint of }\textit{\textcolor{black}{internal
clamp with free microstructure}}\textcolor{black}{{} can provide a second
band gap (additional to the one evidenced in the study of bulk propagation)
which is completely due to the presence of the interface. Even if
we do not explicitly present this case of ``double band-gap'' in
the present paper, we mention this possibility that also allows the
description of local resonances at the level of the microstructure
(see \cite{NonLoc}). }
\end{itemize}

\subsection{Notational agreement}

In this paper, we denote by $\R^{3\times3}$ the set of real $3\times3$
second order tensors, written with capital letters. We denote respectively
by $\cdot\:$, $:$ and $\left.\langle\cdot,\cdot\right.\rangle$
a simple and double contraction and the scalar product between two
tensors of any suitable order\footnote{For example, $(A\cdot v)_{i}=A_{ij}v_{j}$, $(A\cdot B)_{ik}=A_{ij}B_{jk}$,
$A:B=A_{ij}B_{ji}$, $(C\cdot B)_{ijk}=C_{ijp}B_{pk}$, $(C:B)_{i}=C_{ijp}B_{pj}$,
$\left.\langle v,w\right.\rangle=v\cdot w=v_{i}w_{i}$, $\left.\langle A,B\right.\rangle=A_{ij}B_{ij}$
etc.}. Everywhere we adopt the Einstein convention of sum over repeated
indices if not differently specified. The standard Euclidean scalar
product on $\R^{3\times3}$ is given by $\langle{X},{Y}\rangle_{\R^{3\times3}}=\tr({X\cdot Y^{T}})$,
and thus the Frobenius tensor norm is $\|{X}\|^{2}=\langle{X},{X}\rangle_{\R^{3\times3}}$.
In the following we omit the index $\R^{3},\R^{3\times3}$. The identity
tensor on $\R^{3\times3}$ will be denoted by $\id$, so that $\tr({X})=\langle{X},{\id}\rangle$. 

\medskip{}

We consider a body which occupies a bounded open set $B_{L}$ of the
three-dimensional Euclidian space $\R^{3}$ and assume that its boundary
$\partial B_{L}$ is a smooth surface of class $C^{2}$. An elastic
material fills the domain $B_{L}\subset\R^{3}$ and we refer the motion
of the body to rectangular axes $Ox_{i}$. We denote by $\Sigma$
any material surface embedded in $B_{L}.$ We also denote by, $n$
the outward unit normal to $\partial B_{L}$, or to the surface $\Sigma$
embedded in $B_{L}$.

In the following, given any field $a$ defined on the surface $\Sigma$
we will also set 
\begin{equation}
\jump{a}:=a^{+}-a^{-},\label{eq:Jump}
\end{equation}
which defines a measure of the jump of $a$ through the material surface
$\Sigma$, where 
\[
[\cdot]^{-}:=\lim\limits _{\footnotesize{\begin{array}{c}
x\in B_{L}^{-}\setminus\Sigma\\
\ x\rightarrow\Sigma
\end{array}}}\hspace*{0cm}[\cdot],\qquad[\cdot]^{+}:=\lim\limits _{\footnotesize{\begin{array}{c}
x\in B_{L}^{+}\setminus\Sigma\\
\ x\rightarrow\Sigma
\end{array}}}\hspace*{-0.2cm}[\cdot],
\]
and where we denoted by $B_{L}^{-}$ and $B_{L}^{+}$ the two subdomains
which results form dividing the domain $B_{L}$ through the surface
$\Sigma$.

The usual Lebesgue spaces of square integrable functions, vector or
tensor fields on $B_{L}$ with values in $\mathbb{R}$, $\mathbb{R}^{3}$
or $\mathbb{R}^{3\times3}$, respectively will be denoted by $L^{2}(B_{L})$.
Moreover, we introduce the standard Sobolev spaces 
\begin{align}
\begin{array}{ll}
{\rm H}^{1}(B_{L})=\{u\in L^{2}(B_{L})\,|\,{\rm grad}\,u\in L^{2}(B_{L})\}, & \|u\|_{{\rm H}^{1}(B_{L})}^{2}:=\|u\|_{L^{2}(B_{L})}^{2}+\|{\rm grad}\,u\|_{L^{2}(B_{L})}^{2}\,,\vspace{1.5mm}\\
{\rm H}({\rm curl};B_{L})=\{v\in L^{2}(B_{L})\,|\,{\rm curl}\,v\in L^{2}(B_{L})\}, & \|v\|_{{\rm H}({\rm curl};B_{L})}^{2}:=\|v\|_{L^{2}(B_{L})}^{2}+\|{\rm curl}\,v\|_{L^{2}(B_{L})}^{2}\,,\vspace{1.5mm}\\
{\rm H}({\rm div};B_{L})=\{v\in L^{2}(B_{L})\,|\,{\rm div}\,v\in L^{2}(B_{L})\}, & \|v\|_{{\rm H}({\rm div};B_{L})}^{2}:=\|v\|_{L^{2}(B_{L})}^{2}+\|{\rm div}\,v\|_{L^{2}(B_{L})}^{2}\,,
\end{array}
\end{align}
of functions $u$ or vector fields $v$, respectively.

For vector fields $v$ with components in ${\rm H}^{1}(B_{L})$, i.e.
$v=\left(v_{1},v_{2},v_{3}\right)^{T}\,,v_{i}\in{\rm H}^{1}(B_{L}),$
we define \break $\nabla\,v=\left((\nabla\,v_{1})^{T},(\nabla\,v_{2})^{T},(\nabla\,v_{3})^{T}\right)^{T}$,
while for tensor fields $P$ with rows in ${\rm H}({\rm curl}\,;B_{L})$,
resp. ${\rm H}({\rm div}\,;B_{L})$, i.e. \break $P=\left(P_{1}^{T},P_{2}^{T},P_{3}^{T}\right)$,
$P_{i}\in{\rm H}({\rm curl}\,;B_{L})$ resp. $P_{i}\in{\rm H}({\rm div}\,;B_{L})$
we define ${\rm Curl}\,P=\left(({\rm curl}\,P_{1})^{T},({\rm curl}\,P_{2})^{T},({\rm curl}\,P_{3})^{T}\right)^{T},$
${\rm Div}\,P=\left({\rm div}\,P_{1},{\rm div}\,P_{2},{\rm div}\,P_{3}\right)^{T}.$
The corresponding Sobolev-spaces will be denoted by 
\[
{\rm H}^{1}(B_{L}),\qquad{\rm H}^{1}(\Div;B_{L}),\qquad{\rm H}^{1}(\Curl;B_{L}).
\]

\section{\label{sec:Duality_Conditions}Equations of motion and jump duality
conditions}

\textcolor{black}{In the present section, we present, as the result
of the application of a suitable variational principle, the bulk equations
and associated jump duality conditions that have to be verified at
internal interfaces in Cauchy, relaxed micromorphic and Mindlin continua,
respectively. Such preliminary theoretical framework is needed for
the subsequent introduction of }
\begin{itemize}
\item \textcolor{black}{different types of internal connections in Cauchy,
relaxed micromorphic or Mindlin continua (internal surfaces embedded
in such continua)}
\item \textcolor{black}{as a particular case of the previous point, connections
between different media (any combination of Cauchy, Mindlin and relaxed
micromorphic).}
\end{itemize}
\textcolor{black}{\medskip{}
}

\textcolor{black}{Let us consider a fixed time $t_{0}>0$ and a bounded
domain $B_{L}\subset\mathbb{R}^{3}$. We introduce the action functional
of the considered system to be defined as 
\begin{equation}
\mathcal{A}=\int_{0}^{t_{0}}\int_{B_{L}}\left(J-W\right)d\XX\,dt,\label{eq:Action}
\end{equation}
} where $J$ and $W$ are the kinetic and potential energies of the
considered system.

As for the kinetic energy, we consider that it takes the following
form for a Cauchy medium and a micromorphic (standard or relaxed)
medium respectively\footnote{Here and in the sequel we denote by the subscript $t$ the partial
derivative with respect to time of the considered field.} 
\begin{gather}
J=\frac{1}{2}\rho\left\Vert \uu_{,t}\right\Vert ^{2},\qquad J=\frac{1}{2}\rho\left\Vert \uu_{,t}\right\Vert ^{2}+\frac{1}{2}\eta\left\Vert \PP_{,t}\right\Vert ^{2},\label{eq:Kinetic_energy}
\end{gather}
where $\uu$ denotes the classical macroscopic displacement field
and $\PP$ is the micro-distortion tensor which accounts for independent
micro-motions at lower scales. On the other hand, as it will be explained
in the following, the strain energy density $W$ takes a different
form depending whether one considers a classical Cauchy medium, a
relaxed medium or a standard micromorphic one.\medskip{}

In what follows, we will explicitly derive, both for classical Cauchy
and micromorphic (standard and relaxed) media, the equations of motion
in strong form as well as the jump duality conditions which have to
be imposed at material discontinuity surfaces in such media and which
are intrinsically compatible with the least action principle associated
to the action functional (\ref{eq:Action}). Starting from the derived
duality conditions we will deduce different types of possible connections
between Cauchy and generalized media which are automatically compatible
with the associated bulk equations.

\subsection{Classical Cauchy medium}

In this Subsection, since it will be useful in the following, we recall
that the strain energy density for the classical, linear-elastic,
isotropic Cauchy medium takes the form 
\begin{align}
W & =\mu_{\rm macro}\left\Vert \,\mathrm{sym}\,\nabla\uu\,\right\Vert ^{2}+\frac{\lambda_{\rm macro}}{2}\left(\mathrm{tr}\left(\mathrm{sym}\,\nabla\uu\right)\right)^{2},\label{Pot-Cauchy}
\end{align}
where $\lambda_{\rm macro}$ and $\mu_{\rm macro}$ are the classical Lam\'e parameters and
$\uu$ denotes the classical macroscopic displacement field.

The associated equations of motion in strong form, obtained by a classical
least action principle take the usual form\footnote{Here and in the sequel we equivalently write, for the sake of completeness,
our equations both in compact and in index form.} 
\begin{align}
\rho\,\uu_{,tt} & =\mathrm{Div}\:\sis\,,\qquad\rho\,{u}_{i,tt}=\sigma_{ij,j},\label{eq:Motion_Cauchy}
\end{align}
where 
\begin{equation}
\sis\,=2\,\mu_{\rm macro}\,\mathrm{sym}\,\nabla\uu+\lambda_{\rm macro}\mathrm{tr}\left(\mathrm{sym}\,\nabla\uu\right)\mathds1,\qquad\sigma_{ij}=\mu_{\rm macro}\left(u_{i,j}+u_{j,i}\right)+\lambda_{\rm macro}\,u_{k,kj}\,\delta_{ij}\label{eq:constitutive_Cauchy}
\end{equation}
is the classical Cauchy stress tensor for isotropic materials.

\subsubsection{\label{sub:Connections1}Jump duality conditions in classical Cauchy
media}

In Cauchy continua, only force-displacement duality conditions are
possible, and take the form 
\begin{eqnarray}
\left\llbracket \left\langle \ff\,,\delta\uu\right\rangle \right\rrbracket =0,\qquad\left\llbracket f_{i}\,\delta u_{i}\right\rrbracket  & = & 0,\label{eq:JumpCauchy}
\end{eqnarray}
with 
\begin{eqnarray}
\ff\,=\sis\,\cdot\nn\,,\qquad f_{i} & = & \sigma_{ij}\:n_{j}.\label{eq:Cauchy_force}
\end{eqnarray}

\subsection{\label{sub:Relaxed-micromorphic-medium}Relaxed micromorphic medium}

The strain energy density for the relaxed medium is given by

\begin{alignat}{1}
W= & \,\mu_{e}\left\Vert \,\mathrm{sym}\left(\nabla\uu-\PP\right)\,\right\Vert ^{2}+\frac{\lambda_{e}}{2}\left(\mathrm{tr}\left(\nabla\uu-\PP\right)\right)^{2}+\mu_{c}\left\Vert \,\mathrm{skew}\left(\nabla\uu-\PP\right)\,\right\Vert ^{2}\vspace{1.2mm}\label{KinPot}\\
 & +\mu_{\rm micro}\left\Vert \,\mathrm{sym}\,\PP\,\right\Vert ^{2}+\frac{\lambda_{\rm micro}}{2}\left(\mathrm{tr}\,\PP\right)^{2}+\frac{\mu_{e}\,L_{c}^{2}}{2}\left\Vert \,\mathrm{\mathrm{Cur}l}\,\PP\,\right\Vert ^{2}.\nonumber 
\end{alignat}

Imposing the first variation of the action functional to be vanishing
(i. e. $\delta\mathcal{A}=0$), integrating by parts a suitable number
of times and the considering arbitrary variations $\delta\uu$ and
$\delta\PP$ of the basic kinematical fields, we obtain the strong
form of the bulk equations of motion of considered system (see also
\cite{Ghiba,BandGaps1,BandGaps2,Ghiba1}) which read 
\begin{align}
\rho\,\uu_{,tt} & =\mathrm{Div}\:\widetilde{\sis\,},\qquad\qquad\qquad\qquad\qquad\ \ \rho\,{u}_{i,tt}=\widetilde{\sigma}_{ij,j},\vspace{1.2mm}\label{eq:bulk-mod-3}\\
\vspace{1.2mm}\eta\,\PP_{,tt} & =\widetilde{\sis\,}-\gs\,-\mathrm{\mathrm{Cur}l}\:\mm\,,\qquad\qquad\qquad\eta\,{P}_{ij,t}=\widetilde{\sigma}_{ij}-s_{ij}-m_{ik,p}\epsilon_{jpk},\nonumber 
\end{align}
where $\epsilon_{jpk}$ is the Levi-Civita alternator and 
\begin{align}
\widetilde{\sis\,} & =2\,\mu_{e}\,\mathrm{sym}\left(\nabla\uu-\PP\right)+\lambda_{e}\mathrm{tr}\left(\nabla\uu-\PP\right)\mathds1+2\,\mu_{c}\,\mathrm{skew}\left(\nabla\uu-\PP\right),\vspace{1.2mm}\label{eq:Constitutive_relaxed}\\
\gs\, & =2\,\mu_{\rm micro}\,\mathrm{sym}\,\PP+\lambda_{\rm micro}\,\mathrm{tr}\,\PP\:\mathds1,\qquad\mm\,=\mu_{e}\,L_{c}^{2}\,\mathrm{\mathrm{Cur}l}\,\PP,\nonumber 
\end{align}
or equivalently, in index notation: 
\begin{align*}
\widetilde{\sigma}_{ij} & =\mu_{e}\left(u_{i,j}-P_{ij}+u_{j,i}-P_{ji}\right)+\lambda_{e}\left(u_{k,k}-P_{kk}\right)\delta_{ij}+\mu_{c}\left(u_{i,j}-P_{ij}-u_{j,i}+P_{ji}\right)\\
s_{ij} & =\mu_{\rm micro}\left(P_{ij}+P_{ji}\right)+\lambda_{\rm micro}\,P_{kk}\,\delta_{ij},\qquad m_{ik}=\mu_{e}\,L_{c}^{2}\,P_{ia,b}\,\epsilon_{kba}.
\end{align*}
Since it is useful for further calculations, we explicitly note that
the last term in the last equation (\ref{eq:bulk-mod-3}) can be rewritten
in terms of the basic kinematical fields as: 
\begin{flalign*}
\left(\mathrm{\mathrm{Cur}l}\:{m}\right)_{ij}= & \:m_{ik,p}\epsilon_{jpk}=\mu_{e}\,L_{c}^{2}\,P_{ia,bp}\epsilon_{kba}\epsilon_{jpk}=\mu_{e}\,L_{c}^{2}\,P_{ia,bp}\left(\delta_{bj}\delta_{ap}-\delta_{bp}\delta_{aj}\right).\\
= & \:\mu_{e}\,L_{c}^{2}\,\left(P_{ip,jp}-P_{ij,pp}\right)=\mu_{e}\,L_{c}^{2}\,\left(\mathrm{\mathrm{Cur}l}\,\mathrm{\mathrm{Cur}l}\,{P}\right)_{ij}.
\end{flalign*}
 We remark that as soon as in the last balance equation (\ref{eq:bulk-mod-3})
(which is in duality with $\delta P_{ij}$) one sets $j=p=1,2,3$,
then the relaxed term $m_{ik,p}\epsilon_{jpk}=\mu_{e}\,L_{c}^{2}\,\left(P_{ip,jp}-P_{ij,pp}\right)$
is identically vanishing. This means that in the considered relaxed
model, there are three terms which are vanishing compared to the standard
micromorphic one. This is equivalent to say that we are somehow lowering
the order of the considered differential system, so that, as we will
see, less boundary conditions will be necessary in order to have a
well-posed problem.

\subsubsection{\label{sub:Connections2}Jump duality conditions for the relaxed
medium}

Together with the bulk governing equations, the least action principle
simultaneously provides the duality jump conditions which can be imposed
at surfaces of discontinuity of the material properties in relaxed
media which reads 
\[
\ \left\llbracket \left\langle \gt\,,\delta\uu\right\rangle \right\rrbracket =0,\text{\qquad}\left\llbracket \left\langle \gtau\,,\,\delta\PP\right\rangle \right\rrbracket =0,
\]
or, equivalently, in index notation 
\begin{eqnarray}
\left\llbracket t_{i}\,\delta u_{i}\right\rrbracket  & = & 0,\qquad\left\llbracket \tau_{ij}\,\delta P_{ij}\right\rrbracket =0,\label{eq:JumpRelaxed}
\end{eqnarray}
with 
\begin{gather}
\gt\,=\widetilde{\sis\,}\cdot\nn\,,\qquad t_{i}=\widetilde{\sigma}_{ij}n_{j},\qquad\qquad\gtau=-\mm\,\cdot\geps\cdot\nn\,,\text{\qquad\ensuremath{\tau}}_{ij}=-m_{ik}\epsilon_{kjh}n_{h}.\label{eq:Force_DoubForce}
\end{gather}
It can be checked that, given the definition (\ref{eq:Force_DoubForce})
of the double forces $\tau_{ij}=-m_{ik}\epsilon_{kjh}n_{h}=\mu_{e}\,L_{c}^{2}\,\left(P_{ij,h}-P_{ih,j}\right)n_{h}$,
three out of the nine components of the double force $\gtau$ are
identically vanishing. For example, this check is immediate when choosing
the normal oriented along the $x_{1}$ axis: $\nn\,=(1,0,0)$, since
it is straightforward that $\tau_{11}=\tau_{21}=\tau_{31}=0$. This
fact implies that only six of the nine duality jump conditions $\left\llbracket \tau_{ij}\,\delta P_{ij}\right\rrbracket =0$
are actually independent so that one can claim that the set of boundary
conditions to be imposed in a relaxed model is actually underdetermined
with respect to the standard micromorphic model. This result is equivalent
to the one presented in \cite{Ghiba,Ghiba1} in which it is said that
only tangential boundary conditions on $\PP$ must be imposed in order
to prove existence and uniqueness for the relaxed problem.

\subsection{Standard micromorphic model}

The strain energy density for the standard Mindlin's micromorphic
model is 
\begin{align}
W=\: & \mu_{e}\left\Vert \,\mathrm{sym}\left(\nabla\uu-\PP\right)\,\right\Vert ^{2}+\frac{\lambda_{e}}{2}\left(\mathrm{tr}\left(\nabla\uu-\PP\right)\right)^{2}+\mu_{c}\left\Vert \,\mathrm{skew}\left(\nabla\uu-\PP\right)\,\right\Vert ^{2}\vspace{1.2mm}\label{KinPot-1-1}\\
 & +\mu_{\rm micro}\left\Vert \,\mathrm{sym}\,\PP\,\right\Vert ^{2}+\frac{\lambda_{\rm micro}}{2}\left(\mathrm{tr}\,\PP\right)^{2}+\frac{\mu_{e}\,L_{g}^{2}}{2}\left\Vert \nabla\,\PP\right\Vert ^{2}.\nonumber 
\end{align}
The equations of motion obtained by the associated least action principle
are 
\begin{align}
\rho\,\uu_{,tt} & =\mathrm{Div}\:\widetilde{\sis\,},\qquad\qquad\qquad\qquad\ \ \rho\,{u}_{i,tt}=\tilde{\sigma}_{ij,j},\vspace{1.2mm}\label{eq:bulk-mod-3-2-1}\\
\eta\,\PP_{,tt} & =\widetilde{\sis\,}-\gs\,+\mathrm{Div}\:\MM\,,\qquad\qquad\eta\,{P}_{ij,tt}=\tilde{\sigma}_{ij}-s_{ij}M_{ijk,k},\nonumber 
\end{align}
where 
\begin{gather*}
\MM\,=\mu_{e}\,L_{g}^{2}\,\mathrm{\mathrm{\nabla}}\PP,\qquad\qquad M_{ijk}=\mu_{e}\,L_{g}^{2}P_{ij,k}.
\end{gather*}

\subsubsection{\label{sub:Connections3}Jump duality conditions for the standard
micromorphic medium}

The jump duality conditions arising from the least action principle
read 
\[
\left\llbracket \:\left\langle \gt\,,\delta\uu\right\rangle \:\right\rrbracket =0,\text{\qquad}\left\llbracket \:\left\langle \tilde{\gtau}\,,\,\delta\PP\right\rangle \:\right\rrbracket =0,
\]
or, equivalently, in index notation 
\begin{eqnarray*}
\left\llbracket \:t_{i}\,\delta u_{i}\:\right\rrbracket  & = & 0,\qquad\left\llbracket \:\tilde{\tau}_{ij}\,\delta P_{ij}\:\right\rrbracket =0,
\end{eqnarray*}
where the force $\gt\,$ is the same as in the relaxed micromorphic
case, while the double-forces $\tilde{\gtau}$ are defined now as
\begin{equation}
\tilde{\tau}_{ij}=\mu_{e}\,L_{g}^{2}P_{ij,k}n_{k}.\label{eq:Doub_F_Class_microm}
\end{equation}

\subsection{Connections between a classical Cauchy medium and a relaxed micromorphic
(or a standard Mindlin's) medium}

Once that the possible jump duality conditions are established at
surfaces of discontinuity of the m\textcolor{black}{aterial properties
of all the introduced continua (Cauchy, relaxed micromorphic, standard
micromorphic), the subsequent step is to interconnect two different
continua by introducing suitable connections which are compatible
with such jump duality conditions. These connections can be envisaged
by suitably exploiting the jump duality conditions introduced in Sections
\ref{sub:Connections1}-\ref{sub:Connections3}. In particular, connections
between Cauchy/relaxed, relaxed/relaxed, Cauchy/standard-micromorphic,
standard-micromorphic/standard-micromorphic, relaxed/standard-micromorphic
media may be introduced as particular cases of the duality jump conditions
previously presented. Although we present here the general theoretical
framework for establishing all such connections, we postpone to further
investigations the interesting problem of studying wave reflection
and transmission in all these cases, limiting ourselves }to treat
here in detail the case of reflection and transmission of waves at
a surface of discontinuity between a classical Cauchy medium and a
relaxed micromorphic (or standard Mindlin) medium. We chose to focus
our attention on this particular case essentially for two reasons 
\begin{itemize}
\item In a classical Cauchy medium band gaps are not allowed, so that propagation
occurs for any real frequency. This is a good feature if one wants
to control the properties of the incident wave and to be sure to be
able to send a wave at the considered interface for any real frequency
value. 
\item The great majority of engineering materials can be modeled by means
of classical Cauchy continuum theories, so that in view of possible
applications (see \cite{Lucklum,NonLoc}) it is reasonable to start
with a standard material on one side of the discontinuity. 
\end{itemize}
Before studying wave reflection and transmission at a discontinuity
surface between a Cauchy and a relaxed micromorphic medium, we need
to explicitly set up the jump duality conditions that may be established
at such interfaces.

It is possible to check that the duality jump conditions (\ref{eq:JumpCauchy})
and (\ref{eq:JumpRelaxed}) for the Cauchy and relaxed micromorphic
media respectively, allow us to conclude that the following relations
must be verified at the interface between Cauchy and relaxed micromorphic
media 
\[
\left\langle \ff\,,\delta\uu^{-}\right\rangle -\left\langle \gt\,,\delta\uu^{+}\right\rangle =0,\qquad\qquad\left\langle \gtau\,,\,\delta\PP^{+}\right\rangle =0,
\]
or, equivalently, in index notation 
\begin{eqnarray}
f_{i}\,\delta u_{i}^{-}-t_{i}\,\delta u_{i}^{+}=0,\qquad\qquad\tau_{ij}\,\delta P_{ij}^{+}=0.\label{eq:JumpCauchy-1}
\end{eqnarray}
The case of an interface between a Cauchy and a Mindlin medium is
formally equivalent to the case treated here, except that the double
force $\gtau$ for a relaxed medium has 3 vanishing components out
of the 9, while in the standard Mindlin's medium it has all the 9
components which are non-vanishing. We will explain in more detail
this fact in the following sections.

\section{How to consider different connections between two generalized continua}

\textcolor{black}{If one thinks to classical structural mechanics,
it is immediate to understand that the same structural elements can
be interconnected using different constraints (for example, in beam
theory, one can deal with clamps, pivots, rollers, etc.). The reasoning
that we present in this sections is intended to establish analogous
results for generalized continua, in view of the effective modeling
of complex (meta-)structures.}

In this spirit, we present some considerations concerning the possible
choice of different boundary conditions to be imposed at surfaces
of discontinuity between two different generalized media. We start
by considering the case in which we have on the two sides the same
type of medium (Cauchy/Cauchy, relaxed/relaxed, standard/standard).
Such boundary conditions can be subsequently straightforwardly generalized
to the case in which two different media are considered on the two
sides. We start by presenting different connections between two Cauchy
continua, since they are classical and the reader will have an immediate
feeling of the physics which is involved. We will then generalize
such constraints to the case of relaxed media and standard micromorphic
media, trying to understand which are their intrinsic meanings with
a particular attention to their possible physical interpretation.

In other words, in this section we come back to the duality conditions
established in Section \ref{sec:Duality_Conditions} and we list all
the possible jump conditions that one can envisage and which satisfy
these duality conditions. We are going to provide a list of all the
possible sets of jump conditions which are intrinsically compatible
with the strong form of the consider\textcolor{black}{ed equations
of motion due to the fact that both (PDEs and BCs) are deduced from
the same variational principle.}

\textcolor{black}{We remind once again that at the end of the paper
we will be devoted to the numerical study of wave reflection and transmission
at an interface between a classical Cauchy continuum and a relaxed
micromorphic continuum since this can be of use to describe experimental
results of the type presented in \cite{Lucklum}. Nevertheless, for
the sake of completeness, we present here all the possible connections
between different types of continua (Cauchy, relaxed micromorphic
and standard micromorphic) in order to clearly establish the general
theoretical framework for further investigations and generalizations.}

\subsection{\textcolor{black}{\label{sub1}Connections between two Cauchy media}}

In this subsection, we present some well known types of internal constraints
between two Cauchy continua starting from the analysis of the duality
conditions (\ref{eq:JumpCauchy}). As it is classically established,
in order to verify the jump duality conditions (\ref{eq:JumpCauchy})
one can suitably impose displacements and/or forces, so giving rise
to different types of constraints. In the case of classical Cauchy
continua, when considering kinematical constraints, it is only possible
to impose conditions on displacement and not on its space derivatives
as instead happens for second gradient continua (or also for Euler-Bernouilli
beams). It is for this reason that only three types of constraints
can be analyzed in the case of Cauchy continua, namely the internal
clamp, the free boundary and the fixed boundary. When one can impose
continuity of the higher derivatives of displacement (as in second
gradient theory), then one can envisage more complex types of boundary
conditions such as internal hinges and internal rollers (see e.g.
\cite{FdIPlacidi}).

In the following Subsections, we will explicitly set up all the possible
sets of boundary conditions which can be imposed at an interface between
two classical Cauchy media. We will see that, independently of the
type of constraint, we always end up with $3+3=6$ scalar conditions
to be imposed at the aforementioned interface in order to have a well-posed
problem.

We finally recall that in a classical Cauchy medium the constitutive
expression of forces $\ff\,$ is given by Eq. (\ref{eq:Cauchy_force})
together with (\ref{eq:constitutive_Cauchy}).

\subsubsection{Internal clamp}

When imposing continuity of the three components of displacement between
the two sides of the considered discontinuity, one obtains what we
call an internal clamp. This constraint is equivalent to a continuity
constraint and can be used as a check of the used numerical code for
testing that an incident wave continues undisturbed to propagate when
reaching the interface at which the internal clamp is located. As
we said, the conditions to be imposed are given by the continuity
of displacement as follows 
\[
\left\llbracket u_{i}\right\rrbracket =0,\qquad i=1,2,3.
\]
Since the virtual displacements must be compatible with the imposed
boundary conditions, also the virtual displacements must verify such
jump conditions at the considered interface $\left\llbracket \delta u_{1}\right\rrbracket =\left\llbracket \delta u_{2}\right\rrbracket =\left\llbracket \delta u_{3}\right\rrbracket =0$.
Such conditions on the virtual displacements, together with the duality
conditions (\ref{eq:JumpCauchy}), imply for a generalized internal
clamp also the following jump conditions on force must be satisfied

\[
\left\llbracket f_{i}\right\rrbracket =0,\qquad i=1,2,3.
\]

\subsubsection{Free boundary}

We call free boundary that type of connection between the two Cauchy
media corresponding to which the displacements on the two sides of
the considered medium can be completely arbitrary on the two sides.
This corresponds to the physical situation in which the two media
are simply in contact, without any specific constraint. This implies
that also the virtual displacements $\delta u^{+}$ and $\delta u^{-}$
on the two sides of the interface can be completely arbitrary. This
means that, in order to have the duality jump conditions (\ref{eq:JumpCauchy})
to be satisfied, one necessarily has to have 
\[
f_{i}^{+}=f_{i}^{-}=0,\qquad i=1,2,3.
\]
In other words, when assigning arbitrary displacements on the two
sides, one necessarily has to have vanishing forces on the two sides
of the boundary in order to have a well posed problem.

\subsubsection{Fixed boundary}

We call fixed boundary that type of connection in which we set 
\[
u_{i}^{+}=u_{i}^{-}=0,\qquad i=1,2,3.
\]
Such connection corresponds to the physical situation in which the
two Cauchy media are fixed on the two sides of the considered surface.

\subsection{\label{sub2}Connections between two relaxed micromorphic media}

When considering micromorphic media (standard or relaxed) one can
impose more kinematical boundary conditions than in the case of Cauchy
continua. More precisely, one can act on the displacement field $\text{\ensuremath{\uu}}$
and also on the micro-distortion $\text{\ensuremath{\PP}}$. Clearly,
more options are possible with respect to the case of classical Cauchy
continua, so that we can introduce new types of constraints. In particular,
we will show that, for any type of possible connections between two
relaxed micromorphic media, we always have $3+3+6+6=18$ scalar conditions
to be imposed at the considered interface.

We recall that for a relaxed micromorphic medium the constitutive
expression for the force $\gt\,$ and double-force $\gtau$ are given
by Eq. (\ref{eq:Force_DoubForce}) together with (\ref{eq:Constitutive_relaxed}).

\subsubsection{\label{sub:1}Micro/macro internal clamp}

Generalizing what has been done previously, we impose continuity of
the macro-displacement and we also consider continuity of the micro-distortion
as follows 
\[
\left\llbracket u_{i}\right\rrbracket =0,\qquad\left\llbracket P_{ij}\right\rrbracket =0,\qquad i=1,2,3,\ \ j=2,3,
\]
where we remember that the directions $2,3$ are the directions tangent
to the considered surface. Imposing these $3+6=9$ kinematical continuity
conditions we are basically saying that there is no interruption either
in the macroscopic matrix and in the tangent part of the micro-distortion
tensor at the considered interface. Such conditions, together with
the jump duality conditions (\ref{eq:JumpRelaxed}) also imply that
the following conditions on forces and double-forces must be satisfied
\[
\left\llbracket t_{i}\right\rrbracket =0,\qquad\left\llbracket \tau_{ij}\right\rrbracket =0,\qquad i=1,2,3,\ \ j=2,3.
\]
As it has been pointed out before (see Section \ref{sub:Relaxed-micromorphic-medium}),
in a relaxed model only 6 out of the 9 jump conditions on double forces
are actually non-vanishing, so that we let here the subscript $j$
take only the values $2$ and $3$. This choice is coherent to that
of choosing the normal to the surface to take the particular form
$\nn\,=(1,0,0)$ which is what we will do in all the remainder of
this paper.

\subsubsection{\label{sub:2}Free boundary}

In this case the two media are simply in contact, with neither micro-
nor macro-connection across the interface. In this case both micro-distortion
and macro-displacement are arbitrary on both sides of the interface
which implies the following $18$ scalar conditions to be imposed
at the interface 
\[
t_{i}^{+}=t_{i}^{-}=0,\qquad\tau_{ij}^{+}=\tau_{ij}^{-}=0,\qquad i=1,2,3,\ \ j=2,3.
\]
Such conditions imply that all the micro and macro motions are left
arbitrary.

\subsubsection{\label{sub:3}Fixed boundary}

We call fixed relaxed boundary that situation in which 
\[
u_{i}^{+}=u_{i}^{-}=0,\qquad P_{ij}^{+}=P_{ij}^{-}=0,\qquad i=1,2,3,\ \ j=2,3.
\]
This case corresponds to the situation such that macroscopic matrix
and the tangent part of the micro-distortion are blocked on both sides
of the considered interface. Equivalently, this means that forces
and the tangent part of double forces may take arbitrary values on
the two sides of the interface.

\subsubsection{\label{sub:4}Macro internal clamp with free microstructure}

Another type of possible connection between two generalized media
is the case in which the macroscopic matrix is continuous, while the
microstructure is disconnected on the two sides. In formulas, this
case can be formulated by saying that $\delta P_{ij}$ are arbitrary
on the two sides, while the macro-displacements verify the kinematical
jump conditions 
\[
\left\llbracket u_{i}\right\rrbracket =0,\qquad i=1,2,3.
\]
Such kinematical conditions, together with the duality conditions
(\ref{eq:JumpRelaxed}) also imply that

\[
\left\llbracket t_{i}\right\rrbracket =0,\qquad i=1,2,3.
\]
The condition of free microstructure is finally given by the conditions
\[
\tau_{ij}^{+}=\tau_{ij}^{-}=0,\qquad i=1,2,3,\ \ j=2,3
\]
which, in order to respect the duality conditions (\ref{eq:JumpRelaxed}),
actually imply that the micro-distortions $\delta P_{ij}$ are left
arbitrary on the two sides.

\subsubsection{\label{sub:5}Macro internal clamp with fixed microstructure}

The macroscopic clamp is given by the continuity of displacement

\[
\left\llbracket u_{i}\right\rrbracket =0,\qquad i=1,2,3.
\]
Such kinematical conditions, together with the duality conditions
(\ref{eq:JumpRelaxed}) also imply that

\[
\left\llbracket t_{i}\right\rrbracket =0,\qquad i=1,2,3.
\]
The condition of fixed microstructure is given by 
\[
P_{ij}^{+}=P_{ij}^{-}=0,\qquad i=1,2,3,\ \ j=2,3,
\]
that means that the double-force is left arbitrary on the two sides
of the considered surface.

\subsubsection{\label{sub:6}Micro internal clamp with free macrostructure}

This type of constraint considers a connection between the two relaxed
media which is only made through the microstructure. In other words,
the microstructure is continuous across the interface, while the macroscopic
matrix is free to move independently on the two sides. In this case
the kinematical conditions to be imposed are 
\[
\left\llbracket P_{ij}\right\rrbracket =0,\qquad i=1,2,3,\ \ j=2,3.
\]
Such conditions together with the considered duality conditions imply
\[
\left\llbracket \tau_{ij}\right\rrbracket =0,\qquad i=1,2,3,\ \ j=2,3.
\]
The condition of free macro-displacement is instead implied by imposing
vanishing forces on the two sides as follows 
\[
t_{i}^{+}=t_{i}^{-}=0,\qquad i=1,2,3.
\]

\subsubsection{\label{sub:7}Micro internal clamp with fixed macrostructure}

The condition of fixed macrostructure is given by 
\[
u_{i}^{+}=u_{i}^{-}=0,\qquad i=1,2,3.
\]
In addition, we have continuity of the microstructure through the
condition

\[
\left\llbracket P_{ij}\right\rrbracket =0,\qquad i=1,2,3,\ \ j=2,3.
\]
The last jump condition also implies that 
\[
\left\llbracket \tau_{ij}\right\rrbracket =0,\qquad i=1,2,3,\ \ j=2,3.
\]

\subsubsection{\label{sub:8}Fixed macrostructure and free microstructure}

This connection is guaranteed by the relationships 
\[
u_{i}^{+}=u_{i}^{-}=0,\qquad\tau_{ij}^{+}=\tau_{ij}^{-}=0,\qquad i=1,2,3,\ \ j=2,3,
\]
which guarantees that the two sides of the interface remain fixed
and that, on the other hand, the microstructure is able to deform
arbitrarily on the two sides.

\subsubsection{\label{sub:9}Free macrostructure and fixed microstructure}

This is the converse situation with respect to the previous case,
i.e. the macrostructure can move arbitrarily, while the microstructure
cannot move on the two sides of the interface. Such connection is
guaranteed by the conditions 
\[
t_{i}^{+}=t_{i}^{-}=0,\qquad P_{ij}^{+}=P_{ij}^{-}=0,\qquad i=1,2,3,\ \ j=2,3.
\]

\subsection{Connections between two standard micromorphic media}

The types of connections which are possible between two standard micromorphic
media are formally the same as those presented for the relaxed micromorphic
medium. The fundamental difference is that, in this last case, all
the nine components of $\PP$ (and hence all the nine components of
the double-force $\tilde{\gtau}$) are independent, so that one has
to impose $24$ instead of $18$ scalar conditions at the interface.
Hence, to rigorously define the possible constraints between two standard
micromorphic media, it is sufficient to reformulate the jump conditions
given in the previous Subsection by setting now $\gtau=\tilde{\gtau}$
and $j=1,2,3$, instead of $j=2,3$ as in the previous case. This
means that, independently of the specific type of connection, one
must always count $3+3+9+9=24$ scalar conditions to be imposed at
an interface between two standard micromorphic media.

We recall that for a standard micromorphic medium the constitutive
expression for the force $\gt\,$ is the same as in the relaxed micromorphic
case (see Eq. (\ref{eq:Force_DoubForce}) together with (\ref{eq:Constitutive_relaxed}))
while the double-force $\tilde{\gtau}$ is redefined according to
Eq. (\ref{eq:Doub_F_Class_microm}).

\subsection{\label{sub:ConnectionsCauchyRelaxed}Connections between a Cauchy
and a relaxed micromorphic medium}

\textcolor{black}{The study of the possible connections between a
Cauchy and a relaxed micromorphic medium can be performed by suitably
particularizing the results obtained in subsections \ref{sub1} and
\ref{sub2}. In particular, when considering connections between a
Cauchy and a micromorphic medi}um (standard or relaxed) one can impose
more kinematical boundary conditions than in the case of connections
between Cauchy continua (see subsection \textcolor{blue}{\ref{sub1}}),
but less than in the case of connections between two micromorphic
media (see subsection \textcolor{blue}{\ref{sub2}}). More precisely,
one can act on the displacement field $\text{\ensuremath{\uu}}$ (on
both sides of the interface) and also on the micro-deformation $\text{\ensuremath{\PP}}$
(on the side of the interface occupied by the micromorphic continuum).
We will show that for any type of possible connections between a Cauchy
and a micromorphic medium, we always have $3+3+6=12$ scalar conditions
to be imposed at the interface. We recall that, in what follows, we
consider the ``-'' region occupied by the Cauchy continuum and the
``+'' region occupied by the relaxed micromorphic continuum, so
that, accordingly, we set 
\begin{eqnarray*}
f_{i} & = & \sigma_{ij}^{-}n_{j}^{-},\qquad t_{i}=\widetilde{\sigma}_{ij}^{+}\cdot n_{j}^{+},\qquad\tau_{ij}=-m_{ik}^{+}\epsilon_{kjh}n_{h}^{+}.
\end{eqnarray*}

We recall once again that in a classical Cauchy medium the constitutive
expression of forces $\ff\,$ is given by Eq. (\ref{eq:Cauchy_force})
together with (\ref{eq:constitutive_Cauchy}), while for a relaxed
micromorphic medium the constitutive expression for the force $\gt\,$
and double-force $\gtau$ are given by Eq. (\ref{eq:Force_DoubForce})
together with (\ref{eq:Constitutive_relaxed}).

\subsubsection{\label{sub:Macro-clamp-fixed_micro}Macro internal clamp with fixed
microstructure}

In the case in which we have a Cauchy continuum in contact with a
relaxed micromorphic medium the conditions given in Subsections \ref{sub:1}
and \ref{sub:5} collapse in the same set of condition which, in this
particular case, become 
\[
\left\llbracket u_{i}\right\rrbracket =0,\qquad t_{i}-f_{i}=0,\qquad P_{ij}^{+}=0,\qquad i=1,2,3,\ \ j=2,3.
\]
\begin{figure}[H]
\begin{centering}
\includegraphics[scale=0.2]{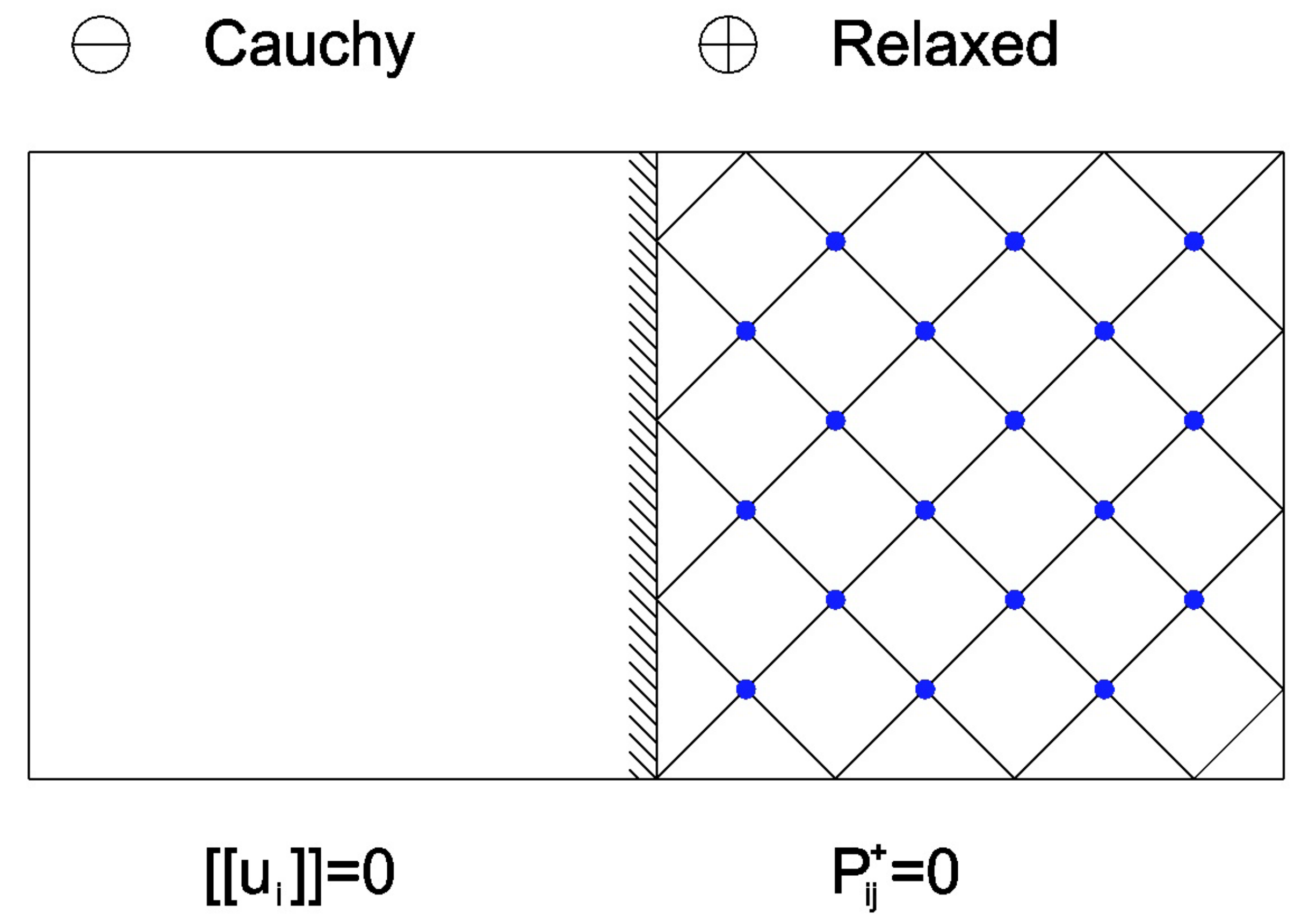} 
\par\end{centering}

\protect\caption{Macro internal clamp with fixed microstructure.}
\end{figure}

As we will see in the following, this will be one of the most representative
connections when studying wave reflection and transmission at an interface
between a Cauchy and a relaxed micromorphic continuum.

\subsubsection{Free boundary}

The conditions of free boundary between a Cauchy and a relaxed micromorphic
medium can be deduced particularizing the conditions given in Subsection
\ref{sub:2} and read 
\[
t_{i}=f_{i}=0,\qquad\tau_{ij}=0,\qquad i=1,2,3,\ \ j=2,3.
\]
Such conditions imply that all the micro and macro motions are left
arbitrary.

\subsubsection{Fixed boundary}

The conditions for a fixed boundary between a Cauchy and a relaxed
micromorphic medium can be obtained from the conditions presented
in Subsections \ref{sub:3} and \ref{sub:7} which, in the considered
particular case, collapse in the following set of conditions

\[
u_{i}^{+}=u_{i}^{-}=0,\qquad P_{ij}^{+}=0,\qquad i=1,2,3,\ \ j=2,3.
\]
This case corresponds to the situation where the macrostructure is
blocked on both sides of the considered interface, while the microstructure
is blocked on the side of the relaxed micromorphic medium.

\subsubsection{\label{sub:Macro-clamp-free_micro}Macro internal clamp with free
microstructure}

Another type of possible connection between two generalized media
is the case in which the macroscopic matrix is continuous, while the
microstructure of the relaxed medium is free to move at the interface.
This type of connection can be obtained suitably particularizing the
jump conditions given in Subsection \ref{sub:4}, which now become
\[
\left\llbracket u_{i}\right\rrbracket =0,\qquad t_{i}-f_{i}=0,\qquad\tau_{ij}=0,\qquad i=1,2,3,\ \ j=2,3.
\]
\begin{figure}[H]
\begin{centering}
\includegraphics[scale=0.2]{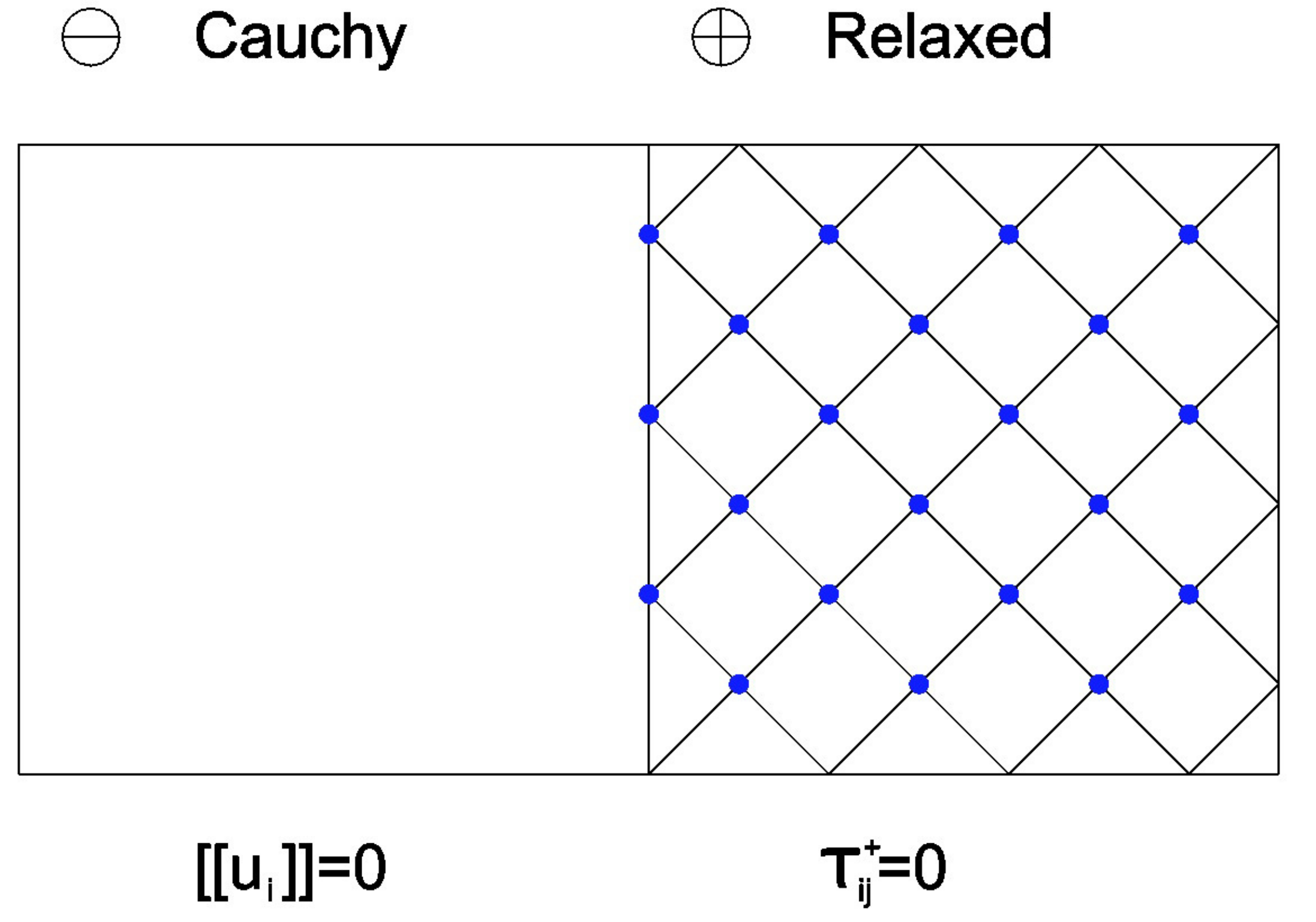} 
\par\end{centering}

\protect\caption{Macro internal clamp with free microstructure.}
\end{figure}

Also for this case, we will see that it is actually one of the more
interesting connections when studying wave reflection and transmission
at an interface between a Cauchy continuum and a relaxed micromorphic
one.

\subsubsection{Free macrostructure and fixed microstructure}

Such connection is a particular case of the more general conditions
\ref{sub:6} and \ref{sub:9}, which in this case collapse in the
same set of conditions, namely 
\[
t_{i}=f_{i}=0,\qquad P_{ij}^{+}=0,\qquad i=1,2,3,\ \ j=2,3.
\]

\subsubsection{Fixed macrostructure and free microstructure}

This connection is a particular case of the conditions given in Subsection
\ref{sub:8} and is provided by the relationships 
\[
u_{i}^{+}=u_{i}^{-}=0,\qquad\tau_{ij}=0,\qquad i=1,2,3,\ \ j=2,3.
\]
This connection guarantees that the two sides of the interface remain
fixed and that, on the other hand, the microstructure is able to deform
arbitrarily on the side occupied by the relaxed medium.

\section{Conservation of total energy}

It is known that if one considers conservative mechanical systems,
like in the present paper, then conservation of total energy must
be verified in the form 
\begin{equation}
E_{,t}+\mathrm{div}\:\HH=0,\label{EnnergyConservation}
\end{equation}
where $E=J+W$ is the total energy of the system and $\HH$ is the
energy flux vector. It is clear that the explicit expressions for
the total energy and for the energy flux are different depending on
whether one considers a classical Cauchy model, a relaxed micromorphic
model or a standard micromorphic model. If the expression of the total
energy $E$ is straightforward for the three mentioned cases (it suffices
to look at the given expressions of $J$ and $W$), the explicit expression
of the energy flux $\HH$ can be more complicated to be obtained.
For this reason we specify the expression of the energy fluxes for
the three cases in the following Subsections and we propose their
detailed deduction. 

\textcolor{black}{The explicit derivation of the energy fluxes for
the considered continua is a fundamental step towards the subsequent
definition of the reflection and transmission coefficients at the
interface between two (meta-)materials.}

\subsection{Classical Cauchy continuum}

In classical Cauchy continua the flux vector $\HH$ can be written
as 
\begin{equation}
\HH=-\sis\,\cdot\uu_{,t},\qquad\qquad H_{k}=-\sigma_{ik}{u}_{k,t},\label{Flux-Cauchy}
\end{equation}
where the Cauchy stress tensor $\sis\,$ has been defined in (\ref{eq:constitutive_Cauchy})
in terms of the displacement field.

Indeed, recalling expressions (\ref{eq:Kinetic_energy}) and (\ref{Pot-Cauchy})
for the kinetic and potential energy densities respectively, together
with definition (\ref{eq:constitutive_Cauchy}) for the stress tensor,
we can write 
\begin{align*}
E_{,t}&=J_{t}+W_{,t}=\rho\left\langle \uu_{,t},\uu_{,tt}\right\rangle +\left\langle \left(2\mu_{\rm macro}\,\mathrm{sym}\nabla\uu+\lambda_{\rm macro}\,\mathrm{tr}(\mathrm{sym}\nabla\uu)\mathds1\right),\mathrm{sym}\nabla\left(\uu_{,t}\right)\right\rangle \notag\\&=\rho\left\langle \uu_{,t},\uu_{,tt}\right\rangle +\left\langle \sis\,,\mathrm{sym}\nabla\left(\uu_{,t}\right)\right\rangle .
\end{align*}
Using the equations of motion (\ref{eq:Motion_Cauchy}) for replacing
the quantity $\rho\,u_{,tt}$, using the fact that $\left\langle \sis\,,\mathrm{sym}\nabla\left(\uu_{,t}\right)\right\rangle =\mathrm{Div}\left(\sis\cdot\uu_{,t}\right)-\mathrm{Div}\sis\,\cdot\uu_{,t}$
and simplifying we have: 
\begin{align*}
E_{,t}=\mathrm{Div}\sis\,\cdot\uu_{,t}+\mathrm{Div}\left(\sis\cdot\uu_{,t}\right)-\mathrm{Div}\sis\,\cdot\uu_{,t}=\mathrm{Div}\left(\sis\,\cdot\uu_{,t}\right).
\end{align*}
This last expression for the time derivative of the total energy $E$
yields expression (\ref{Flux-Cauchy}) for the energy flux $\text{\ensuremath{\HH}}$.

\subsection{Relaxed micromorphic model}

As for the relaxed micromorphic continuum, the energy flux vector
$\HH$ is defined as

\begin{equation}
\HH=-\widetilde{\sis\,}^{T}\cdot\uu_{,t}-(\mm\,^{T}\cdot\PP_{,t}):\geps,\qquad\qquad H_{k}=-{u}_{i,t}\widetilde{\sigma}_{ik}-m_{ih}\,{P}_{ij,t}\,\epsilon_{jhk},\label{Flux}
\end{equation}
where the stress tensor $\widetilde{\sis\,}$ and the hyper-stress
tensor $\mm\,$ have been defined in (\ref{eq:Constitutive_relaxed})
in terms of the basic kinematical fields.

To prove that in relaxed micromorphic media the energy flux takes
the form (\ref{Flux}), we start noticing that, using eqs. (\ref{eq:Kinetic_energy})
and (\ref{KinPot}), the time derivative of the total energy $E$
can be computed as 
\begin{align*}
E_{,t}= & \:\rho\left\langle \uu_{,t},\uu_{,tt}\right\rangle +\eta\left\langle \PP_{,t},\PP_{,tt}\right\rangle +\left\langle 2\,\mu_{e}\,\mathrm{sym}\left(\nabla\uu-\PP\right),\,\mathrm{sym}\left(\nabla\uu_{,t}-\PP_{,t}\right)\right\rangle \vspace{1.2mm}\\
 & +\left\langle \lambda_{e}\mathrm{tr}\left(\nabla\uu-\PP\right)\mathds1,\,\nabla\uu_{,t}-\PP_{,t}\right\rangle +\left\langle 2\,\mu_{c}\,\mathrm{skew}\left(\nabla\uu-\PP\right),\,\mathrm{skew}\left(\nabla\uu_{,t}-\PP_{,t}\right)\right\rangle \vspace{1.2mm}\\
 & +\left\langle 2\,\mu_{\rm micro}\:\mathrm{sym}\,\PP,\mathrm{sym}\,\PP_{,t}\right\rangle +\left\langle \lambda_{\rm micro}\,(\mathrm{tr}\,\PP)\,\mathds1,\,\PP_{,t}\right\rangle +\left\langle \mu_{e}L_{c}^{2}\,\mathrm{Curl}\,\PP,\,\mathrm{Curl}\,\PP_{,t}\right\rangle ,
\end{align*}
or equivalently, using definitions (\ref{eq:Constitutive_relaxed})
for $\widetilde{\sis\,}$, $\gs\,$ and $\mm\,$ 
\begin{align*}
E_{,t}=\: & \rho\left\langle \uu_{,t},\uu_{,tt}\right\rangle +\eta\left\langle \PP_{,t},\PP_{,tt}\right\rangle +\left\langle 2\,\mu_{e}\,\mathrm{sym}\left(\nabla\uu-\PP\right)+\lambda_{e}\,\mathrm{tr}\left(\nabla\uu-\PP\right)\mathds1+2\,\mu_{c}\,\mathrm{skew}\left(\nabla\uu-\PP\right),\,\nabla\uu_{,t}-\PP_{,t}\right\rangle \vspace{1.2mm}\\
 & +\left\langle 2\,\mu_{\rm micro}\,\mathrm{sym}\,\PP+\lambda_{\rm micro}\:(\mathrm{tr}\,\PP)\mathds1,\,\PP_{,t}\right\rangle \vspace{1.2mm}\\
=\: & \rho\left\langle \uu_{,t},\uu_{,tt}\right\rangle +\eta\left\langle \PP_{,t},\PP_{,tt}\right\rangle +\left\langle \widetilde{\sis\,},\nabla\uu_{,t}\right\rangle -\left\langle \widetilde{\sis\,}-\gs\,,\PP_{,t}\right\rangle +\left\langle \mm\,,\,\mathrm{Curl}\,\PP_{,t}\right\rangle .
\end{align*}
Using now the equations of motion (\ref{eq:bulk-mod-3}) to replace
the quantities $\rho\,u_{,tt}$ and $\eta\,P_{,tt}$, recalling that
$\left\langle \mm\,,\,\mathrm{Curl}\,\PP_{,t}\right\rangle =\mathrm{Div}\left(\,\left(\mm\,^{T}\cdot\PP_{,t}\right):\geps\,\right)-\left\langle \mathrm{Curl}\,\mm\,,\,\PP_{,t}\right\rangle $,
manipulating and simplifying, it can be recognized that 
\begin{align*}
E_{,t} & =\left\langle \uu_{,t},\mathrm{Div}\sis\,\right\rangle +\left\langle \PP_{,t},\widetilde{\sis\,}-\gs\,-\mathrm{Curl}\,\mm\,\right\rangle +\mathrm{Div}\left(\uu_{,t}\cdot\widetilde{\sis\,}\right)-\left\langle \uu_{,t},\mathrm{Div}\widetilde{\sis\,}\right\rangle -\left\langle \widetilde{\sis\,}-\gs\,,\PP_{,t}\right\rangle +\left\langle \mm\,,\,\mathrm{Curl}\,\PP_{,t}\right\rangle \vspace{1.2mm}\\
 & =\mathrm{Div}\left(\widetilde{\sis\,}^{T}\cdot\uu_{,t}\right)-\left\langle \mathrm{Curl}\,\mm\,,\,\PP_{,t}\right\rangle +\mathrm{Div}\left(\,\left(\mm\,^{T}\cdot\PP_{,t}\right):\geps\,\right)+\left\langle \mathrm{Curl}\,\mm\,,\,\PP_{,t}\right\rangle \vspace{1.2mm}\\
 & =\mathrm{Div}\left(\,\widetilde{\sis\,}^{T}\cdot\uu_{,t}+\left(\mm\,^{T}\cdot\PP_{,t}\right):\geps\,\right).
\end{align*}

\subsection{Standard micromorphic model}

Concerning the standard micromorphic model, the energy flux $\HH$
can be introduced as

\begin{equation}
\HH=-\widetilde{\sis\,}^{T}\cdot\uu_{,t}-\PP_{,t}^{T}:\MM\,,\qquad\qquad H_{k}=-{u}_{i,t}\widetilde{\sigma}_{ik}-{P}_{ij,t}M_{ijk}.\label{Flux-Mindlin}
\end{equation}
Analogously to what done for the relaxed micromorphic model we can
compute the first time derivative of the total energy as follows 
\begin{align*}
E_{,t} & =\rho\left\langle \uu_{,t},\uu_{,tt}\right\rangle +\eta\left\langle \PP_{,t},\PP_{,tt}\right\rangle +\left\langle \widetilde{\sis\,},\nabla\uu_{,t}\right\rangle -\left\langle \widetilde{\sis\,}-\gs\,,\PP_{,t}\right\rangle +\left\langle \MM\,,\,\nabla\PP_{,t}\right\rangle \vspace{1.2mm}\\
 & =\left\langle \uu_{,t},\mathrm{Div}\widetilde{\sis\,}\right\rangle +\left\langle \PP_{,t},\widetilde{\sis\,}-\gs\,+\mathrm{Div}\,\MM\,\right\rangle +\mathrm{Div}\left(\widetilde{\sis\,}^{T}\cdot\uu_{,t}\right)-\left\langle \uu_{,t},\mathrm{Div}\widetilde{\sis\,}\right\rangle -\left\langle \widetilde{\sis\,}-\gs\,,\PP_{,t}\right\rangle +\left\langle \MM\,,\,\mathrm{\nabla}\PP_{,t}\right\rangle \vspace{1.2mm}\\
 & =\mathrm{Div}\left(\widetilde{\sis\,}^{T}\cdot\uu_{,t}\right)+\left\langle \mathrm{Div}\,\MM\,,\,\PP_{,t}\right\rangle +\mathrm{Div}\left(\PP_{,t}^{T}:\MM\,\right)-\left\langle \mathrm{Div}\,\MM\,,\,\PP_{,t}\right\rangle =\mathrm{Div}\left(\uu_{,t}\cdot\widetilde{\sis\,}+\PP_{,t}^{T}:\MM\,\right),
\end{align*}
where the equations of motion have been used to replace the quantities
to replace the quantities $\rho\,u_{,tt}$ and $\eta\,P_{,tt}$, together
with the identity $\left\langle \MM\,,\,\mathrm{\nabla}\PP_{,t}\right\rangle =\mathrm{Div}\left(\PP_{,t}^{T}:\MM\,\right)-\left\langle \mathrm{Div}\,\MM\,,\,\PP_{,t}\right\rangle $.
The obtained expression for the time derivative of the total energy
$E_{,t}$ proves the expression (\ref{Flux-Mindlin}) for the energy
flux in a standard micromorphic medium.

\section{Plane wave ansatz: simplification of the governing equations and
boundary conditions}

\textcolor{black}{In order to proceed towards the numerical exploitation
of some of the theoretical results presented in the first part of
the paper and in view of suitable applications to cases of real interest
\cite{Lucklum,NonLoc}, we particularize to the case of plane waves
the previously derived equilibrium equations and associated }jump
conditions for the relaxed micromorphic, the standard micromorphic
and the Cauchy continua. In what follows we call ``plane wave ansatz''
the hypothesis according to which the unknown fields ($\uu$ and $\PP$)
are supposed to depend only on one component $x_{1}$ of the space
variable $\text{\ensuremath{\XX\,}}=(x_{1},x_{2},x_{3})$ which is
also supposed to be the direction of propagation of the considered
waves. We will also refer to such simplified framework as ``1D case''
if no confusion can arise.

The simplified partial differential equations obtained by means of
this hypothesis will be solved using different types of possible jump
conditions in order to study reflection and transmission of plane
waves at a Cauchy-relaxed interface and at a Cauchy-standard-micromorphic
interface. Since it will be needed to determine the reflection and
transmission coefficients, we also derive the particularized 1D form
of the energy flux $\HH$ in all such media.

\subsection{Classical Cauchy medium}

We briefly recall that in the case of plane waves (the displacement
field $\uu=(u_{1},u_{2},u_{3}$) is supposed to depend only on the
first component $x_{1}$ of the space variable $\XX=(x_{1},x_{2},x_{3})$),
the equations of motion in strong form take the form,

\begin{gather}
{u}_{1,tt}=\frac{\lambda_{\rm macro}+2\mu_{\rm macro}}{\rho}\,u_{1,11},\qquad{u}_{2,tt}=\frac{\mu_{\rm macro}}{\rho}\,u_{2,11},\qquad{u}_{3,tt}=\frac{\mu_{\rm macro}}{\rho}\,u_{3,11}.\label{CauchyAS}
\end{gather}
It can be noticed that the first equation only involves the longitudinal
displacement, while the last two equations involve transverse displacements
in the $2$ and $3$ directions.

\subsubsection{Jump conditions in the 1D case}

Considering the unit normal to be directed along the $x_{1}$ axis,
the jump duality conditions (\ref{eq:JumpCauchy}) simplify into 
\begin{eqnarray*}
\left\llbracket f_{1}\,\delta u_{1}\right\rrbracket =0,\qquad\left\llbracket f_{2}\,\delta u_{2}\right\rrbracket =0,\qquad\left\llbracket f_{3}\,\delta u_{3}\right\rrbracket =0
\end{eqnarray*}
with 
\[
f_{1}=(\lambda_{\rm macro}+2\mu_{\rm macro})\,u_{1,1},\quad f_{2}=\mu_{\rm macro}\,u_{2,1},\quad f_{3}=\mu_{\rm macro}\,u_{3,1}.
\]

\subsubsection{Energy flux for the Cauchy model in the 1D case}

The first component of the energy flux vector given in Eq. (\ref{Flux-Cauchy}),
simplifies in the 1D case into 
\begin{equation}
H_{1}=-{u}_{1,t}\left[\left(\lambda_{\rm macro}+2\mu_{\rm macro}\right)\:u_{1,1}\right]-{u}_{2,t}\left[\mu_{\rm macro}\,u_{2,1}\right]-{u}_{3,t}\left[\mu_{\rm macro}\,u_{3,1}\right].\label{Flux-Cauchy-1}
\end{equation}

\subsection{Relaxed micromorphic continuum}

We first introduce the decomposition of $\PP$ such that $P_{ij}=P_{ij}^{D}+P^{S}\delta_{ij}$
and consider the new variables 
\begin{equation}
P^{S}=\frac{1}{3}P_{kk},\qquad P_{ij}^{D}=P_{ij}-P^{S}\delta_{ij}=({\rm dev\,}\,P)_{ij}.\label{SpherDev}
\end{equation}
We also denote the symmetric and anti-symmetric part of $\PP$ as
\begin{equation}
P_{(ij)}=\frac{P_{ij}+P_{ji}}{2}=({\rm sym}\,P)_{ij},\qquad P_{[ij]}=\frac{P_{ij}-P_{ji}}{2}=({\rm skew}\,P)_{ij}\label{SymSkew}
\end{equation}
and introduce the variable 
\begin{equation}
P^{V}=P_{22}-P_{33}.\label{VolumeVar}
\end{equation}
Conversely, we explicitly remark that the following relationships
hold which relate the components of the micro-distortion tensor $\PP$
to some of the new introduced variables: 
\begin{align}
P_{11} & =P^{S}+P_{11}^{D},\qquad\qquad\qquad P_{22}=\frac{1}{2}\left(P^{V}+2P^{S}-P_{11}^{D}\right),\qquad\qquad P_{33}=\frac{1}{2}\left(2P^{S}-P^{V}-P_{11}^{D}\right),\vspace{1.2mm}\nonumber \\
P_{23} & =P_{\left(23\right)}+P_{\left[23\right]},\qquad\qquad\quad\!P_{1\alpha}=P_{\left(1\alpha\right)}+P_{\left[1\alpha\right]},\ \alpha=2,3,\vspace{1.2mm}\label{eq:P_change_var}\\
P_{32} & =P_{\left(23\right)}-P_{\left[23\right]},\qquad\qquad\quad\!P_{\alpha1}=P_{\left(1\alpha\right)}-P_{\left[1\alpha\right]},\ \alpha=2,3.\nonumber 
\end{align}
As done for Cauchy continua, also for relaxed micromorphic media we
limit ourselves to the case of plane waves. In other words, we suppose
that the space dependence of all the introduced kinematical fields
$u_{i}$ and $P_{ij}$ is limited only to the component $x_{1}$ of
$\XX\,$ which we also will suppose to be the direction of propagation
of the considered plane wave. Since it will be useful in the following,
let us collect some of the new variables of our problem as 
\begin{equation}
\vv_{1}:=\left(u_{1},P_{11}^{D},P^{S}\right),\qquad\vv_{2}:=\left(u_{2},P_{(12)},P_{[12]}\right),\qquad\vv_{3}:=\left(u_{3},P_{(13)},P_{[13]}\right),\label{eq:Unknowns1}
\end{equation}
and, for having a homogeneous notation, let us set 
\begin{equation}
\vv_{4}:=P_{\left(23\right)},\qquad\vv_{5}:=P_{[23]},\qquad\vv_{6}:=P^{V}.\label{eq:Unknowns2}
\end{equation}

Using the plane wave ansatz and with the proposed new choice of variables
we are able to rewrite the governing equations (\ref{eq:bulk-mod-3})
for a relaxed micromorphic continuum as different uncoupled sets of
equations (see \cite{BandGaps1,BandGaps2} for extended calculations),
namely: 
\begin{align}
{v}_{1,tt} & =\gA\,_{1}^{R}\cdot\vv''_{1}+\BB\,_{1}^{R}\cdot\vv_{1}'+\CC\,_{1}^{R}\cdot\vv_{1},\qquad{v}_{\alpha,tt}=\gA\,_{\alpha}^{R}\cdot\vv''_{\alpha}+\BB\,_{\alpha}^{R}\cdot\vv'_{\alpha}+\CC\,_{\alpha}^{R}\cdot\vv_{\alpha},\qquad\alpha=2,3\vspace{1.2mm}\label{eq:BulkEq}\\
{\vv}_{4,tt} & =A_{4}^{R}\vv_{4}''+C_{4}^{R}\vv_{4},\qquad{\vv}_{5,tt}=A_{5}^{R}\vv_{5}''+C_{5}^{R}\vv_{5},\qquad{\vv}_{6,tt}=A_{6}^{R}\vv_{6}''+C_{6}^{R}\vv_{6},\nonumber 
\end{align}
where $\left(\cdot\right)'$ denoted the derivative of the quantity
$\left(\cdot\right)$ with respect to $x_{1}$ and we set 
\begin{align}
 & \gA\,_{1}^{R}=\begin{pmatrix}\frac{\lambda_{e}+2\mu_{e}}{\rho} & 0 & 0\vspace{1.2mm}\\
0 & \frac{\mu_{e}\,L_{c}^{2}}{3\eta} & -\frac{2\,\mu_{e}\,L_{c}^{2}}{3\eta}\vspace{1.2mm}\\
0 & -\frac{\mu_{e}\,L_{c}^{2}}{3\eta} & \frac{2\,\mu_{e}\,L_{c}^{2}}{3\eta}
\end{pmatrix},\qquad\gA\,_{\alpha}^{R}=\begin{pmatrix}\frac{\mu_{e}+\mu_{c}}{\rho} & 0 & 0\vspace{1.2mm}\\
0 & \frac{\mu_{e}\,L_{c}^{2}}{2\eta} & \frac{\mu_{e}\,L_{c}^{2}}{2\eta}\vspace{1.2mm}\\
0 & \frac{\mu_{e}\,L_{c}^{2}}{2\eta} & \frac{\mu_{e}\,L_{c}^{2}}{2\eta}
\end{pmatrix},\vspace{1.2mm}\nonumber \\
 & \BB\,_{1}^{R}=\begin{pmatrix}0 & -\frac{2\mu_{e}}{\rho} & -\frac{3\lambda_{e}+2\mu_{e}}{\rho}\vspace{1.2mm}\\
\frac{4}{3}\,\frac{\mu_{e}}{\eta} & 0 & 0\vspace{1.2mm}\\
\frac{3\lambda_{e}+2\mu_{e}}{3\eta} & 0 & 0
\end{pmatrix},\qquad\BB\,_{\alpha}^{R}=\begin{pmatrix}0 & -\frac{2\mu_{e}}{\rho} & \frac{2\mu_{c}}{\rho}\vspace{1.2mm}\\
\frac{\mu_{e}}{\eta} & 0 & 0\vspace{1.2mm}\\
-\frac{\mu_{c}}{\eta} & 0 & 0
\end{pmatrix},\vspace{1.2mm}\label{eq:DefMatr}\\
 & \CC\,_{1}^{R}=\begin{pmatrix}0 & 0 & 0\vspace{1.2mm}\\
0 & -\frac{2\left(\mu_{e}+\mu_{\rm micro}\right)}{\eta} & 0\vspace{1.2mm}\\
0 & 0 & -\frac{\left(3\lambda_{e}+2\mu_{e}\right)+\left(3\lambda_{\rm micro}+2\mu_{\rm micro}\right)}{\eta}
\end{pmatrix},\text{\qquad}\CC\,_{\alpha}^{R}=\begin{pmatrix}0 & 0 & 0\vspace{1.2mm}\\
0 & -\frac{2\left(\mu_{e}+\mu_{\rm micro}\right)}{\eta} & 0\vspace{1.2mm}\\
0 & 0 & -\frac{2\mu_{c}}{\eta}
\end{pmatrix},\ \alpha=2,3.\vspace{1.2mm}\nonumber \\
 & A_{4}^{R}=A_{5}^{R}=A_{6}^{R}=\frac{\mu_{e}\,L_{c}^{2}}{\eta},\qquad C_{4}^{R}=C_{6}^{R}=-\frac{2\left(\mu_{e}+\mu_{\rm micro}\right)}{\eta},\qquad C_{5}^{R}=-\frac{2\mu_{c}}{\eta}.\nonumber 
\end{align}
These 12 scalar partial differential equations can be used to study
plane wave propagation in the relaxed micromorphic media (as done
in \cite{BandGaps1,BandGaps2}). We remark again that, when studying
bulk propagation of waves in unbounded relaxed micromorphic media
by means of equations (\ref{eq:BulkEq}), there are six different
problems which can be studied separately for the variables $\vv_{1},\vv_{2},\vv_{3},\vv_{4},\vv_{5},\vv_{6}$
respectively. More precisely, the bulk differential equations (\ref{eq:BulkEq})
associated to each of such six variables are completely uncoupled
(see also \cite{BandGaps1,BandGaps2}). As we will see, on the other
hand, the problem becomes partially coupled when considering the bulk
equations together with the jump conditions at a Cauchy/relaxed interface.

\subsubsection{Jump duality conditions in the 1D case}

Considering that the fields $u_{i}$ and $P_{ij}$ only depend on
the variable $x_{1}$ and supposing that the unit normal vector is
directed along the $x_{1}$ axis, i.e. $\nn\,=(1,0,0)$ which is also
the direction of propagation of waves, one can check that the jump
duality conditions (\ref{eq:JumpRelaxed}) simplify into

\begin{eqnarray}
\left\llbracket \,\left\langle \gt\,,\delta\uu\right\rangle \,\right\rrbracket =0,\qquad\left\llbracket \,\left\langle \gtau,\delta\PP\right\rangle \,\right\rrbracket =0,\qquad i=1,2,3,\ \ j=1,2,3,\label{eq:JumpRelaxed-1-2}
\end{eqnarray}
or equivalently, in index notation 
\begin{eqnarray}
\left\llbracket \,t_{i}\,\delta u_{i}\,\right\rrbracket =0,\qquad\left\llbracket \,\tau_{ij}\,\delta P_{ij}\,\right\rrbracket =0,\qquad i=1,2,3,\ \ j=1,2,3,\label{eq:JumpRelaxed-1}
\end{eqnarray}
with 
\begin{align}
t_{1} & =\begin{pmatrix}\lambda_{e}+2\mu_{e}\vspace{1.2mm}\\
0\vspace{1.2mm}\\
0
\end{pmatrix}\cdot\vv_{1}'+\begin{pmatrix}0\vspace{1.2mm}\\
-2\mu_{e}\vspace{1.2mm}\\
-(3\lambda_{e}+2\mu_{e})
\end{pmatrix}\cdot\vv_{1}\vspace{1.2mm}\label{eq:JumpRelaxed-1_1}\\
t_{\alpha} & =\begin{pmatrix}\mu_{e}+\mu_{c}\vspace{1.2mm}\\
0\vspace{1.2mm}\\
0
\end{pmatrix}\cdot\vv_{\alpha}'+\begin{pmatrix}0\vspace{1.2mm}\\
-2\mu_{e}\vspace{1.2mm}\\
2\mu_{c}
\end{pmatrix}\cdot\vv_{\alpha},\quad\alpha=2,3\nonumber 
\end{align}
and 
\begin{align}
\tau_{11} & =0, & \tau_{12}= & \begin{pmatrix}0\vspace{1.2mm}\\
\mu_{e}L_{c}^{2}\vspace{1.2mm}\\
\mu_{e}L_{c}^{2}
\end{pmatrix}\cdot\vv_{2}', & \tau_{13}= & \begin{pmatrix}0\vspace{1.2mm}\\
\mu_{e}L_{c}^{2}\vspace{1.2mm}\\
\mu_{e}L_{c}^{2}
\end{pmatrix}\cdot\vv_{3}',\vspace{1.2mm}\nonumber \\
\tau_{21} & =0, & \tau_{22}= & \begin{pmatrix}0\vspace{1.2mm}\\
-\frac{\mu_{e}L_{c}^{2}}{2}\vspace{1.2mm}\\
\mu_{e}L_{c}^{2}
\end{pmatrix}\cdot\vv_{1}'+\frac{\mu_{e}L_{c}^{2}}{2}\vv_{6}', & \tau_{23}= & \,\mu_{e}L_{c}^{2}\left(\vv_{4}'+\vv_{5}'\right),\vspace{1.2mm}\label{eq:BC_Relaxed_Ax_Double_Force}\\
\tau_{31} & =0, & \tau_{32}= & \,\mu_{e}L_{c}^{2}\left(\vv_{4}'-\vv_{5}'\right), & \tau_{33}= & \begin{pmatrix}0\vspace{1.2mm}\\
-\frac{\mu_{e}L_{c}^{2}}{2}\vspace{1.2mm}\\
\mu_{e}L_{c}^{2}
\end{pmatrix}\cdot\vv_{1}'-\frac{\mu_{e}L_{c}^{2}}{2}\vv_{6}'.\nonumber 
\end{align}
We can remark once again that, since three components of the relaxed
double force are vanishing, 3 out of the 12 jump conditions (\ref{eq:JumpRelaxed-1})
are automatically satisfied. This means that only 9 out of the 12
jump conditions (\ref{eq:JumpRelaxed-1}) are independent. This is
coherent with what has been underlined in \cite{Ghiba1} where it
is proven that only tangential conditions on the micro-distortion
tensor $\PP$ can be imposed in a relaxed model.

Moreover, it can be remarked that, due to the expression of double
forces given in Eqs. (\ref{eq:BC_Relaxed_Ax_Double_Force}), the jump
conditions (\ref{eq:JumpRelaxed-1-2}) actually produce a partial
coupling of the considered problem. More precisely, if the bulk equations
for the introduced unknown variables $\vv_{1},\vv_{2},\vv_{3},\vv_{4},\vv_{5},\vv_{6}$
are completely uncoupled, a coupling between the variables $\vv_{1},\vv_{6}$
and $\vv_{4},\vv_{5}$ intervenes through the jump conditions on double
forces. This means that the study of reflection and transmission of
waves at surfaces of discontinuity of the material properties cannot
be studied without accounting for such coupling phenomena.

\subsubsection{Energy flux for the relaxed model in the 1D case}

When considering conservation of total energy, it can be checked that
the first component of the energy flux (\ref{Flux}) can be rewritten
in terms of the new variables as

\begin{equation}
H_{1}=H_{1}^{1}+H_{1}^{2}+H_{1}^{3}+H_{1}^{4}+H_{1}^{5}+H_{1}^{6},\label{eq:FluxRelaxed}
\end{equation}
with

\begin{flalign}
H_{1}^{1}= & \:{\vv}_{1,t}\cdot\left[\begin{pmatrix}-\left(\lambda_{e}+2\mu_{e}\right) & 0 & 0\vspace{1.2mm}\\
0 & -\frac{\mu_{e}\,L_{c}^{2}}{2} & \mu_{e}\,L_{c}^{2}\vspace{1.2mm}\\
0 & \mu_{e}\,L_{c}^{2} & -2\,\mu_{e}\,L_{c}^{2}
\end{pmatrix}\text{\ensuremath{\cdot}}\vv_{1}'+\begin{pmatrix}0 & 2\mu_{e} & \left(3\lambda_{e}+2\mu_{e}\right)\vspace{1.2mm}\\
0 & 0 & 0\vspace{1.2mm}\\
0 & 0 & 0
\end{pmatrix}\text{\ensuremath{\cdot}}\vv_{1}\right],\vspace{1.2mm}\nonumber \\
H_{1}^{2}= & \:{\vv}_{2,t}\cdot\left[\begin{pmatrix}-\left(\mu_{e}+\mu_{c}\right) & 0 & 0\vspace{1.2mm}\\
0 & -\mu_{e}\,L_{c}^{2} & -\mu_{e}\,L_{c}^{2}\vspace{1.2mm}\\
0 & -\mu_{e}\,L_{c}^{2} & -\mu_{e}\,L_{c}^{2}
\end{pmatrix}\text{\ensuremath{\cdot}}\vv_{2}'+\begin{pmatrix}0 & 2\mu_{e} & -2\text{\ensuremath{\mu}}_{c}\vspace{1.2mm}\\
0 & 0 & 0\vspace{1.2mm}\\
0 & 0 & 0
\end{pmatrix}\text{\ensuremath{\cdot}}\vv_{2}\right],\vspace{1.2mm}\label{FlLong}\\
H_{1}^{3}=\: & {\vv}_{3,t}\cdot\left[\begin{pmatrix}-\left(\mu_{e}+\mu_{c}\right) & 0 & 0\vspace{1.2mm}\\
0 & -\mu_{e}\,L_{c}^{2} & -\mu_{e}\,L_{c}^{2}\vspace{1.2mm}\\
0 & -\mu_{e}\,L_{c}^{2} & -\mu_{e}\,L_{c}^{2}
\end{pmatrix}\text{\ensuremath{\cdot}}\vv_{3}'+\begin{pmatrix}0 & 2\mu_{e} & -2\text{\ensuremath{\mu}}_{c}\vspace{1.2mm}\\
0 & 0 & 0\vspace{1.2mm}\\
0 & 0 & 0
\end{pmatrix}\text{\ensuremath{\cdot}}\vv_{3}\right],\vspace{1.2mm}\nonumber \\
H_{1}^{4}= & -2\mu_{e}L_{c}^{2}\left(\vv_{4}\right)_{,1}{\vv}_{4,t},\qquad H_{1}^{5}=-2\mu_{e}L_{c}^{2}\left(\vv_{5}\right)_{,1}{\vv}_{5,t},\qquad H_{1}^{6}=-\frac{\mu_{e}\,L_{c}^{2}}{2}\left(\vv_{6}\right)_{,1}{\vv}_{6,t}.\nonumber 
\end{flalign}

\subsection{Standard micromorphic}

The governing equations (\ref{eq:bulk-mod-3}) can be rewritten in
terms of the new variables as 
\begin{align}
{v}_{1,tt} & =\gA\,_{1}^{C}\cdot\vv''_{1}+\BB\,_{1}^{C}\cdot\vv_{1}'+\CC\,_{1}^{C}\cdot\vv_{1},\qquad{v}_{\alpha,tt}=\gA\,_{\alpha}^{C}\cdot\vv''_{\alpha}+\BB\,_{\alpha}^{C}\cdot\vv'_{\alpha}+\CC\,_{\alpha}^{C}\cdot\vv_{\alpha},\quad\alpha=2,3\ \vspace{1.2mm}\label{eq:BulkEq-1}\\
{\vv}_{4,tt} & =A_{4}^{C}\vv_{4}''+C_{4}^{C}\vv_{4},\qquad{\vv}_{5,tt}=A_{5}^{C}\vv_{5}''+C_{5}^{C}\vv_{5},\qquad{\vv}_{6,tt}=A_{6}^{C}\vv_{6}''+C_{6}^{C}\vv_{6},\nonumber 
\end{align}
where 
\begin{align*}
 & \gA\,_{1}^{C}=\begin{pmatrix}\frac{\lambda_{e}+2\mu_{e}}{\rho} & 0 & 0\vspace{1.2mm}\\
0 & \frac{\mu_{e}\,L_{g}^{2}}{\eta} & 0\vspace{1.2mm}\\
0 & 0 & \frac{\mu_{e}\,L_{g}^{2}}{\eta}
\end{pmatrix},\qquad\gA\,_{\alpha}^{C}=\begin{pmatrix}\frac{\mu_{e}+\mu_{c}}{\rho} & 0 & 0\vspace{1.2mm}\\
0 & \frac{\mu_{e}L_{g}^{2}}{\eta} & 0\vspace{1.2mm}\\
0 & 0 & \frac{\mu_{e}L_{g}^{2}}{\eta}
\end{pmatrix}\vspace{1.2mm},\\
 & \BB\,_{1}^{C}=\BB\,_{1}^{R},\qquad\BB\,_{\alpha}^{R}=\BB\,_{\alpha}^{R},\qquad\CC\,_{1}^{C}=\CC\,_{1}^{R},\text{\qquad}\CC\,_{\alpha}^{C}=\CC\,_{\alpha}^{R},\quad\alpha=2,3,\vspace{1.2mm}\\
 & A_{4}^{C}=A_{5}^{C}=A_{6}^{C}=\frac{\mu_{e}\,L_{g}^{2}}{\eta},\qquad C_{4}^{C}=C_{6}^{C}=C_{4}^{R},\qquad C_{5}^{C}=C_{5}^{R}.
\end{align*}

\subsubsection{Jump duality conditions in the 1D case}

Considering that the fields $u_{i}$ and $P_{ij}$ depend only on
the variable $x_{1}$ and that the unit normal vector is directed
along the $x_{1}$ axis, one can check that the jump duality conditions
(\ref{eq:JumpRelaxed}) simplify into 
\begin{eqnarray}
\left\llbracket \,\left\langle \gt\,,\delta\uu\right\rangle \,\right\rrbracket =0,\qquad\left\llbracket \,\left\langle \tilde{\gtau},\delta\PP\right\rangle \,\right\rrbracket =0,\qquad i=1,2,3,\ \ j=1,2,3,\label{eq:JumpRelaxed-1-2-1}
\end{eqnarray}
or equivalently, in index notation

\begin{eqnarray}
\left\llbracket \,t_{i}\,\delta u_{i}\,\right\rrbracket =0,\qquad\left\llbracket \tilde{\tau}_{ij}\,\delta P_{ij}\right\rrbracket =0,\qquad i=1,2,3,\ \ j=1,2,3,\label{eq:JumpRelaxed-1-1}
\end{eqnarray}
where the components of the force $\gt\,$ are the same compared to
the previous case (see Eqs. ~(\ref{eq:JumpRelaxed-1_1})), while
the components of the double force are given by 
\begin{align}
\tilde{\tau}_{11}= & \begin{pmatrix}0\vspace{1.2mm}\\
\mu_{e}L_{g}^{2}\vspace{1.2mm}\\
\mu_{e}L_{g}^{2}
\end{pmatrix}\cdot\vv_{1}' & \tilde{\tau}_{12}= & \begin{pmatrix}0\vspace{1.2mm}\\
\mu_{e}\,L_{g}^{2}\vspace{1.2mm}\\
\mu_{e}\,L_{g}^{2}
\end{pmatrix}\cdot\vv_{2}' & \tilde{\tau}_{13}= & \begin{pmatrix}0\vspace{1.2mm}\\
\mu_{e}\,L_{g}^{2}\vspace{1.2mm}\\
\mu_{e}\,L_{g}^{2}
\end{pmatrix}\cdot\vv_{3}',\vspace{1.2mm}\nonumber \\
\tilde{\tau}_{21}= & \begin{pmatrix}0\vspace{1.2mm}\\
\mu_{e}\,L_{g}^{2}\vspace{1.2mm}\\
-\mu_{e}\,L_{g}^{2}
\end{pmatrix}\cdot\vv_{2}', & \tilde{\tau}_{22}= & \begin{pmatrix}0\vspace{1.2mm}\\
-\frac{\mu_{e}L_{g}^{2}}{2}\vspace{1.2mm}\\
\mu_{e}\,L_{g}^{2}
\end{pmatrix}\cdot\vv_{1}'+\frac{\mu_{e}\,L_{g}^{2}}{2}\vv_{6}', & \tilde{\tau}_{23}= & \,\mu_{e}\,L_{g}^{2}\left(\vv_{4}'+\vv_{5}'\right),\vspace{1.2mm}\label{eq:BC_Relaxed_Ax_Double_Force-1}\\
\tilde{\tau}_{31}= & \begin{pmatrix}0\vspace{1.2mm}\\
\mu_{e}\,L_{g}^{2}\vspace{1.2mm}\\
-\mu_{e}\,L_{g}^{2}
\end{pmatrix}\cdot\vv_{3}', & \tilde{\tau}_{32}= & \,\mu_{e}\,L_{g}^{2}\left(\vv_{4}'-\vv_{5}'\right), & \tilde{\tau}_{33}= & \begin{pmatrix}0\vspace{1.2mm}\\
-\frac{\mu_{e}L_{g}^{2}}{2}\vspace{1.2mm}\\
\mu_{e}\,L_{g}^{2}
\end{pmatrix}\cdot\vv_{1}'-\frac{\mu_{e}\,L_{g}^{2}}{2}\vv_{6}'.\nonumber 
\end{align}
We notice that in the case of standard micromorphic medium, all the
nine components of the double force are no vanishing, which is equivalent
to say that the set of boundary conditions is not under-determined
as in the case of the relaxed model. Moreover, a coupling between
the variables $\vv_{1},\vv_{6}$ and $\vv_{4},\vv_{5}$ is introduced
also in this case through the jump conditions.

\subsubsection{Energy flux for the standard micromorphic medium in the 1D case}

When considering conservation of total energy, it can be checked that
the first component of the energy flux (\ref{Flux-Mindlin}) may be
rewritten in terms of the new variables as

\[
\widetilde{H}_{1}=\widetilde{H}_{1}^{1}+\widetilde{H}_{1}^{2}+\widetilde{H}_{1}^{3}+\widetilde{H}_{1}^{4}+\widetilde{H}_{1}^{5}+\widetilde{H}_{1}^{6},
\]
with

\begin{align}
\widetilde{H}_{1}^{1}= & \:{\vv}_{1,t}\cdot\left[\begin{pmatrix}-\left(\lambda_{e}+2\mu_{e}\right) & 0 & 0\vspace{1.2mm}\\
0 & -\frac{3\mu_{e}\,L_{g}^{2}}{2} & 0\vspace{1.2mm}\\
0 & 0 & -3\mu_{e}\,L_{g}^{2}
\end{pmatrix}\text{\ensuremath{\cdot}}\vv_{1}'+\begin{pmatrix}0 & 2\mu_{e} & \left(3\lambda_{e}+2\mu_{e}\right)\vspace{1.2mm}\\
0 & 0 & 0\vspace{1.2mm}\\
0 & 0 & 0
\end{pmatrix}\text{\ensuremath{\cdot}}\vv_{1}\right],\vspace{1.2mm}\nonumber \\
\widetilde{H}_{1}^{2}= & \:{\vv}_{2,t}\cdot\left[\begin{pmatrix}-\left(\mu_{e}+\mu_{c}\right) & 0 & 0\vspace{1.2mm}\\
0 & -2\mu_{e}\,L_{g}^{2} & 0\vspace{1.2mm}\\
0 & 0 & -2\mu_{e}\,L_{g}^{2}
\end{pmatrix}\text{\ensuremath{\cdot}}\vv_{2}'+\begin{pmatrix}0 & 2\mu_{e} & -2\text{\ensuremath{\mu}}_{c}\vspace{1.2mm}\\
0 & 0 & 0\vspace{1.2mm}\\
0 & 0 & 0
\end{pmatrix}\text{\ensuremath{\cdot}}\vv_{2}\right],\vspace{1.2mm}\label{FlLong-1}\\
\widetilde{H}_{1}^{3}=\: & {\vv}_{3,t}\cdot\left[\begin{pmatrix}-\left(\mu_{e}+\mu_{c}\right) & 0 & 0\vspace{1.2mm}\\
0 & -2\mu_{e}\,L_{g}^{2} & 0\vspace{1.2mm}\\
0 & 0 & -2\mu_{e}\,L_{g}^{2}
\end{pmatrix}\text{\ensuremath{\cdot}}\vv_{3}'+\begin{pmatrix}0 & 2\mu_{e} & -2\text{\ensuremath{\mu}}_{c}\vspace{1.2mm}\\
0 & 0 & 0\vspace{1.2mm}\\
0 & 0 & 0
\end{pmatrix}\text{\ensuremath{\cdot}}\vv_{3}\right],\vspace{1.2mm}\nonumber \\
\widetilde{H}_{1}^{4}= & -2\mu_{e}\,L_{g}^{2}\left(\vv_{4}\right)_{,1}{\vv}_{4,t},\qquad\widetilde{H}_{1}^{5}=-2\mu_{e}\,L_{g}^{2}\left(\vv_{5}\right)_{,1}{\vv}_{5,t},\qquad\widetilde{H}_{1}^{6}=-\frac{\mu_{e}\,L_{g}^{2}}{2}\left(\vv_{6}\right)_{,1}{\vv}_{6,t}.\nonumber 
\end{align}

\section{Planar wave propagation in semi-infinite media}

\textcolor{black}{Before approaching the problem of reflection and
transmission at specific Cauchy/relaxed and Cauchy/Mindlin interfaces,
we present in the present section the problem of the study of wave
propagation in semi-infinite Cauchy, relaxed micromorphic and Mindlin
continua. In particular, we consider bulk propagation of plane waves
and we show the dispersion relations for classical Cauchy, relaxed
micromorphic and standard micromorphic continua. This will allow to
establish again the result provided in \cite{BandGaps1,BandGaps2}
according to which the relaxed micromorphic continuum is the only
generalized model which allows the description of band-gaps in a continuum
framework. Additionally to a summary of the results already presented
in \cite{BandGaps1,BandGaps2}, we propose a systematic study of the
asymptotic properties of the dispersion curves for the relaxed micromorphic
continuum. In fact, the characteristic behavior of the dispersion
curves for $k\rightarrow\infty$ is fundamental for the assessment
of the band-gap existence in the framework of the relaxed micromorphic
model. Analogous results are briefly presented also for the Cauchy
and Mindlin continua, so showing that the relaxed micromorphic model
is in fact the only one featuring band-gaps in the conventional, linear-elastic
micromorphic framework.}

\textcolor{black}{To study bulk wave propagation, we assume in what
follows that the involved unknown variables take the harmonic form
\begin{equation}
u_{1}=\alpha_{1}\,e^{i(kx_{1}-\omega t)},\quad\quad u_{2}=\alpha_{2}\,e^{i(kx_{1}-\omega t)},\quad\quad u_{3}=\alpha_{3}\,e^{i(kx_{1}-\omega t)},\label{eq:WaveFormCauchy}
\end{equation}
}on the left side occupied by the Cauchy medium and 
\begin{alignat}{2}
\vv_{1}=\beta_{1}\,e^{i(kx_{1}-\omega t)},\qquad & \vv_{2}=\beta_{2}\,e^{i(kx_{1}-\omega t)},\qquad & \vv_{3}=\beta_{3}\,e^{i(kx_{1}-\omega t)},\nonumber \\
\label{eq:WaveForm}\\
\vv_{4}=\beta_{4}\,e^{i(kx_{1}-\omega t)},\qquad & \vv_{5}=\beta_{5}\,e^{i(kx_{1}-\omega t)},\qquad & \vv_{6}=\beta_{6}\,e^{i(kx_{1}-\omega t)}.\nonumber 
\end{alignat}
on the right side occupied by the relaxed micromorphic medium.

\subsection{Classical Cauchy media}

\begin{figure}[H]
\begin{centering}
\begin{tabular}{ccc}
\includegraphics{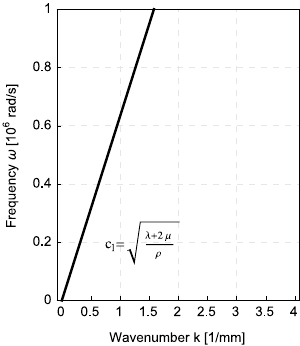}  &  & \includegraphics{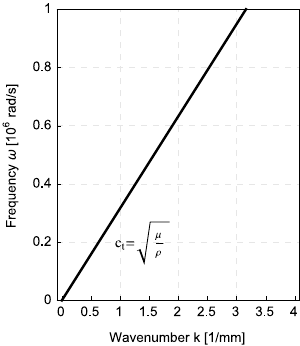}\tabularnewline
$\quad$(a)  &  & $\quad$(b)\tabularnewline
\end{tabular}
\par\end{centering}

\protect\caption{\label{fig:DispersionCauchy}Dispersion relations for a Cauchy continuum:
(a) longitudinal waves and (b) transverse waves. The dispersion relations
are straight lines (non-dispersive behavior) and $c_{l}$ and $c_{t}$
are their slopes for longitudinal and transverse waves respectively.}
\end{figure}

As far as bulk propagation in Cauchy media is concerned, we can find
the standard dispersion relations for Cauchy media simply replacing
the expressions (\ref{eq:WaveFormCauchy}) in the equations of motion
(\ref{CauchyAS}) and simplifying, so obtaining 
\begin{gather}
\omega^{2}=\frac{\lambda_{\rm macro}+2\mu_{\rm macro}}{\rho}k^{2},\qquad\omega^{2}=\frac{\mu_{\rm macro}}{\rho}k^{2},\qquad\omega^{2}=\frac{\mu_{\rm macro}}{\rho}k^{2},\label{CauchyAS-1}
\end{gather}
or, equivalently 
\begin{gather}
\omega=\pm c_{l}k,\qquad\omega=\pm c_{t}k,\qquad\omega=\pm c_{t}k,\label{CauchyAS-1-1}
\end{gather}
where we denoted by 
\[
c_{l}=\sqrt{\frac{\lambda_{\rm macro}+2\mu_{\rm macro}}{\rho}},\qquad c_{t}=\sqrt{\frac{\mu_{\rm macro}}{\rho}},
\]
the characteristic speeds in classical Cauchy media of longitudinal
and transverse waves, respectively. Here, we have assumed 
\begin{align}
\rho>0,\qquad\mu_{\rm macro}\geq0,\qquad\lambda_{\rm macro}+2\mu_{\rm macro}\geq0.
\end{align}
The dispersion relations (\ref{CauchyAS-1-1}) can be traced in the
plane $(\omega,k)$, giving rise to the standard non-dispersive behavior
for a classical Cauchy continuum (see Figure \ref{fig:DispersionCauchy})

It can be noticed in Figure \ref{fig:DispersionCauchy} that we have
one straight line for longitudinal waves and two for transverse waves
(there are two superimposed lines in picture (b), according to Eqs.
(\ref{CauchyAS-1-1})).

On the other hand, the equations (\ref{CauchyAS-1-1}) can be inverted
and we can state that the expressions of the wavenumbers as function
of the frequency $\omega$ can be found as 
\begin{equation}
k_{1}\left(\omega\right)=\pm\frac{1}{c_{l}}\omega,\qquad k_{2}\left(\omega\right)=\pm\frac{1}{c_{t}}\omega,\qquad k_{3}\left(\omega\right)=\pm\frac{1}{c_{t}}\omega.\label{eq:Eigen_Cauchy}
\end{equation}
Based on equations (\ref{eq:WaveFormCauchy}), the solution for the
displacement field can hence be rewritten as\footnote{We explicitly remark that the positive or negative roots for the wavenumbers
$k_{i}$ must be chosen in Eq. (\ref{eq:WaveFormCauchy-2}) depending
whether the wave travels in the $x_{1}$ or $-x_{1}$ direction.} 
\begin{equation}
\quad u_{1}=\alpha_{1}\,e^{i(k_{1}(\omega)x_{1}-\omega t)},\quad\quad u_{2}=\alpha_{2}\,e^{i(k_{2}(\omega)x_{1}-\omega t)},\quad\quad u_{3}=\alpha_{3}\,e^{i(k_{3}(\omega)x_{1}-\omega t)}.\label{eq:WaveFormCauchy-2}
\end{equation}
Such relationships establish the solution for the displacement field
for any real frequency $\omega$ if the amplitudes $\alpha_{1}$,
$\alpha_{2}$, and $\alpha_{3}$ are known. Such amplitudes will be
calculated in the following by imposing suitable boundary conditions.

We explicitly remark that, as far as second gradient continua are
concerned, the dispersion relations are similar to those presented
in figure \ref{fig:DispersionCauchy}, since the underlying kinematics
is the same in both Cauchy and second gradient media (only the macroscopic
displacement field). On the other hand, contrarily to Cauchy continua,
second gradient media may exhibit dispersive behaviors in the sense
that the dispersion relations analogous to those presented in figure
\ref{fig:DispersionCauchy} are not straight lines anymore (see \cite{BandGaps1,FdIPlacidi}).

\subsection{Relaxed micromorphic media}

As far as the relaxed micromorphic model is concerned, we proceed
in an analogous way and we replace the wave-forms (\ref{eq:WaveForm})
for the unknown fields in the bulk equations (\ref{eq:BulkEq}), so
obtaining 
\begin{align}
\left(k^{2}\gA\,_{1}^{R}-\omega^{2}\:\mathds1-i\,k\BB\,_{1}^{R}-\CC\,_{1}^{R}\right)\cdot\beta_{1}=0\qquad\qquad & \text{longitudinal waves}\vspace{1.2mm}\label{eq:Long}\\
\left(k^{2}\gA\,_{\alpha}^{R}-\omega^{2}\:\mathds1-i\,k\BB\,_{\alpha}^{R}-\CC\,_{\alpha}^{R}\right)\cdot\beta_{\alpha}=0,\qquad\alpha=2,3,\qquad & \text{transverse waves}\vspace{1.2mm}\label{eq:Transv}\\
\omega^{2}=A_{4}^{R}\,k^{2}-C_{4}^{R},\qquad\omega^{2}=A_{5}^{R}\,k^{2}-C_{5}^{R},\qquad\omega^{2}=A_{6}^{R}\,k^{2}-C_{6}^{R},\qquad & \text{uncoupled waves}.\label{eq:Uncouplde}
\end{align}

\subsubsection{Uncoupled waves.}

The specific behavior of uncoupled waves can be easily studied, since,
starting from Eq. (\ref{eq:Uncouplde}), the frequency is explicitly
obtained as a function of the wavenumber as: 
\begin{equation}
\omega(k)=\pm\sqrt{\omega_{s}^{2}+c_{m}^{2}k^{2}},\qquad\omega(k)=\pm\sqrt{\omega_{r}^{2}+c_{m}^{2}k^{2}},\qquad\omega(k)=\pm\sqrt{\omega_{s}^{2}+c_{m}^{2}k^{2}},\label{eq:dispersion_uncoupled}
\end{equation}
where we introduced the characteristic speed $c_{m}$ and the characteristic
frequencies $\omega_{s}$ and $\omega_{r}$ as 
\[
c_{m}=\sqrt{A_{4}^{R}}=\sqrt{A_{5}^{R}}=\sqrt{A_{6}^{R}}=\sqrt{\frac{\mu_{e}\,L_{c}^{2}}{\eta}},\quad\omega_{s}=\sqrt{-C_{4}^{R}}=\sqrt{-C_{6}^{R}}=\sqrt{\frac{2\left(\mu_{e}+\mu_{\rm micro}\right)}{\eta}},\quad\omega_{r}=\sqrt{-C_{5}^{R}}=\sqrt{\frac{2\mu_{c}}{\eta}}.
\]
We have assumed 
\begin{align}
\eta>0,\qquad\mu_{e}\geq0,\qquad\mu_{\rm micro}\geq0,\qquad\mu_{c}\geq0.
\end{align}
We start remarking that, since the equations for $\vv_{4}$ and $\vv_{6}$
are formally the same, we have two coinciding dispersion relations.
Moreover, it is clear that, when $k=0$, one has that 
\begin{equation}
\omega(0)=\omega_{s},\qquad\omega(0)=\omega_{r},\qquad\omega(0)=\omega_{s},\label{eq:dispersion_uncoupled-1}
\end{equation}
which allows to determine the cut-off frequencies $\omega_{s}$ and
$\omega_{r}$ for the uncoupled waves.

\begin{figure}[H]
\begin{centering}
\begin{tabular}{ccccc}
\includegraphics[scale=0.75]{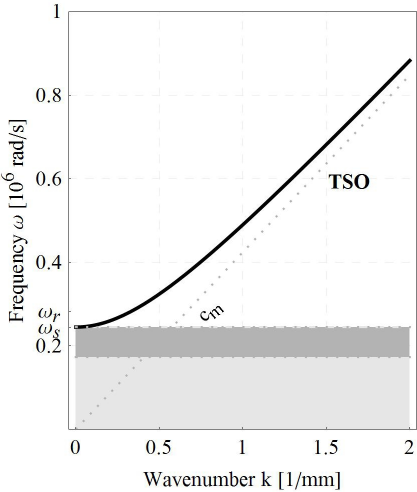}  &  & \includegraphics[scale=0.75]{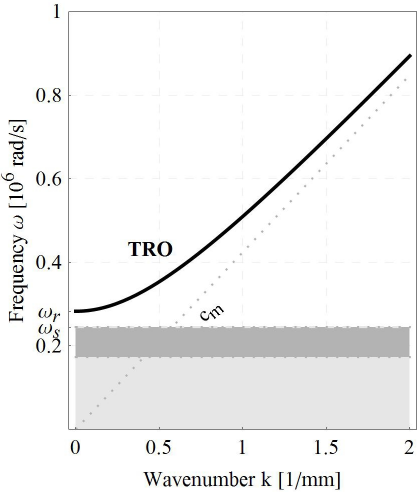}  &  & \includegraphics[scale=0.75]{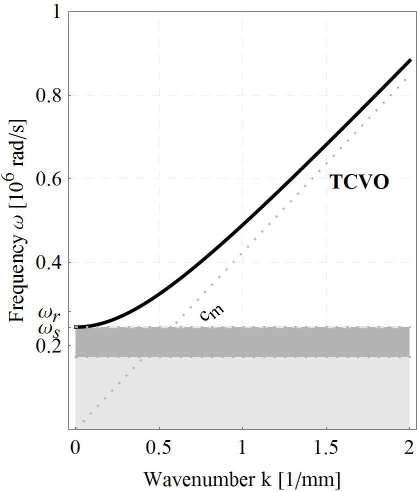}\tabularnewline
(a)  &  & (b)  &  & (c)\tabularnewline
\end{tabular}
\par\end{centering}

\protect\caption{\label{fig:DispersionUncoupled}Dispersion curves for uncoupled waves.
Such waves have cut-off frequencies $\omega_{s}$ and $\omega_{r}$
and they all have the same asymptote of slope $c_{m}$. The curves
(a) and (c) are superimposed.}
\end{figure}

On the other hand, when $k\rightarrow\infty$, one has 
\begin{equation}
\omega(k)\sim c_{m}k,\qquad\omega(k)\sim c_{m}k,\qquad\omega(k)\sim c_{m}k,\label{eq:dispersion_uncoupled-2}
\end{equation}
which means that the dispersion curves for the uncoupled waves all
have an asymptote on $c_{m}$. The behavior of the dispersion curves
for uncoupled waves is depicted in Figure \ref{fig:DispersionUncoupled}.

The solutions (\ref{eq:dispersion_uncoupled}) can be inverted for
$\omega>\max\{\omega_{s},\omega_{r}\}$, so finding 
\begin{equation}
k_{s}(\omega)=\pm\frac{1}{c_{m}}\sqrt{\omega^{2}-\omega_{s}^{2}},\qquad k_{r}(\omega)=\pm\frac{1}{c_{m}}\sqrt{\omega^{2}-\omega_{r}^{2}},\qquad k_{s}(\omega)=\pm\frac{1}{c_{m}}\sqrt{\omega^{2}-\omega_{s}^{2}},\label{eq:dispersion_uncoupled-4}
\end{equation}
where we remark that the wavenumbers $k_{s}$ and $k_{r}$ become
purely imaginary for real frequencies $\omega<\omega_{s}$ and $\omega<\omega_{r}$,
respectively.

Based on the wave-form assumption (\ref{eq:WaveForm}), the solution
for the uncoupled waves can hence be written as\footnote{We explicitly remark that the positive or negative roots for the wavenumbers
$k_{s}$ and $k_{r}$ must be chosen in Eq. (\ref{eq:WaveSolUnc})
depending wether the wave travels in the $x_{1}$ or $-x_{1}$ direction.
Moreover, when $\omega<\omega_{s}$ an exponential decaying with $x_{1}$
appears so that the solution is not periodic anymore and so-called
standing waves appear for $\vv_{4}$ and $\vv_{6}$. The same happens
for the variable $\vv_{5}$ when $\omega<\omega_{r}$.} 
\begin{equation}
\vv_{4}=\beta_{4}\,e^{i(k_{s}(\omega)x_{1}-\omega t)},\quad\quad\vv_{5}=\beta_{5}\,e^{i(k_{r}(\omega)x_{1}-\omega t)},\quad\quad\vv_{6}=\beta_{6}\,e^{i(k_{s}(\omega)x_{1}-\omega t)}.\label{eq:WaveSolUnc}
\end{equation}
The unknown amplitudes $\beta_{4}$, $\beta_{5}$ and $\beta_{6}$
can be calculated by imposing suitable boundary conditions as it will
be shown later on.

\subsubsection{\label{sub:Long_asymptotics}Longitudinal waves.}

We start by noticing that replacing the wave form (\ref{eq:WaveForm})
in (\ref{eq:Long}) we get 
\begin{equation}
\gA\,_{1}\cdot\beta_{1}=\left(k^{2}\,\gA\,_{1}^{R}-\omega^{2}\:\mathds1-i\,k\,\BB\,_{1}^{R}-\CC\,_{1}^{R}\right)\cdot\beta_{1}=0.\label{eq:DispersionLong}
\end{equation}
In order to have non-trivial solutions of this algebraic system we
must have 
\[
\mathrm{det}\gA\,_{1}=0,
\]
that will furnish the solutions $\omega=\omega(k$) which are usually
known as \textit{dispersion relations}. It can be checked that $\mathrm{det}\gA\,_{1}$
is a polynomial of the 6$^{th}$ order in $\omega$ and of the 4$^{th}$
order in $k$ and that only even powers of both $\omega$ and $k$
appear.

We assume 
\begin{align}
\lambda_{e}+2\,\mu_{e}\geq0,\qquad\lambda_{\rm micro}+2\,\mu_{\rm micro}\geq0.
\end{align}
In order to get some preliminary information concerning the function
$\mathrm{det}\gA\,_{1}$, we can start looking at its behaviour for
$k\rightarrow0$ and $k\rightarrow\infty$.

When $k=0$ the system of algebraic equations (\ref{eq:DispersionLong})
simplifies into 
\[
\left(\omega^{2}\:\mathds1+\CC\,_{1}^{R}\right)\cdot\beta_{1}=0,
\]
which, recalling the definition (\ref{eq:DefMatr}) for $\CC\,_{1}^{R}$,
equivalently reads 
\[
\begin{pmatrix}\omega^{2} & 0 & 0\vspace{1.2mm}\\
0 & \omega^{2}-\omega_{s}^{2} & 0\vspace{1.2mm}\\
0 & 0 & \omega^{2}-\omega_{p}^{2}
\end{pmatrix}\cdot\beta_{1}=0,
\]
where we introduced the new characteristic frequency 
\[
\omega_{p}=\sqrt{\frac{\left(3\lambda_{e}+2\mu_{e}\right)+\left(3\lambda_{\rm micro}+2\mu_{\rm micro}\right)}{\eta}}.
\]
This means that for very small values of $k$ the three dispersion
equations become uncoupled and have cut-off frequencies which are
respectively $\omega=0$, $\omega=\omega_{s}$ and $\omega=\omega_{r}$.

On the other hand, letting $k\rightarrow\infty$, and considering
the case for which the ratio $k/\omega$ remains finite, instead of
studying the whole system of equations (\ref{eq:DispersionLong}),
we can consider the reduced system 
\begin{equation}
\left(k^{2}\,\gA\,_{1}^{R}-\omega^{2}\:\mathds1\right)\cdot\beta_{1}=0,\label{eq:DispersionLong-1}
\end{equation}
which, given the expression (\ref{eq:DefMatr}) for $\gA\,_{1}^{R}$,
equivalently reads 
\[
\begin{pmatrix}c_{p}^{2}\,k^{2}-\omega^{2} & 0 & 0\vspace{1.2mm}\\
0 & \frac{1}{3}c_{m}^{2}\:k^{2}-\omega^{2} & -\frac{2}{3}c_{m}^{2}\,k^{2}\vspace{1.2mm}\\
0 & -\frac{1}{3}c_{m}^{2}\,k^{2} & \frac{2}{3}c_{m}^{2}\,k^{2}-\omega^{2}
\end{pmatrix}\cdot\beta_{1}=0,
\]
where we introduced the new characteristic speed $c_{p}$ as 
\[
c_{p}=\sqrt{\frac{\lambda_{e}+2\mu_{e}}{\rho}}.
\]
It can be seen from this last system of algebraic equations that for
high values of $k$, the first equation becomes uncoupled from the
other two and implies that 
\[
\omega=c_{p}\,k.
\]
This means, in other words, that one of the longitudinal waves will
have an asymptote in $c_{p}$. The last two equations remain coupled
and, in order to let them have a non trivial solution, it must be
set 
\[
\mathrm{det}\begin{pmatrix}\frac{1}{3}c_{m}^{2}\:k^{2}-\omega^{2} & -\frac{2}{3}c_{m}^{2}\,k^{2}\vspace{1.2mm}\\
-\frac{1}{3}c_{m}^{2}\,k^{2} & \frac{2}{3}c_{m}^{2}\,k^{2}-\omega^{2}
\end{pmatrix}=\omega^{4}-c_{m}^{2}\,k^{2}\,\omega^{2}=0.
\]
This implies a solution $\omega=0$ (which has to be excluded since
it does not fall in the case $k/\omega$ finite for $k\rightarrow\text{\ensuremath{\infty}}$)
and a solution $\omega=c_{m}\,k.$ This means that for large values
of $k$ one of the longitudinal dispersion curves will have an asymptote
in $c_{m}$.

\begin{figure}[H]
\begin{centering}
\begin{tabular}{ccccc}
\includegraphics[scale=0.75]{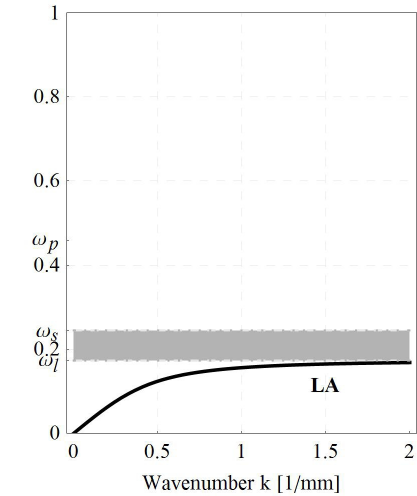}  &  & \includegraphics[scale=0.75]{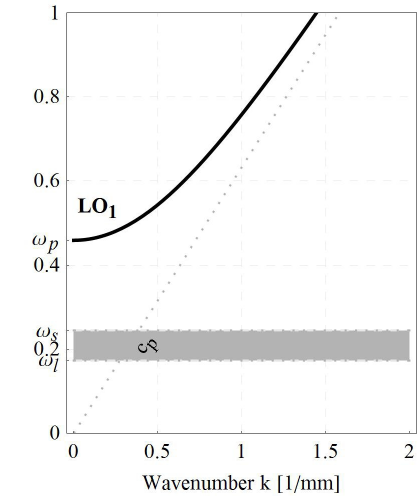}  &  & \includegraphics[scale=0.75]{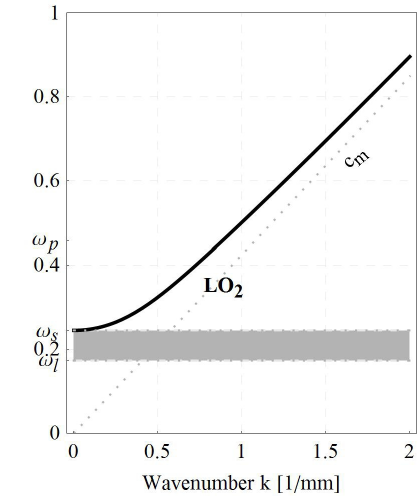}\tabularnewline
(a)  &  & (b)  &  & (c)\tabularnewline
\end{tabular}
\par\end{centering}

\protect\caption{\label{fig:DispersionLong}Dispersion relations for longitudinal waves.
Two asymptotes of slope $c_{p}$ and $c_{m}$ are identified, together
with a horizontal asymptote at $\omega=\omega_{l}$. Two cut-off frequencies
$\omega_{p}$ and $\omega_{s}$ are identified for the optic waves.}
\end{figure}

Once that the two asymptotes $c_{m}$ and $c_{p}$ are known, we can
look for the horizontal asymptote $\omega_{l}$ by noticing that (for
$k\rightarrow\text{\ensuremath{\infty}}$) the function $\mathrm{det}\gA\,_{1}$
can be factorized 
\[
\mathrm{det}\gA\,_{1}=(\omega^{2}-c_{m}^{2}\,k^{2})(\omega^{2}-c_{p}^{2}\,k^{2})(\omega^{2}-\omega_{l}^{2})=0.
\]
It can be checked that the horizontal asymptote is found to be 
\[
\omega_{l}=\sqrt{\frac{\lambda_{\rm micro}+2\mu_{\rm micro}}{\eta}}.
\]
The three dispersion relations for longitudinal waves are separately
shown in Figure \ref{fig:DispersionLong}.

On the other hand, given the structure of the polynomial $\mathrm{det}\gA\,_{1}$,
if it is solved in terms of $k=k(\omega)$, four solutions are found
which can be formally written as\footnote{To the sake of simplicity, we do not show the explicit form of the
functions $k_{1}^{1}(\omega)$ and $k_{2}^{1}(\omega)$. } 
\[
\pm k_{1}^{1}(\omega),\qquad\text{\ensuremath{\pm}}k_{1}^{2}(\omega).
\]
We denote by $h_{1}^{1}$ and $h_{1}^{2}$ the eigenvectors associated
to the eigenvalues $k_{1}^{1}(\omega)$ and $k_{1}^{2}(\omega)$ and
verifying Eq. (\ref{eq:DispersionLong}) with $\beta_{1}=h_{1}^{1}$
and $\beta_{1}=h_{1}^{2}$, respectively. According to Eq. (\ref{eq:WaveForm}),
thel solution for $\vv_{1}$ can finally be written as 
\begin{equation}
\vv_{1}=\beta_{1}^{1}h_{1}^{1}e^{i(\pm k_{1}^{1}(\omega)x_{1}-\omega t)}+\beta_{1}^{2}h_{1}^{2}e^{i(\pm k_{2}^{1}(\omega)x_{1}-\omega t)},\label{eq:WaveSolLong}
\end{equation}
where the $+$ or $-$ sign must be chosen for the wavenumbers $k_{1}^{1}$
and $k_{2}^{1}$ depending whether the considered wave travels in
the $x_{1}$ or $-x_{1}$ respectively. The two scalar unknowns $\beta_{1}^{1}$
and $\beta_{1}^{2}$ can be found by imposing suitable boundary conditions
as it will be explained later on.

\subsubsection{\label{sub:asymptotics_transv}Transverse waves.}

\begin{figure}[H]
\begin{centering}
\begin{tabular}{ccccc}
\includegraphics[scale=0.75]{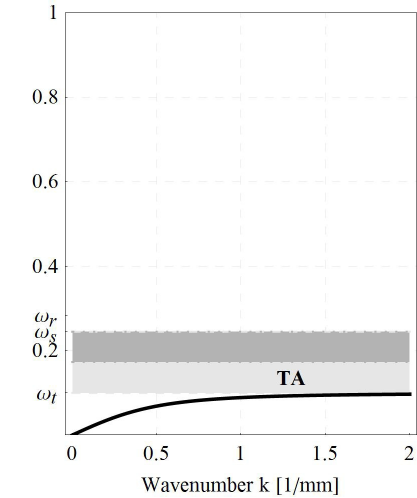}  &  & \includegraphics[scale=0.75]{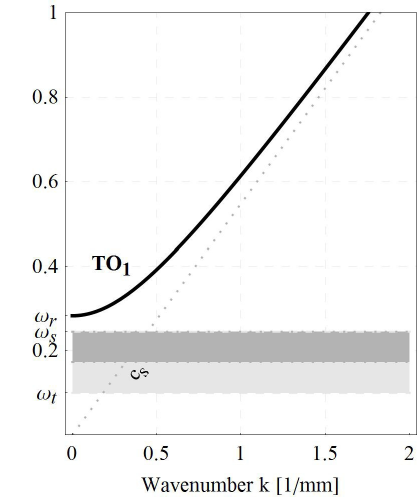}  &  & \includegraphics[scale=0.75]{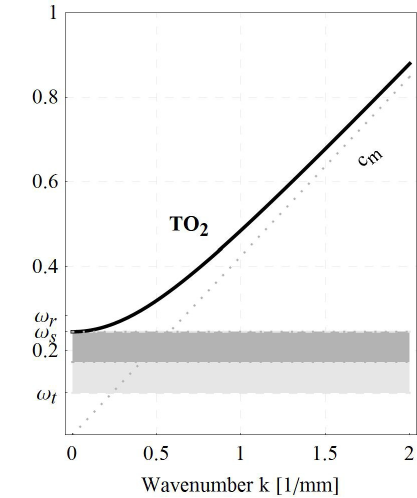}\tabularnewline
(a)  &  & (b)  &  & (c)\tabularnewline
\end{tabular}
\par\end{centering}

\protect\caption{\label{fig:DispersionTransv}Dispersion relations for transverse waves.
Two asymptotes of slope $c_{s}$ and $c_{m}$ are identified, together
with a horizontal asymptote at $\omega=\omega_{t}$. Two cut-off frequencies
$\omega_{r}$ and $\omega_{s}$ are identified for the optic waves.}
\end{figure}

We start by noticing that replacing the wave form (\ref{eq:WaveForm})
in (\ref{eq:Transv}) we get 
\begin{equation}
\gA\,_{\alpha}\cdot\beta_{\alpha}=\left(k^{2}\,\gA\,_{\alpha}^{R}-\omega^{2}\:\mathds1-i\,k\,\BB\,_{\alpha}^{R}-\CC\,_{\alpha}^{R}\right)\cdot\beta_{\alpha}=0.\label{eq:DispersionTransv}
\end{equation}
In order to have non-trivial solutions of this algebraic system we
must have 
\[
\mathrm{det}\gA\,_{\alpha}=0.
\]
Following an equivalent reasoning with respect to what done for longitudinal
waves, it can be checked that the solutions $\omega=\omega(k)$ of
this equations are such that they have two asymptotes of slope $c_{s}$
and $c_{m}$ respectively and a horizontal asymptote at $\omega=\omega_{t}$,
where we set 
\[
c_{s}=\sqrt{\frac{\mu_{e}+\mu_{c}}{\rho}},\qquad\omega_{t}=\sqrt{\frac{{\mu_{\rm micro}}}{\eta}}.
\]
The dispersion curves for transverse waves are depicted in Figure
\ref{fig:DispersionTransv}.

As done for the longitudinal waves, given the structure of the polynomial
$\mathrm{det}\gA\,_{\alpha}$, if it is solved in terms of $k=k(\omega)$,
four solutions are found which can be formally written as\footnote{To the sake of simplicity, we do not show the explicit form of the
functions $k_{\alpha}^{1}(\omega)$ and $k_{\alpha}^{2}(\omega)$. } 
\[
\pm k_{\alpha}^{1}(\omega),\qquad\text{\ensuremath{\pm}}k_{\alpha}^{2}(\omega),\qquad\alpha=2,3.
\]
We denote by $h_{\alpha}^{1}$ and $h_{\alpha}^{2}$ the eigenvectors
associated to the eigenvalues $k_{\alpha}^{1}(\omega)$ and $k_{\alpha}^{2}(\omega)$
and verifying Eq. (\ref{eq:DispersionTransv}) with $\beta_{\alpha}=h_{\alpha}^{1}$
and $\beta_{\alpha}=h_{\alpha}^{2}$, respectively. According to Eq.
(\ref{eq:WaveForm}), the final solution for $\vv_{\alpha}$ can be
finally written as

\begin{equation}
\vv_{\alpha}=\beta_{\alpha}^{1}h_{\alpha}^{1}e^{i(\pm k_{\alpha}^{1}(\omega)x_{1}-\omega t)}+\beta_{\alpha}^{2}h_{\alpha}^{2}e^{i(\pm k_{\alpha}^{2}(\omega)x_{1}-\omega t)},\qquad\text{\ensuremath{\alpha}=}2,3.\label{eq:WaveSolTransv}
\end{equation}
where the $+$ or $-$ sign must be chosen for the wavenumbers $k_{\alpha}^{1}$
and $k_{\alpha}^{2}$ depending whether the considered wave travels
in the $x_{1}$ or $-x_{1}$ respectively. The scalar unknowns $\beta_{\alpha}^{1}$
and $\beta_{\alpha}^{2}$ can be found by imposing suitable boundary
conditions as it will be explained in the next section.

\medskip{}

\subsubsection{A summary concerning dispersion curves in relaxed micromorphic media}

\begin{figure}[H]
\begin{centering}
\begin{tabular}{ccccc}
\includegraphics[scale=0.75]{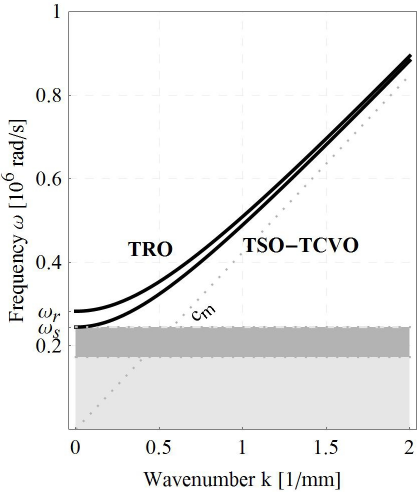}  &  & \includegraphics[scale=0.75]{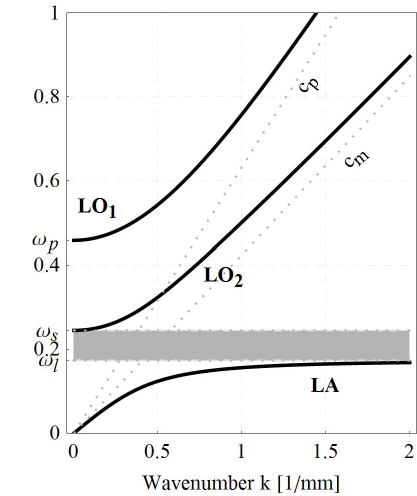}  &  & \includegraphics[scale=0.75]{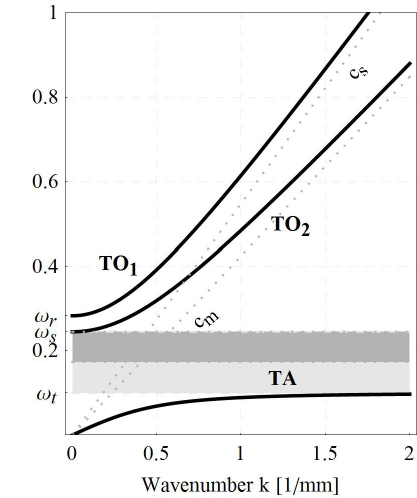}\tabularnewline
(a)  &  & (b)  &  & (c)\tabularnewline
\end{tabular}
\par\end{centering}

\protect\caption{\label{fig:DispersionTotal}Dispersion relations for all the considered
uncoupled, longitudinal and transverse waves. Identification of the
complete band-gap.}
\end{figure}

We now want to summarize our findings concerning the characteristic
behavior of the dispersion curves in relaxed micromorphic media. To
do so, we start presenting in figure \ref{fig:DispersionTotal} all
the obtained dispersion curve diagrams for uncoupled, longitudinal
and transverse waves, so highlighting the presence of a complete band
gap\footnote{\textcolor{black}{We explicitly remark that the relaxed micromorphic
model is intrinsically a macroscopic model which is able to account
for the presence of the microstructure in an ï¿œaveragedï¿œ sense at the
continuum level. This means that it can be usefully employed as far
as the wavelength of the considered waves is bigger than the characteristic
size of the unit cell of the underlying microstructure. For this reason,
the dispersion relations shown in Fig. \ref{fig:DispersionTotal}
can be used for the description of band-gap metamaterials as far as
the wavenumber $k$ is smaller than a threshold value corresponding
to the situation in which the wavelength of the traveling wave is
equal to the characteristic size of the unit cell of the considered
metamaterial. }}. 

Moreover, we summarize in the following formulas all the characteristic
quantities appearing in Figure \ref{fig:DispersionTotal} whose expressions
in terms of the constitutive parameters have been identified in the
preceding subsections: 
\begin{align}
\omega_{s} & =\sqrt{\frac{2\,(\mu_{e}+\mu_{\rm micro})}{\eta}},\qquad\omega_{r}=\sqrt{\frac{2\,\mu_{c}}{\eta}},\qquad\omega_{p}=\sqrt{\frac{(3\,\lambda_{e}+2\,\mu_{e})+(3\,\lambda_{\rm micro}+2\,\mu_{\rm micro})}{\eta}},\vspace{1.2mm}\nonumber \\
\omega_{l} & =\sqrt{\frac{\lambda_{\rm micro}+2\mu_{\rm micro}}{\eta}},\qquad\quad\omega_{t}=\sqrt{\frac{2\mu_{\rm micro}}{\eta}},\vspace{1.2mm}\label{eq:characteristic_quantities}\\
c_{m} & =\sqrt{\frac{\mu_{e}\,L_{c}^{2}}{\eta}},\qquad\qquad\ \ c_{p}=\sqrt{\frac{\lambda_{e}+2\mu_{e}}{\rho}},\qquad c_{s}=\sqrt{\frac{\mu_{e}+\mu_{c}}{\rho}},\nonumber 
\end{align}
assuming a priori that\footnote{The considered conditions on the parameters are dictated by the positive
definiteness of the strain energy density as well as by the fact of
considering positive macro and micro mass densities.} 
\begin{align}
\rho>0,\qquad\eta>0,\qquad\mu_{e} & >0,\qquad\mu_{\rm micro}>0,\qquad\mu_{c}\geq0,\qquad\lambda_{e}+2\,\mu_{e}>0,\qquad\lambda_{\rm micro}+2\,\mu_{\rm micro}>0.\label{eq:DefPos}
\end{align}

\subsubsection{The degenerate limit case $L_{c}=0$: internal variable model}

It is immediate to notice that as far as we set $L_{c}=0$ in the
strain energy density (\ref{KinPot}), higher order terms due to the
space derivatives of the micro-distortion tensor $\PP$ are lost.
This means that non-local effects are not accounted for any-more and
the considered model reduces to what is sometimes called an internal
variable model. The repercussions on the dispersion curves shown in
Figure \ref{fig:DispersionTotal} are that the oblique asymptote $c_{m}$
becomes a horizontal one, with the effect that the associated dispersion
curves flatten in order to become horizontal for $k\rightarrow\infty$.
A non-negligible class of band-gap metamaterials can, at least in
a first approximation, be modeled by means of such an internal variable
model. Suggestive results in this sense have been provided in the
recent paper \cite{Geers1} in which numerical homogenization techniques
are applied in order to obtain an internal variable continuum model
at the homogenized scale. This homogenization procedure makes use
of so-called ``separation of scale hypothesis'' which basically
means that all micro-resonant phenomena remain confined inside the
single elementary cells. Non-local effects are implicitly neglected,
and this is the main reason for which an internal variable continuum
model is obtained instead of a relaxed micromorphic one. The relaxed
micromorphic model must be the target for describing band-gap metamaterials
with non-local effect.

It can be checked that, as soon as $L_{c}=0$ (or equivalently $c_{m}=0$): 
\begin{itemize}
\item The cut-off frequencies $\omega_{p}$, $\omega_{s}$ and $\omega_{r}$
remain the same as in the general case $L_{c}\neq0$, 
\item the uncoupled waves all become horizontal straight lines (see eq.
\eqref{eq:dispersion_uncoupled}) and, in particular 
\begin{equation}
\omega(k)=\omega_{s},\qquad\omega(k)=\omega_{r}\qquad\omega(k)=\omega_{s}.\label{eq:dispersion_uncoupled-5}
\end{equation}

\item the characteristic equation $\mathrm{det}\,\gA\,_{1}=0$ for longitudinal
waves is modified in the sense that it becomes of the second order
in $k$ (instead that of the 4th order for the case $L_{c}\neq0$).
This means that the oblique asymptote $c_{p}$ of the previous case
is preserved, while two horizontal asymptotes $\omega_{l}^{1}$ and
$\omega_{l}^{2}$ can be found that are defined as 
\begin{gather*}
\omega_{l}^{1,2}=\sqrt{\frac{a\pm\sqrt{a^{2}-b}}{2\,\eta\,(\lambda_{e}+2\,\mu_{e})}},
\end{gather*}
with
\begin{align*}
a&=6\,\lambda_{\rm micro}\,\mu_{e}+4\mu_{e}(\mu_{e}+2\,\mu_{\rm micro})+\lambda_{e}(3\,\lambda_{\rm micro}+6\,\mu_{e}+4\,\mu_{\rm micro}),\\
b&=8\left(\lambda_{e}+2\,\mu_{e}\right)\left[\,\lambda_{e}\,\left(3\,\lambda_{\rm micro}\,(\mu_{e}+\mu_{\rm micro})+2\,\mu_{\rm micro}\,(3\,\mu_{e}+\mu_{\rm micro})\right)\right.\\&\ \ \ \left. +2\,\mu_{e}\left(2\,\mu_{\rm micro}\,(\mu_{e}+\mu_{\rm micro})+\lambda_{\rm micro}\,(\mu_{e}+3\mu_{\rm micro})\right)\,\right]
\end{align*}
Although not immediately evident, it can be checked that under the
considered hypotheses on the constitutive parameters (definite positiveness
of the energy and positive macro and micro mass densities), the obtained
values of the longitudinal asymptotes are real and positive.
\item the characteristic equation $\mathrm{det}\,\gA\,_{\alpha}=0$ for
transverse waves is modified in the sense that it also becomes of
the second order in $k$ (instead that of the 4th order for the case
$L_{c}\neq0$). This means that the oblique asymptotes $c_{s}$ of
the previous case is preserved, while two horizontal asymptotes $\omega_{t}^{1}$
and $\omega_{t}^{2}$ can be found that are defined as 
\[
\omega_{t}^{1,2}=\sqrt{\frac{2\,\mu_{c}\,\mu_{e}+(\mu_{c}+\mu_{e})\,\mu_{\rm micro}\pm\sqrt{\left(2\,\mu_{c}\,\mu_{e}+(\mu_{c}+\mu_{e})\,\mu_{\rm micro}\right)^{2}-4\,\mu_{c}\,\mu_{e}\,\mu_{\rm micro}\,(\mu_{c}+\mu_{e})}}{\eta\,(\mu_{c}+\mu_{e})}}.
\]
Although not immediately evident, it can be checked that under the
considered hypotheses on the constitutive parameters (definite positiveness
of the energy and positive macro and micro mass densities), the obtained
values of the transverse asymptotes are real and positive.
\end{itemize}
This internal variable model, seen as a degenerate singular limit
case of the relaxed micromorphic model, will be discussed in more
detail in future contributions in which the interest of using a relaxed
micromorphic model at the continuum level will be discussed. It is
clear that, whenever possible, the use of a relaxed micromorphic model
for the description of band-gap metamaterials is preferable to the
use of an internal variable model, at least for two reasons: 
\begin{itemize}
\item it allows to account for non-local effects, i.e. for the description
of metamaterials in which the unit cells can interact 
\item the relationships between the characteristic frequencies and velocities
which appear as basic quantities in the dispersion curves are more
easily related to the constitutive parameters than in the case of
the internal variable model. 
\end{itemize}
In summary, we can say that the case $L_{c}=0$ is a degenerate case
in the sense that the response of the material with respect to wave
propagation becomes completely different from the response that can
be observed in the relaxed micromorphic model with infinitely small,
but non-vanishing, $L_{c}$. We leave to a subsequent work the task
of showing in more detail how the degenerate case $L_{c}=0$ differs
from the limit for very small $L_{c}$ of our relaxed micromorphic
model. We limit here ourselves to remark that non-local effects might
be taken into account when dealing with metamaterials with heterogeneous
microstructures and strong contrasts in the micro-mechanical properties,
even if in some particular cases they may be very small.

\subsubsection{The degenerate case $\mu_{c}=0$ and $L_{c}=0$}

In our previous work \cite{BandGaps1} we have shown that as soon
as non-local effects are present ($L_{c}\neq0$), then the Cosserat
couple modulus $\mu_{c}$ must be non-vanishing (and larger than a
given threshold) in order to have a complete band-gap appearing. The
case with $L_{c}>0$ and $\mu_{c}=0$ still remains well-posed but
no band-gaps are allowed in this case. We explicitly remark here that
if we consider the degenerate case $\mu_{c}=0$ and simultaneously
$L_{c}=0$, then the skew-symmetric part of the micro-distortion tensor
$\PP$ cannot be controlled anymore since the associated governing
equations reduce to $\mathrm{skew}\,\PP_{,tt}=0$. \textit{Mutatis
mutandis}, we claim that such degenerate model is not able to describe
the rotational vibrations in band-gaps metamaterials, even if the
model could be able to provide band-gap behaviors. The phenomenological
consistency of such degenerate model is thus questionable, since one
should individuate a material in which microscopic rotations are forbidden
inside the unit cell in order to let it be applicable. On the other
hand, the case $\mu_{c}\neq0$ and $L_{c}=0$ presented in the previous
Subsection can describe a certain class of band-gap metamaterials
in the very special case in which non-local which allow for rotational
vibrations at relatively high frequencies, but not for non-local effects.

\subsection{The standard micromorphic continuum}

For the standard micromorphic model, we proceed in an analogous way
and we replace the wave-forms (\ref{eq:WaveForm}) for the unknown
fields in the bulk equations (\ref{eq:BulkEq-1}), so obtaining 
\begin{align}
\left(k^{2}\gA\,_{1}^{C}-\omega^{2}\:\mathds1-i\,k\BB\,_{1}^{C}-\CC\,_{1}^{C}\right)\cdot\beta_{1}=0\qquad\qquad & \text{longitudinal waves}\vspace{1.2mm}\label{eq:Long-1}\\
\left(k^{2}\gA\,_{\alpha}^{C}-\omega^{2}\:\mathds1-i\,k\BB\,_{\alpha}^{C}-\CC\,_{\alpha}^{C}\right)\cdot\beta_{\alpha}=0,\qquad\alpha=2,3,\qquad & \text{transverse waves}\vspace{1.2mm}\label{eq:Transv-1}\\
\omega^{2}=A_{4}^{C}\,k^{2}-C_{4}^{C},\qquad\omega^{2}=A_{5}^{C}\,k^{2}-C_{5}^{C},\qquad\omega^{2}=C_{6}^{R}\,k^{2}-C_{6}^{C},\qquad & \text{uncoupled waves}.\label{eq:Uncouplde-1}
\end{align}
The assumptions on the considered constitutive coefficients are the
same given in eq. \ref{eq:DefPos} for the relaxed micromorphic case.

\subsubsection{Uncoupled waves.}

The dispersion relations for uncoupled waves in standard micromorphic
media are the same as those obtained for the relaxed micromorphic
continuum and read: 
\begin{equation}
\omega(k)=\sqrt{\omega_{s}^{2}+c_{g}^{2}k^{2}},\qquad\omega(k)=\sqrt{\omega_{r}^{2}+c_{g}^{2}k^{2}},\qquad\omega(k)=\sqrt{\omega_{s}^{2}+c_{g}^{2}k^{2}},\label{eq:dispersion_uncoupled-3}
\end{equation}
where we set 
\[
c_{g}=\sqrt{\frac{\mu_{e}\,L_{g}^{2}}{\eta}}.
\]
On the basis of what has been explained before, such curves have cut-off
frequencies $\omega_{s}$ and $\omega_{r}$ and they all share the
same asymptote of slope $c_{g}$.

\subsubsection{Longitudinal waves.}

We start by noticing that replacing the wave form (\ref{eq:WaveForm})
in (\ref{eq:Long}) we get 
\begin{equation}
\tilde{\gA\,}_{1}=\left(k^{2}\,\gA\,_{1}^{C}-\omega^{2}\:\mathds1-i\,k\,\BB\,_{1}^{C}-\CC\,_{1}^{C}\right)\cdot\beta_{1}=0.\label{eq:DispersionLong-3}
\end{equation}
Following the same steps presented in Subsection \ref{sub:Long_asymptotics},
we can establish that for $k\rightarrow0$ the longitudinal waves
have the same cut-off frequencies than those evaluated for the relaxed
micromorphic case, in particular: $\omega=0$, $\omega=\omega_{s}$
and $\omega=\omega_{r}$.

In the case $k\rightarrow\infty$ with $k/\omega$ finite, on the
other hand, one finds that in order to have non-trivial solutions
it must be 
\[
\mathrm{det}\left(k^{2}\,\gA\,_{1}^{C}-\omega^{2}\:\mathds1\right)=\mathrm{det}\begin{pmatrix}c_{p}^{2}\,k^{2}-\omega^{2} & 0 & 0\vspace{1.2mm}\\
0 & c_{g}^{2}\:k^{2}-\omega^{2} & 0\vspace{1.2mm}\\
0 & 0 & c_{g}^{2}\:k^{2}-\omega^{2}
\end{pmatrix}=0.
\]
This means that the three dispersion curves become uncoupled for big
wavenumbers and they have asymptotes $c_{p}$, $c_{g}$ and $c_{g}$
respectively, where we define 
\[
c_{g}=\sqrt{\frac{\mu_{e}L_{g}^{2}}{\eta}}.
\]
No horizontal asymptote is found for longitudinal waves in standard
micromorphic media.

On the other hand, given the structure of the polynomial $\mathrm{det}\tilde{\gA\,}_{1}$,
if it is solved in terms of $k=k(\omega)$, six solutions are found
which can be formally written as\footnote{To the sake of simplicity, we do not show the explicit form of the
functions $\tilde{k}_{1}^{1}(\omega)$, $\tilde{k}_{1}^{2}(\omega)$
and $\tilde{k}_{1}^{3}(\omega)$. } 
\[
\pm\tilde{k}_{1}^{1}(\omega),\qquad\text{\ensuremath{\pm}}\tilde{k}_{1}^{2}(\omega),\qquad\text{\ensuremath{\pm}}\tilde{k}_{1}^{3}(\omega).
\]
We denote by $\tilde{h}_{1}^{1}$ and $\tilde{h}_{2}^{1}$ the eigenvectors
associated to the eigenvalues $\tilde{k}_{1}^{1}(\omega)$,$\tilde{k}_{1}^{2}(\omega)$
and $\tilde{k}_{1}^{3}(\omega)$ and verifying (\ref{eq:DispersionLong})
with $\beta_{1}=\tilde{h}_{1}^{1}$ and $\beta_{1}=\tilde{h}_{2}^{1}$,
respectively. According to (\ref{eq:WaveForm}), the final solution
for $\tilde{h}_{1}$ can be finally written as 
\begin{equation}
\tilde{\vv}_{1}=\tilde{\beta}_{1}^{1}\tilde{h}_{1}^{1}e^{i(\pm\tilde{k}_{1}^{1}(\omega)x_{1}-\omega t)}+\tilde{\beta}_{1}^{2}\tilde{h}_{1}^{2}e^{i(\pm\tilde{k}_{1}^{2}(\omega)x_{1}-\omega t)}+\tilde{\beta}_{1}^{3}\tilde{h}_{1}^{3}e^{i(\pm\tilde{k}_{1}^{3}(\omega)x_{1}-\omega t)},\label{eq:WaveSolLong-1}
\end{equation}
where the $+$ or $-$ sign must be chosen for the wavenumber depending
whether the considered wave travels in the $x_{1}$ or $-x_{1}$ direction,
respectively. The three scalar unknowns $\tilde{\beta}_{1}^{1}$,
$\tilde{\beta}_{1}^{2}$ and $\tilde{\beta}_{1}^{3}$ can be found
by imposing suitable boundary conditions as it will be explained later
on.

\subsubsection{Transverse waves.}

We start by noticing that replacing the wave form (\ref{eq:WaveForm})
in (\ref{eq:Transv}) we get 
\begin{equation}
\tilde{\gA\,}_{\alpha}=\left(k^{2}\,\gA\,_{\alpha}^{C}-\omega^{2}\:\mathds1-i\,k\,\BB\,_{\alpha}^{C}-\CC\,_{\alpha}^{C}\right)\cdot\beta_{\alpha}=0.\label{eq:DispersionLong-2-1}
\end{equation}
Following the same steps as before, it is found that the transverse
waves have asymptotes $c_{s}$, $c_{g}$ and $c_{g}$ respectively
and no horizontal asymptote is identified.

\medskip{}
 We conclude that, according to what has been presented in \cite{BandGaps1},
no complete band gaps are possible in standard micromorphic continua
due to the loss of the horizontal asymptotes for longitudinal and
transverse waves.

As done for the longitudinal waves, given the structure of the polynomial
$\mathrm{det}\tilde{\gA\,}_{\alpha}$, if it is solved in terms of
$k=k(\omega)$, six solutions are found which can be formally written
as\footnote{To the sake of simplicity, we do not show the explicit form of the
functions $\tilde{k}_{\alpha}^{1}(\omega)$, $\tilde{k}_{\alpha}^{2}(\omega)$
and $\tilde{k}_{\alpha}^{3}(\omega)$. } 
\[
\pm\tilde{k}_{\alpha}^{1}(\omega),\qquad\text{\ensuremath{\pm}}\tilde{k}_{\alpha}^{2}(\omega),\qquad\text{\ensuremath{\pm}}\tilde{k}_{\alpha}^{3}(\omega),\qquad\alpha=2,3.
\]

We denote $\tilde{h}_{\alpha}^{1}$, $\tilde{h}_{\alpha}^{2}$ and
$\tilde{h}_{\alpha}^{3}$ the eigenvectors associated to the eigenvalues
$\tilde{k}_{\alpha}^{1}(\omega)$, $\tilde{k}_{\alpha}^{2}(\omega)$
and $\tilde{k}_{\alpha}^{3}(\omega)$ and verifying (\ref{eq:DispersionLong-2-1})
with $\beta_{\alpha}=\tilde{h}_{\alpha}^{1}$, $\beta_{\alpha}=\tilde{\vv}_{\alpha}^{2}$
and $\beta_{\alpha}=\tilde{h}_{\alpha}^{3}$, respectively. According
to (\ref{eq:WaveForm}), the final solution for $\tilde{\vv}_{\alpha}$
can be finally written as

\begin{equation}
\tilde{\vv}_{\alpha}=\tilde{\beta}_{\alpha}^{1}\tilde{h}_{\alpha}^{1}e^{i(\pm\tilde{k}_{\alpha}^{1}(\omega)x_{1}-\omega t)}+\tilde{\beta}_{\alpha}^{2}\tilde{h}_{\alpha}^{2}e^{i(\pm\tilde{k}_{\alpha}^{2}(\omega)x_{1}-\omega t)}+\tilde{\beta}_{\alpha}^{3}\tilde{h}_{\alpha}^{3}e^{i(\pm\tilde{k}_{\alpha}^{2}(\omega)x_{1}-\omega t)},\qquad\text{\ensuremath{\alpha}=}2,3.\label{eq:WaveSolTransv-1}
\end{equation}
where the $+$ or $-$ sign must be chosen for the wavenumber depending
whether the considered wave travels in the $x_{1}$ or $-x_{1}$ respectively.
The scalar unknowns $\tilde{\beta}_{\alpha}^{1}$, $\tilde{\beta}_{\alpha}^{2}$
and $\tilde{\beta}_{\alpha}^{3}$ can be found by imposing suitable
boundary conditions as it will be explained in the next section.\medskip{}

\begin{figure}[H]
\begin{centering}
\begin{tabular}{ccccc}
\includegraphics[scale=0.275]{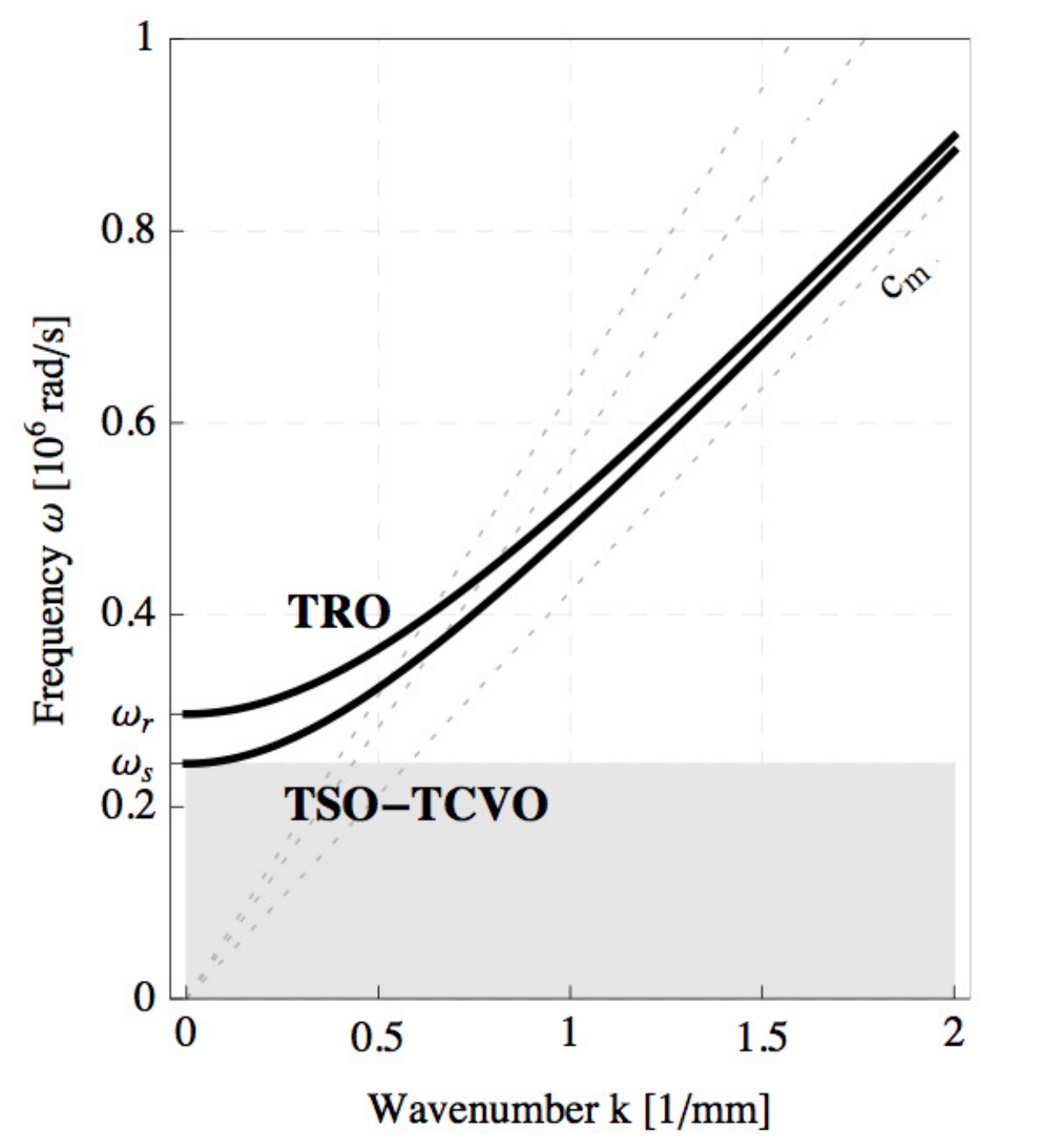}  &  & \includegraphics[scale=0.26]{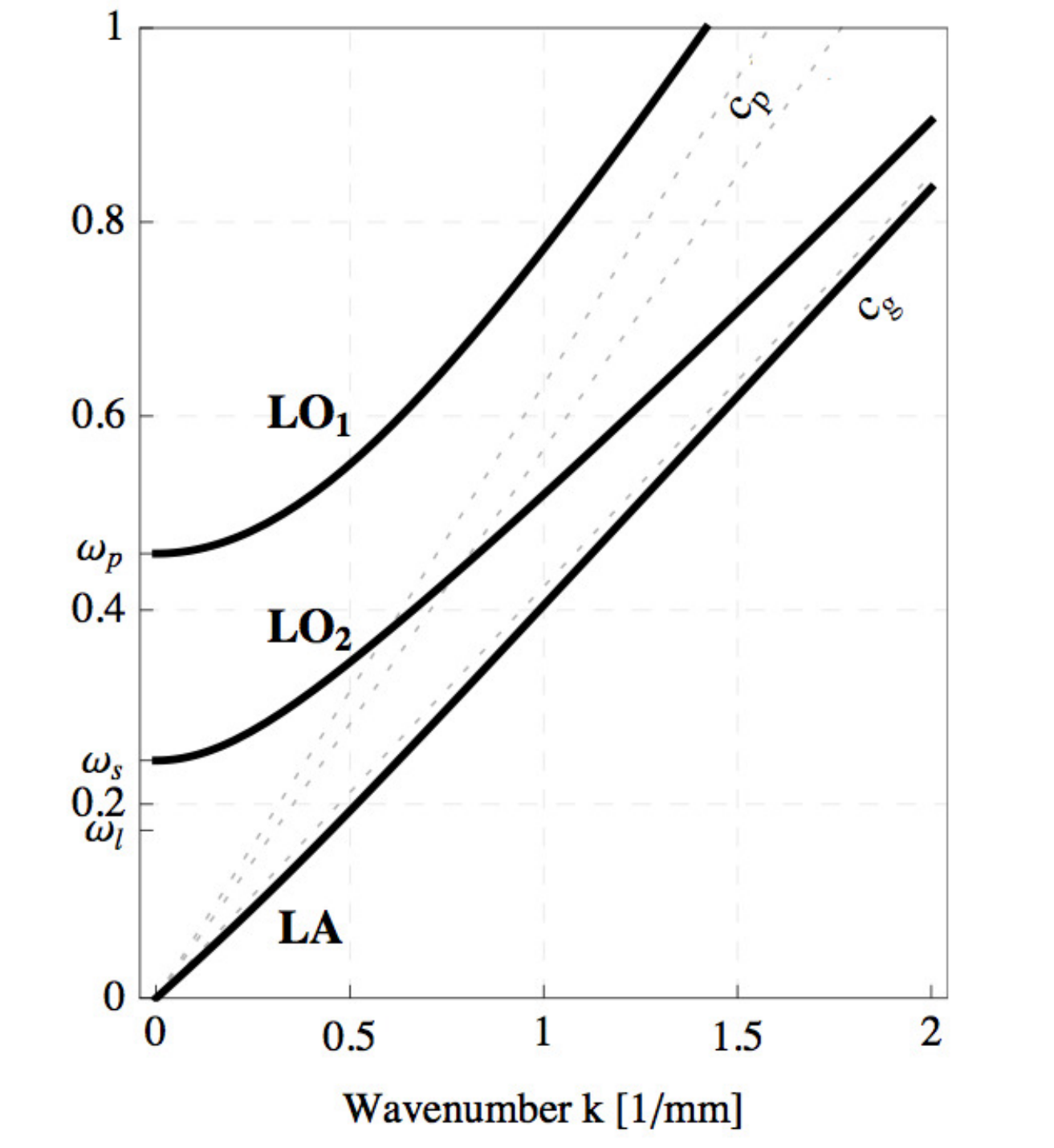}  &  & \includegraphics[scale=0.25]{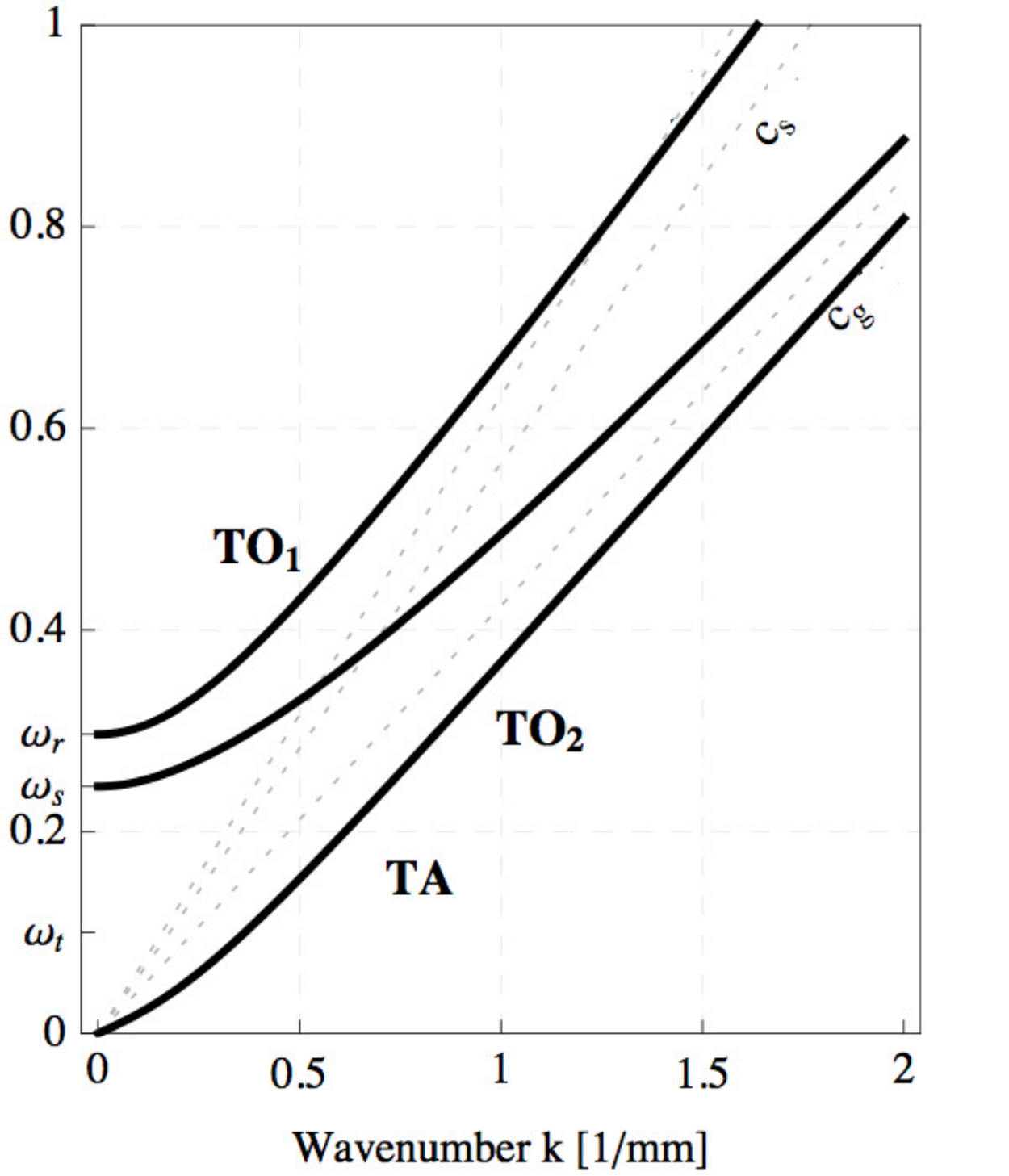}\tabularnewline
(a)  &  & (b)  &  & (c)\tabularnewline
\end{tabular}
\par\end{centering}

\protect\caption{\label{fig:DispersionTotal-Mind}Dispersion relations for all the
considered uncoupled, longitudinal and transverse waves in the case
of standard Mindlin micromorphic continua. No complete band-gap can
be identified.}
\end{figure}

In Figure (\ref{fig:DispersionTotal-Mind}) we show the dispersion
relations for classical Mindlin media grouped as uncoupled, longitudinal
and transverse waves.

\section{Reflection and transmission at\textcolor{black}{{} Cauchy/relaxed-micromorphic
and Cauchy/Mindlin interfaces}}

\textcolor{black}{In order to perform suitable numerical simulations
based on the theoretical framework developed in the previous sections,
w}e study reflection and transmission of plane waves at a surface
of discontinuity between a classical Cauchy continuum and a micromorphic
(relaxed or standard) one which is located at $x_{1}=0$. To this
purpose we introduce the quantities 
\[
J_{i}=\frac{1}{\tau}\int_{0}^{\tau}H_{i}\left(0,t\right)dt,\qquad J_{r}=\frac{1}{\tau}\int_{0}^{\tau}H_{r}\left(0,t\right)dt,\qquad J_{t}=\frac{1}{\tau}\int_{0}^{\tau}H_{t}\left(0,t\right)dt,
\]
where $\tau$ is the period of the traveling plane wave and $H_{i}\,$,
$H_{r}$ and $H_{t}$ are the energy fluxes of the incident, reflected
and transmitted energies, respectively. The reflection ($R$) and
transmission ($T$) coefficients can hence be defined as 
\begin{equation}
R=\frac{J_{r}}{J_{i}},\qquad T=\frac{J_{t}}{J_{i}}.\label{eq:Reflection_coefficient}
\end{equation}
Since the considered system is conservative, one must have $R+T=1$.

For all the numerical simulations presented in this section we chose
the set of parameters shown in Table \ref{tab:Numerical-values} if
not differently specified. 
\begin{table}[H]
\begin{centering}
\begin{tabular}{|c|c|c|c|c|c|c|c|c|c|c|}
\hline 
$\rho$  & $\eta$  & $\mu_{c}$  & $\mu_{e}$  & $\mu_{\rm micro}$  & $\lambda_{\rm micro}$  & $\lambda_{e}$  & $\lambda_{\rm macro}$  & $\mu_{\rm macro}$  & $L_{c}$  & $L_{g}$\tabularnewline
\hline 
\hline 
{[}$Kg/m^{3}${]}  & {[}$Kg/m${]}  & {[}$Pa${]}  & {[}$Pa${]}  & {[}$Pa${]}  & {[}$Pa${]}  & {[}$Pa${]}  & {[}$Pa${]}  & {[}$Pa${]}  & {[}$m${]}  & {[}$m${]}\tabularnewline
\hline 
$2000$  & $10^{-2}$  & $2\times10^{9}$  & $2\times10^{8}$  & $10^{8}$  & $10^{8}$  & $4\times10^{8}$  & $4\times10^{8}$  & $2\times10^{8}$  & $10^{-2}$  & $10^{-2}$\tabularnewline
\hline 
\end{tabular}
\par\end{centering}

\protect\caption{\label{tab:Numerical-values}Numerical values of the constitutive
parameters used for the numerical simulations.}
\end{table}

\subsection{Interface between a Cauchy and a relaxed micromorphic medium}

When considering a classical Cauchy continuum on the $-$ side and
a relaxed micromorphic one on the $+$ side, we have that, recalling
Eqs. (\ref{eq:WaveFormCauchy-2}) (\ref{eq:WaveSolLong}), (\ref{eq:WaveSolTransv}),
(\ref{eq:WaveSolUnc}) the solution can be written in terms of the
calculated eigenvalues and eigenvectors as\footnote{We suppose that the amplitudes $\bar{\alpha}_{1}$, $\bar{\alpha}_{2}$
and $\bar{\alpha}_{3}$ of the incident waves traveling in the Cauchy
continuum are known.} 
\begin{equation}
u_{1}^{-}=u_{1}^{i}+u_{1}^{r},\quad\quad u_{2}^{-}=u_{2}^{i}+u_{2}^{r},\quad\quad u_{3}^{-}=u_{3}^{i}+u_{3}^{r},\label{eq:WaveFormCauchy-1}
\end{equation}
where we set for compactness of notation 
\begin{gather*}
u_{1}^{i}=\bar{\alpha}_{1}\,e^{i(\omega/c_{l}\,x_{1}-\omega t)},\quad\quad u_{1}^{r}=\alpha_{1}\,e^{i(-\omega/c_{l}\,x_{1}-\omega t)},\vspace{1.2mm}\\
u_{2}^{i}=\bar{\alpha}_{2}\,e^{i(\omega/c_{t}\,x_{1}-\omega t)},\quad\quad u_{2}^{r}=\alpha_{2}\,e^{i(-\omega/c_{t}\,x_{1}-\omega t)},\vspace{1.2mm}\\
u_{3}^{i}=\bar{\alpha}_{3}\,e^{i(\omega/c_{t}\,x_{1}-\omega t)},\quad\quad u_{3}^{r}=\alpha_{3}\,e^{i(-\omega/c_{t}\,x_{1}-\omega t)}
\end{gather*}
and moreover 
\begin{gather}
\vv_{1}^{+}=\beta_{1}^{1}h_{1}^{1}e^{i(k_{1}^{1}(\omega)x_{1}-\omega t)}+\beta_{1}^{2}h_{1}^{2}e^{i(k_{1}^{2}(\omega)x_{1}-\omega t)},\quad\quad\vv_{\alpha}^{+}=\beta_{\alpha}^{1}h_{\alpha}^{1}e^{i(k_{\alpha}^{1}(\omega)x_{1}-\omega t)}+\beta_{\alpha}^{2}h_{\alpha}^{2}e^{i(k_{\alpha}^{2}(\omega)x_{1}-\omega t)},\quad\quad\alpha=2,3,\vspace{1.2mm}\nonumber \\
\vspace{1.2mm}\label{eq:WaveForm-1}\\
\vv_{4}^{+}=\beta_{4}\,e^{i(1/c_{m}\:\sqrt{\omega^{2}-\omega_{s}^{2}}\:x_{1}-\omega t)},\quad\quad\vv_{5}^{+}=\beta_{5}\,e^{i(1/c_{m}\:\sqrt{\omega^{2}-\omega_{r}^{2}}\:x_{1}-\omega t)},\quad\quad\vv_{6}^{+}=\beta_{6}\,e^{i(1/c_{m}\:\sqrt{\omega^{2}-\omega_{s}^{2}}\:x_{1}-\omega t)}.\nonumber 
\end{gather}
Given the frequency $\omega$ of the traveling wave, the solution
is hence known except for the amplitudes $\alpha_{i}$ and $\beta_{i}$.
If we count, we have 12 unknown amplitudes to be determined. We focus
here on the study of wave reflection and transmission concerning two
of out of the six possible types of connection between such media
(see Section (\ref{sub:ConnectionsCauchyRelaxed})), because the remaining
four constraints always give rise to complete reflection. In particular,
we will study wave transmission and reflection at surfaces of discontinuity
between a Cauchy medium and a relaxed micromorphic one connected with
the following constraints 
\begin{itemize}
\item macro internal clamp with fixed microstructure, 
\item macro internal clamp with free microstructure. 
\end{itemize}
With the introduced notations, the flux associated to the incident,
reflected and transmitted waves can be computed recalling Eqs. (\ref{Flux-Cauchy-1}),
(\ref{eq:WaveFormCauchy-1}), (\ref{eq:FluxRelaxed}) as 
\begin{flalign}
H_{i}= & -{u}_{1,t}^{i}\left[\left(\lambda_{\rm macro}+2\mu_{\rm macro}\right)\:u_{1,1}^{i}\right]-{u}_{2,t}^{i}\left[\mu_{\rm macro}\,u_{2,1}^{i}\right]-{u}_{3,t}\left[\mu_{\rm macro}\,u_{3,1}^{i}\right],\vspace{1.2mm}\label{Flux-Cauchy-incident}\\
H_{r}= & -{u}_{1,t}^{r}\left[\left(\lambda_{\rm macro}+2\mu_{\rm macro}\right)\:u_{1,1}^{r}\right]-{u}_{2,t}^{r}\left[\mu_{\rm macro}\,u_{2,1}^{r}\right]-{u}_{3,t}\left[\mu_{\rm macro}\,u_{3,1}^{r}\right],\vspace{1.2mm}\label{Flux-Cauchy-incident-1}\\
H_{t}=\: & H_{1}^{1}+H_{1}^{2}+H_{1}^{3}+H_{1}^{4}+H_{1}^{5}+H_{1}^{6},\label{Flux-relaxed-transmitted}
\end{flalign}
where we explicitly notice that Eqs. (\ref{FlLong}) and (\ref{eq:WaveForm-1})
are used for the computation of the transmitted flux.

\subsubsection{Macro internal clamp with fixed microstructure}

We recall from Subsection (\ref{sub:Macro-clamp-fixed_micro}) that
the 12 jump conditions to be imposed at $x_{1}=0$ for the considered
connection are 
\[
\left\llbracket u_{i}\right\rrbracket =0,\qquad t_{i}-f_{i}=0,\qquad P_{ij}^{+}=0,\qquad i=1,2,3,\ \ j=2,3,
\]
which in terms of the new variables (see Eqs. (\ref{eq:Cauchy_force}),
(\ref{eq:JumpRelaxed-1_1}), (\ref{eq:P_change_var}), (\ref{eq:Unknowns1})and
(\ref{eq:Unknowns2})), recalling that $\nn\,=(1,0,0)$ and introducing
the tangent vectors $\gtau_{1}=(0,1,0)$ and $\gtau_{2}=(0,0,1)$,
read\medskip{}
 
\[
\vv_{1}^{+}\cdot\nn\,-u_{1}^{-}=0,\quad\vv_{2}^{+}\cdot\nn\,-u_{2}^{-}=0,\quad\vv_{3}^{+}\cdot\nn\,-u_{3}^{-}=0,
\]
\begin{gather}
\begin{pmatrix}\lambda_{e}+2\mu_{e}\vspace{1.2mm}\\
0\vspace{1.2mm}\\
0
\end{pmatrix}\cdot\left(\vv_{1}^{+}\right)'+\begin{pmatrix}0\vspace{1.2mm}\\
-2\mu_{e}\vspace{1.2mm}\\
-(3\lambda_{e}+2\mu_{e})
\end{pmatrix}\cdot\vv_{1}^{+}=(\lambda_{\rm macro}+2\mu_{\rm macro})\,\left(u_{1}^{-}\right)'\vspace{1.2mm},\nonumber \\
\vspace{1.2mm}\label{eq:JumpRelaxed-1_1-1}\\
\begin{pmatrix}\mu_{e}+\mu_{c}\vspace{1.2mm}\\
0\vspace{1.2mm}\\
0
\end{pmatrix}\cdot\left(\vv_{2}^{+}\right)'+\begin{pmatrix}0\vspace{1.2mm}\\
-2\mu_{e}\vspace{1.2mm}\\
2\mu_{c}
\end{pmatrix}\cdot\vv_{2}^{+}=\mu_{\rm macro}\,\left(u_{2}^{-}\right)',\quad\quad\begin{pmatrix}\mu_{e}+\mu_{c}\vspace{1.2mm}\\
0\vspace{1.2mm}\\
0
\end{pmatrix}\cdot\left(\vv_{3}^{+}\right)'+\begin{pmatrix}0\vspace{1.2mm}\\
-2\mu_{e}\vspace{1.2mm}\\
2\mu_{c}
\end{pmatrix}\cdot\vv_{3}^{+}=\mu_{\rm macro}\,\left(u_{3}^{-}\right)',\nonumber 
\end{gather}
\medskip{}
 
\begin{flalign}
P_{22}= & \,\frac{1}{2}\left(\vv_{6}^{+}+2\,\vv_{1}^{+}\cdot\gtau_{2}-\vv_{1}^{+}\cdot\gtau_{1}\right)=0, & P_{23}= & \,\vv_{4}^{+}+\vv_{5}^{+}=0, & P_{32}= & \,\vv_{4}^{+}-\vv_{5}^{+}=0,\vspace{1.2mm}\nonumber \\
\vspace{1.2mm}\label{eq:P_change_var-1}\\
P_{33}= & \,\frac{1}{2}\left(-\vv_{6}^{+}+2\,\vv_{1}^{+}\cdot\gtau_{2}-\vv_{1}^{+}\cdot\gtau_{1}\right)=0, & P_{12}= & \,\vv_{2}^{+}\cdot\gtau_{1}+\vv_{2}^{+}\cdot\gtau_{2}=0, & P_{13}= & \,\vv_{^{+}3}\cdot\gtau_{1}+\vv_{3}^{+}\cdot\gtau_{2}=0.\nonumber 
\end{flalign}
Inserting the wave-form solutions (\ref{eq:WaveFormCauchy-1}) and
(\ref{eq:WaveForm-1}) with $x_{1}=0$ in such expressions for the
jump conditions the unknown amplitudes can be determined. We do not
report here their explicit form since it is rather complex and it
does not add essential information to the present treatise. Nevertheless,
we can easily notice that from the boundary conditions $P_{23}=P_{32}=0,$
it follows that $\beta_{4}=\beta_{5}=0$ and hence ${v_{4}=\vv_{5}=0}\ \forall x_{1}$.
This means that the modes $\vv_{4}$ and $\vv_{5}$ cannot be activated
at the considered interface between a relaxed micromorphic medium
and a Cauchy continuum.

\medskip{}
 Once the amplitudes (and hence the solution) have been determined,
the incident, reflected and transmitted flux can be computed according
to equations (\ref{Flux-Cauchy-incident})-(\ref{Flux-relaxed-transmitted}),
in which the amplitudes calculated for the constraint considered in
this Subsection are used. The reflection and transmission coefficients
$R$ and $T$ can then be computed according to Eqs. (\ref{eq:Reflection_coefficient})

\begin{figure}[H]
\begin{centering}
\includegraphics[scale=0.7]{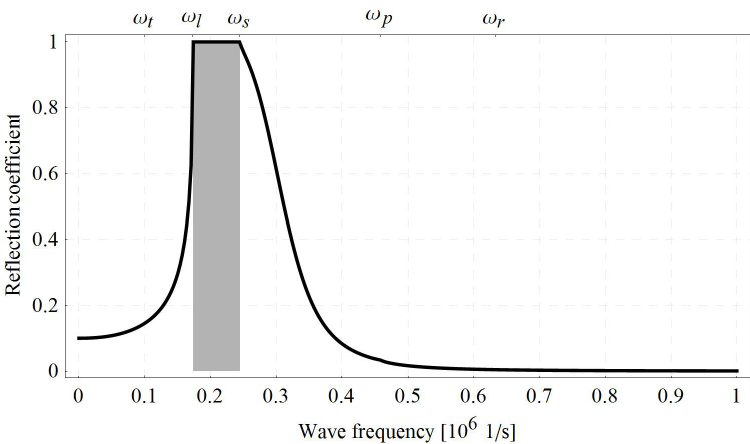} 
\par\end{centering}

\protect\caption{\label{fig:RP}\textcolor{black}{Cauchy/relaxed-micromorphic interface:
macro clamp with fixed microstructure. }Reflection coefficient as
function of frequency for incident P waves ($\bar{\alpha}_{1}=1,\ \bar{\alpha}_{2}=\bar{\alpha}_{3}=0$).
Complete reflection is triggered in the frequency range for which
band-gaps are known to occur in bulk wave propagation.}
\end{figure}

\begin{figure}[H]
\begin{centering}
\includegraphics[scale=0.7]{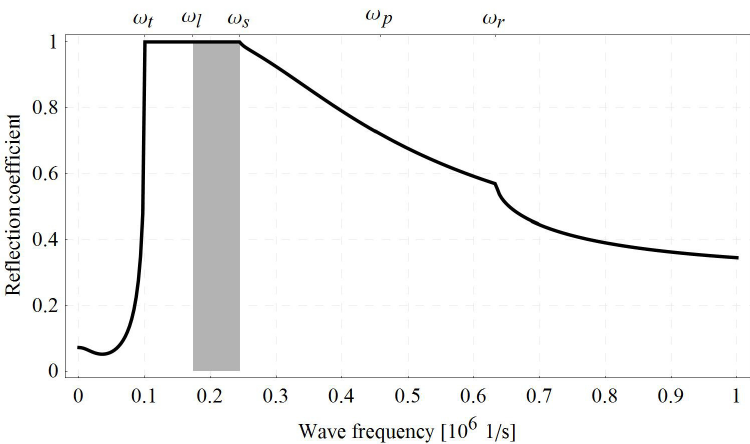} 
\par\end{centering}

\protect\caption{\label{fig:RS}\textcolor{black}{Cauchy/relaxed-micromorphic interface:
macro clamp with fixed microstructure.} Reflection coefficient as
function of frequency for incident S waves ($\bar{\alpha}_{1}=0,\ \bar{\alpha}_{2}=1,\ \bar{\alpha}_{3}=0$,
or equivalently $\bar{\alpha}_{1}=0,\ \bar{\alpha}_{2}=0,\ \bar{\alpha}_{3}=1$).
Complete reflection is triggered in the frequency range for which
band-gaps are known to occur in bulk wave propagation.}
\end{figure}

\begin{figure}[H]
\begin{centering}
\includegraphics[scale=0.7]{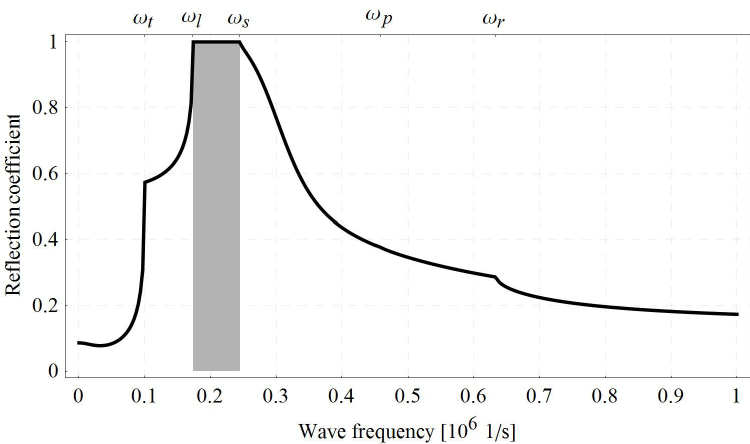} 
\par\end{centering}

\protect\caption{\label{fig:RPS}\textcolor{black}{Cauchy/relaxed-micromorphic interface:
macro clamp with fixed microstructure.} Reflection coefficient as
function of frequency for incident P+S wave ($\bar{\alpha}_{1}=1,\ \bar{\alpha}_{2}=1,\ \bar{\alpha}_{3}=1$).
Complete reflection is triggered in the frequency range for which
band-gaps are known to occur in bulk wave propagation.}
\end{figure}

Figures \ref{fig:RP}, \ref{fig:RS} and \ref{fig:RPS} show the reflection
coefficient $R$ plotted as a function of the frequency $\omega$
for the case of incident longitudinal, transverse or generic (longitudinal
+ transverse) waves and for the considered constraint. We explicitly
comment only the plot referring to the generic incident wave (Figure
\ref{fig:RPS}), being the remarks concerning pure longitudinal and
pure transverse incident waves completely analogous. Hence, with reference
to Figure \ref{fig:RPS} it can be noticed that for the frequency
range $[\omega_{l},\omega_{s}]$ a complete reflection takes place.
This is quite sensible considered that the frequency interval $[\omega_{l},\omega_{s}]$
is the one for which a complete band gap occurs when considering bulk
wave propagation (see Figure \ref{fig:DispersionTotal}): since no
wave can propagate in the relaxed medium in such frequency range,
all the energy carried by the incident wave is completely reflected
and travels back in the Cauchy medium. On the other hand, it can be
inferred that, outside the band-gap frequency interval, the amount
of reflected energy drops drastically, which means that a non-negligible
amount of the energy initially transported by the incident wave is
transmitted in the relaxed medium. In other words we can say that,
outside the range of frequencies for which band-gaps occur, the continuity
of the macroscopic displacement allows the connection between the
Cauchy medium and the relaxed micromorphic medium in such a way that
a considerable amount of energy is transferred to the relaxed continuum.

\subsubsection{\label{sub:freeMicro}Macro internal clamp with free microstructure}

We recall from Subsection (\ref{sub:Macro-clamp-fixed_micro}) that
the 12 jump conditions to be imposed at $x_{1}=0$ for the considered
connection are 
\[
\left\llbracket u_{i}\right\rrbracket =0,\qquad t_{i}-f_{i}=0,\qquad\tau_{ij}^{+}=0,\qquad i=1,2,3,\ \ j=2,3,
\]
which in terms of the new variables (see Eqs. (\ref{eq:Cauchy_force}),
(\ref{eq:JumpRelaxed-1_1}), (\ref{eq:P_change_var}), (\ref{eq:Unknowns1})
and (\ref{eq:Unknowns2})) and introducing the tangent vectors $\gtau_{1}=(0,1,0)$
and $\gtau_{2}=(0,0,1)$, read\medskip{}
 
\[
\left\llbracket u_{1}\right\rrbracket =0,\quad\left\llbracket u_{2}\right\rrbracket =0,\quad\left\llbracket u_{3}\right\rrbracket =0,
\]
\begin{gather}
\begin{pmatrix}\lambda_{e}+2\mu_{e}\vspace{1.2mm}\\
0\vspace{1.2mm}\\
0
\end{pmatrix}\cdot\left(\vv_{1}^{+}\right)'+\begin{pmatrix}0\vspace{1.2mm}\\
-2\mu_{e}\vspace{1.2mm}\\
-(3\lambda_{e}+2\mu_{e})
\end{pmatrix}\cdot\vv_{1}^{+}=(\lambda_{\rm macro}+2\mu_{\rm macro})\,\left(u_{1}^{-}\right)'\vspace{1.2mm},\nonumber \\
\vspace{1.2mm}\label{eq:JumpRelaxed-1_1-1-1}\\
\begin{pmatrix}\mu_{e}+\mu_{c}\vspace{1.2mm}\\
0\vspace{1.2mm}\\
0
\end{pmatrix}\cdot\left(\vv_{\alpha}^{+}\right)'+\begin{pmatrix}0\vspace{1.2mm}\\
-2\mu_{e}\vspace{1.2mm}\\
2\mu_{c}
\end{pmatrix}\cdot\vv_{\alpha}^{+}=\mu_{\rm macro}\,\left(u_{\alpha}^{-}\right)',\qquad\alpha=2,3,\nonumber 
\end{gather}
\medskip{}
 
\begin{flalign}
\tau_{22}= & \begin{pmatrix}0\vspace{1.2mm}\\
-\frac{\mu_{e}L_{c}^{2}}{2}\vspace{1.2mm}\\
\mu_{e}L_{c}^{2}
\end{pmatrix}\cdot\left(\vv_{1}^{+}\right)'+\frac{\mu_{e}L_{c}^{2}}{2}\left(\vv_{6}^{+}\right)'=0, & \tau_{23}=\  & \mu_{e}L_{c}^{2}\left(\left(\vv_{4}^{+}\right)'+\left(\vv_{5}^{+}\right)'\right), & \tau_{12}= & \begin{pmatrix}0\vspace{1.2mm}\\
\mu_{e}L_{c}^{2}\vspace{1.2mm}\\
\mu_{e}L_{c}^{2}
\end{pmatrix}\cdot\left(\vv_{2}^{+}\right)'=0,\vspace{1.2mm}\nonumber \\
\vspace{1.2mm}\label{eq:P_change_var-1-1}\\
\tau_{33}= & \begin{pmatrix}0\vspace{1.2mm}\\
-\frac{\mu_{e}L_{c}^{2}}{2}\vspace{1.2mm}\\
\mu_{e}L_{c}^{2}
\end{pmatrix}\cdot\left(\vv_{1}^{+}\right)'-\frac{\mu_{e}L_{c}^{2}}{2}\left(\vv_{6}^{+}\right)'=0 & \tau_{32}=\  & \mu_{e}L_{c}^{2}\left(\left(\vv_{4}^{+}\right)'-\left(\vv_{5}^{+}\right)'\right)=0, & \tau_{13}= & \begin{pmatrix}0\vspace{1.2mm}\\
\mu_{e}L_{c}^{2}\vspace{1.2mm}\\
\mu_{e}L_{c}^{2}
\end{pmatrix}\cdot\left(\vv_{3}^{+}\right)'=0.\nonumber 
\end{flalign}
Inserting the wave-form solutions (\ref{eq:WaveFormCauchy-1}) and
(\ref{eq:WaveForm-1}) with $x_{1}=0$ in these expressions for the
jump conditions the unknown amplitudes can be determined. As for the
previous case, we do not report here their explicit form for the sake
of conciseness. Nevertheless, we can notice that from the boundary
conditions $\tau_{23}=\tau_{32}=0,$ it follows that $\beta_{4}=\beta_{5}=0$
and hence ${v_{4}=\vv_{5}=0}$ for all $x_{1}$. This means that the
modes $\vv_{4}$ and $\vv_{5}$ cannot be activated at the considered
interface between a relaxed micromorphic medium and a Cauchy continuum.\medskip{}

Once that the amplitudes (and hence the solution) have been determined,
the incident, reflected and transmitted flux can be computed according
to equations (\ref{Flux-Cauchy-incident})-(\ref{Flux-relaxed-transmitted}),
in which the amplitudes calculated for the constraint considered in
this Subsection are used.

\begin{figure}[H]
\begin{centering}
\includegraphics[scale=0.7]{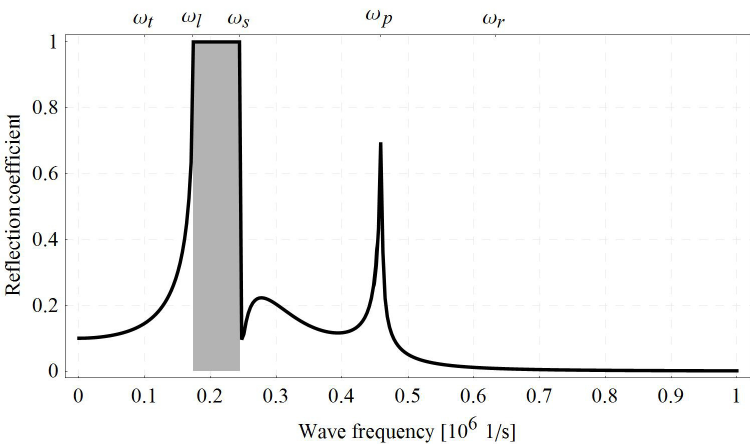} 
\par\end{centering}

\protect\caption{\label{fig:RP1}\textcolor{black}{Cauchy/relaxed-micromorphic interface:
macro clamp with free microstructure. }Reflection coefficient as function
of frequency for incident P waves ($\bar{\alpha}_{1}=1,\ \bar{\alpha}_{2}=\bar{\alpha}_{3}=0$).
Complete reflection is triggered in the frequency range for which
band-gaps are known to occur in bulk wave propagation. A local resonance
frequency can be recognized at $\omega_{p}$ due to the fact that
the microstructure is free to vibrate.}
\end{figure}

\begin{figure}[H]
\begin{centering}
\includegraphics[scale=0.7]{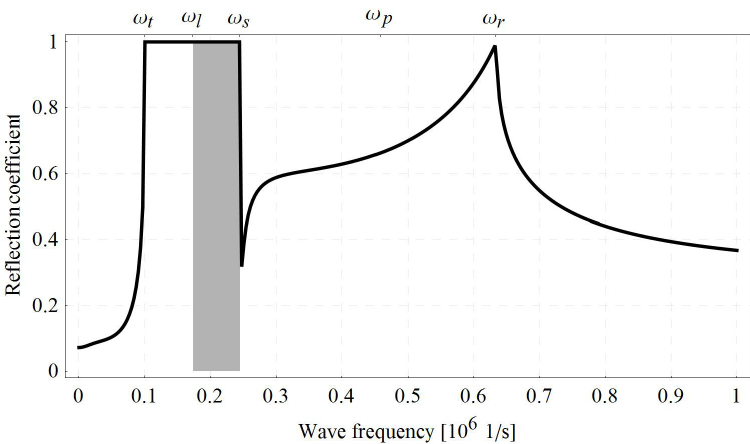} 
\par\end{centering}

\protect\caption{\label{fig:RS1}\textcolor{black}{Cauchy/relaxed-micromorphic interface:
macro clamp with free microstructure. }Reflection coefficient as function
of frequency for incident S waves ($\bar{\alpha}_{1}=0,\ \bar{\alpha}_{2}=1,\ \bar{\alpha}_{3}=0$,
or equivalently $\bar{\alpha}_{1}=0,\ \bar{\alpha}_{2}=0,\ \bar{\alpha}_{3}=1$).
Complete reflection is triggered in the frequency range for which
band-gaps are known to occur in bulk wave propagation. A local resonance
frequency can be recognized at $\omega_{r}$ due to the fact that
the microstructure is free to vibrate.}
\end{figure}

\begin{figure}[H]
\begin{centering}
\includegraphics[scale=0.7]{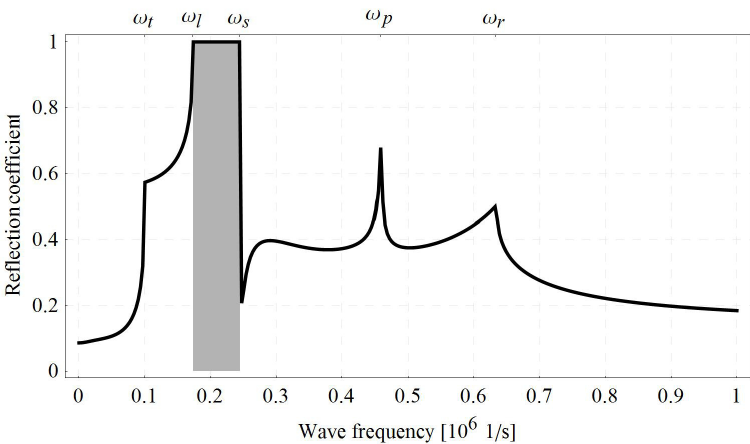} 
\par\end{centering}

\protect\caption{\label{fig:RPS1}\textcolor{black}{Cauchy/relaxed-micromorphic interface:
macro clamp with free microstructure. }Reflection coefficient as function
of frequency for incident P+S wave ($\bar{\alpha}_{1}=1,\ \bar{\alpha}_{2}=1,\ \bar{\alpha}_{3}=1$).
Complete reflection is triggered in the frequency range for which
band-gaps are known to occur in bulk wave propagation. Two local resonance
frequencies can be recognized at $\omega_{p}$ and $\omega_{r}$ due
to the fact that the microstructure is free to vibrate.}
\end{figure}

In Figures \ref{fig:RP1}, \ref{fig:RS1} and \ref{fig:RPS1}, we
show the behavior of the reflection coefficient $R$ as a function
of frequency for longitudinal, transverse and generic (longitudinal
+ transverse) incident waves. Once again, being the case of pure longitudinal
and pure transverse incident wave completely analogous, we only comment
here the behavior of the reflection coefficient for generic (longitudinal
+ transverse) waves, with reference to Figure \ref{fig:RPS1}. It
can be seen that a complete reflection can be observed in the band-gap
frequency interval. On the other hand, some phenomena of localized
resonances occur at $\omega_{p}$ and $\omega_{r}$ for longitudinal
and transverse waves, respectively. This means that the cut-off frequencies
$\omega_{p}$ and $\omega_{r}$ are indeed resonance frequencies for
the considered free microstructure. Such peaks of reflected energy
can hence be completely associated to the characteristics of the considered
microstructures and to their characteristic resonant behaviors. 

\textcolor{black}{We explicitly remark that, suitably tuning the values
of the parameters shown in Table \ref{tab:Numerical-values}, the
local resonant behaviors presented in Figures \ref{fig:RP1}-\ref{fig:RPS1}
can be modified and may eventually lead to the formation of a second
band gap entirely due to the presence of the interface. The fact of
creating a second band gap can be directly related to both i) the
values of the constitutive parameters presented in table \ref{tab:Numerical-values}
and ii) the type of the chosen connection, namely the }\textit{\textcolor{black}{internal
clamp with free microstructure}}\textcolor{black}{. Suitably tuning
the parameters of the relaxed micromorphic model can thus lead to
the description of such double band gap profiles which, as shown in
\cite{NonLoc}, may be of help for the modeling of real phononic crystals
of the type studied in \cite{Lucklum}.}

\subsection{Interface between a Cauchy and a standard micromorphic medium}

When considering a classical Cauchy continuum on the $-$ side and
a standard micromorphic one on the $+$ side, we have that, recalling
Eqs. (\ref{eq:WaveFormCauchy-2}) (\ref{eq:WaveSolLong}), (\ref{eq:WaveSolTransv}),
(\ref{eq:WaveSolUnc}) that the solution can be written in terms of
the calculated eigenvalues and eigenvectors as\footnote{We suppose that the amplitudes $\bar{\alpha}_{1}$, $\bar{\alpha}_{2}$
and $\bar{\alpha}_{3}$ of the incident waves traveling in the Cauchy
continuum are known.} 
\begin{equation}
u_{1}^{-}=u_{1}^{i}+u_{1}^{r},\quad\quad u_{2}^{-}=u_{2}^{i}+u_{2}^{r},\quad\quad u_{3}^{-}=u_{3}^{i}+u_{3}^{r},\label{eq:WaveFormCauchy-1-1}
\end{equation}
where we set for compactness of notation 
\begin{gather*}
u_{1}^{i}=\bar{\alpha}_{1}\,e^{i(\omega/c_{l}\,x_{1}-\omega t)},\quad\quad u_{1}^{r}=\alpha_{1}\,e^{i(-\omega/c_{l}\,x_{1}-\omega t)},\vspace{1.2mm}\\
\vspace{1.2mm}\\
u_{2}^{i}=\bar{\alpha}_{2}\,e^{i(\omega/c_{l}\,x_{1}-\omega t)},\quad\quad u_{2}^{r}=\alpha_{2}\,e^{i(-\omega/c_{l}\,x_{1}-\omega t)},\vspace{1.2mm}\\
\vspace{1.2mm}\\
u_{3}^{i}=\bar{\alpha}_{3}\,e^{i(\omega/c_{l}\,x_{1}-\omega t)},\quad\quad u_{3}^{r}=\alpha_{3}',e^{i(-\omega/c_{l}\,x_{1}-\omega t)},
\end{gather*}
and moreover 
\begin{gather}
\tilde{\vv}_{1}^{+}=\tilde{\beta}_{1}^{1}\tilde{\vv}_{1}^{1}e^{i(\tilde{k}_{1}^{1}(\omega)x_{1}-\omega t)}+\tilde{\beta}_{1}^{2}\tilde{\vv}_{1}^{2}e^{i(\tilde{k}_{1}^{2}(\omega)x_{1}-\omega t)}+\tilde{\beta}_{1}^{3}\tilde{\vv}_{1}^{3}e^{i(\tilde{k}_{1}^{3}(\omega)x_{1}-\omega t)}\vspace{1.2mm},\nonumber \\
\vspace{1.2mm}\nonumber \\
\tilde{\vv}_{\alpha}^{+}=\tilde{\beta}_{\alpha}^{1}\tilde{h}_{\alpha}^{1}e^{i(\tilde{k}_{\alpha}^{1}(\omega)x_{1}-\omega t)}+\tilde{\beta}_{\alpha}^{2}\tilde{h}_{\alpha}^{2}e^{i(\tilde{k}_{\alpha}^{2}(\omega)x_{1}-\omega t)}+\tilde{\beta}_{\alpha}^{3}\tilde{h}_{\alpha}^{3}e^{i(\tilde{k}_{\alpha}^{3}(\omega)x_{1}-\omega t)},\quad\alpha=2,3,\vspace{1.2mm}\label{eq:WaveForm-1-1}\\
\vspace{1.2mm}\nonumber \\
\tilde{\vv}_{4}^{+}=\tilde{\beta}_{4}\,e^{i(1/c_{m}\:\sqrt{\omega^{2}-\omega_{s}^{2}}\:x_{1}-\omega t)},\quad\quad\tilde{\vv}_{5}^{+}=\tilde{\beta}_{5}\,e^{i(1/c_{m}\:\sqrt{\omega^{2}-\omega_{r}^{2}}\:x_{1}-\omega t)},\quad\quad\tilde{\vv}_{6}^{+}=\tilde{\beta}_{6}\,e^{i(1/c_{m}\:\sqrt{\omega^{2}-\omega_{s}^{2}}\:x_{1}-\omega t)}.\nonumber 
\end{gather}
Given the frequency $\omega$ of the traveling wave, the solution
is hence known except for the amplitudes $\alpha_{i}$ and $\tilde{\beta}_{i}$.
If we count, we have 15 unknown amplitudes to be determined. We focus
here on the study of wave reflection and transmission concerning two
out of the six possible types of connection between such media (see
Section (\ref{sub:ConnectionsCauchyRelaxed})), because the remaining
four constraints always give rise to complete reflection. In particular,
we will study wave transmission and reflection at surfaces of discontinuity
between a Cauchy medium and a relaxed micromorphic one connected with
the following constraints 
\begin{itemize}
\item macro internal clamp with fixed microstructure, 
\item macro internal clamp with free microstructure. 
\end{itemize}
With the introduced notations, the flux associated to the incident,
reflected and transmitted waves can be computed recalling Eqs. (\ref{Flux-Cauchy-1}),
(\ref{eq:WaveFormCauchy-1-1}), (\ref{eq:FluxRelaxed}), as 
\begin{flalign}
H_{i}= & -{u}_{1,t}^{i}\left[\left(\lambda_{\rm macro}+2\mu_{\rm macro}\right)\:u_{1,1}^{i}\right]-{u}_{2,t}^{i}\left[\mu_{\rm macro}\,u_{2,1}^{i}\right]-{u}_{3,t}\left[\mu_{\rm macro}\,u_{3,1}^{i}\right],\vspace{1.2mm}\label{Flux-Cauchy-incident-2}\\
H_{r}= & -{u}_{1,t}^{r}\left[\left(\lambda_{\rm macro}+2\mu_{\rm macro}\right)\:u_{1,1}^{r}\right]-{u}_{2,t}^{r}\left[\mu_{\rm macro}\,u_{2,1}^{r}\right]-{u}_{3,t}\left[\mu_{\rm macro}\,u_{3,1}^{r}\right],\vspace{1.2mm}\label{Flux-Cauchy-incident-1-1}\\
H_{t}= & \ H_{1}^{1}+H_{1}^{2}+H_{1}^{3}+H_{1}^{4}+H_{1}^{5}+H_{1}^{6},\label{Flux-relaxed-transmitted-1}
\end{flalign}
where we explicitly notice that Eqs. (\ref{FlLong}) and (\ref{eq:WaveForm-1-1})
are used for the computation of the transmitted flux.

\subsubsection{Macro internal clamp with fixed microstructure}

We recall from Subsection (\ref{sub:Macro-clamp-fixed_micro}) that
the 12 jump conditions to be imposed at $x_{1}=0$ for the considered
connection are 
\[
\left\llbracket u_{i}\right\rrbracket =0,\qquad t_{i}-f_{i}=0,\qquad P_{ij}^{+}=0,\qquad i=1,2,3,\ \ j=2,3,
\]
which in terms of the new variables (see Eqs. (\ref{eq:Cauchy_force}),
(\ref{eq:JumpRelaxed-1_1}), (\ref{eq:P_change_var}), (\ref{eq:Unknowns1})and
(\ref{eq:Unknowns2})), recalling that $\nn\,=(1,0,0)$ and introducing
the tangent vectors $\gtau_{1}=(0,1,0)$ and $\gtau_{2}=(0,0,1)$,
read\medskip{}
 
\[
\vv_{1}^{+}\cdot\nn\,-u_{1}^{-}=0,\quad\quad\vv_{2}^{+}\cdot\nn\,-u_{2}^{-}=0,\quad\quad\vv_{3}^{+}\cdot\nn\,-u_{3}^{-}=0,
\]
\medskip{}
 
\begin{gather}
\begin{pmatrix}\lambda_{e}+2\mu_{e}\vspace{1.2mm}\\
0\vspace{1.2mm}\\
0
\end{pmatrix}\cdot\left(\vv_{1}^{+}\right)'+\begin{pmatrix}0\vspace{1.2mm}\\
-2\mu_{e}\vspace{1.2mm}\\
-(3\lambda_{e}+2\mu_{e})
\end{pmatrix}\cdot\vv_{1}^{+}=(\lambda_{\rm macro}+2\mu_{\rm macro})\,\left(u_{1}^{-}\right)'\vspace{1.2mm},\nonumber \\
\vspace{1.2mm}\label{eq:JumpRelaxed-1_1-1-2}\\
\begin{pmatrix}\mu_{e}+\mu_{c}\vspace{1.2mm}\\
0\vspace{1.2mm}\\
0
\end{pmatrix}\cdot\left(\vv_{2}^{+}\right)'+\begin{pmatrix}0\vspace{1.2mm}\\
-2\mu_{e}\vspace{1.2mm}\\
2\mu_{c}
\end{pmatrix}\cdot\vv_{2}^{+}=\mu_{\rm macro}\,\left(u_{2}^{-}\right)',\quad\quad\begin{pmatrix}\mu_{e}+\mu_{c}\vspace{1.2mm}\\
0\vspace{1.2mm}\\
0
\end{pmatrix}\cdot\left(\vv_{3}^{+}\right)'+\begin{pmatrix}0\vspace{1.2mm}\\
-2\mu_{e}\vspace{1.2mm}\\
2\mu_{c}
\end{pmatrix}\cdot\vv_{3}^{+}=\mu_{\rm macro}\,\left(u_{3}^{-}\right)',\nonumber 
\end{gather}
\medskip{}
 
\begin{flalign}
P_{22}= & \,\frac{1}{2}\left(\vv_{6}^{+}+2\,\vv_{1}^{+}\cdot\gtau_{2}-\vv_{1}^{+}\cdot\gtau_{1}\right)=0, & P_{23}= & \vv_{4}^{+}+\vv_{5}^{+}=0 & P_{32}= & \,\vv_{4}^{+}-\vv_{5}^{+}=0,\vspace{1.2mm}\nonumber \\
\vspace{1.2mm}\label{eq:P_change_var-1-2}\\
P_{33}= & \,\frac{1}{2}\left(-\vv_{6}^{+}+2\,\vv_{1}^{+}\cdot\gtau_{2}-\vv_{1}^{+}\cdot\gtau_{1}\right)=0, & P_{12}= & \vv_{2}^{+}\cdot\gtau_{1}+\vv_{2}^{+}\cdot\gtau_{2}=0, & P_{13}= & \,\vv_{^{+}3}\cdot\gtau_{1}+\vv_{3}^{+}\cdot\gtau_{2}=0.\nonumber 
\end{flalign}
\medskip{}
Inserting the wave-form solutions (\ref{eq:WaveFormCauchy-1-1}) and
(\ref{eq:WaveForm-1-1}) with $x_{1}=0$ in such expressions for the
jump conditions the unknown amplitudes can be determined. We do not
report here their explicit form since it is rather complex and it
does not add essential information to the present treatise. Nevertheless,
we can easily notice that from the boundary conditions $P_{23}=P_{32}=0,$
it follows that $\beta_{4}=\beta_{5}=0$ and hence ${v_{4}=\vv_{5}=0}$
for all $x_{1}$. This means that the modes $\vv_{4}$ and $\vv_{5}$
cannot be activated at the considered interface between a relaxed
micromorphic medium and a Cauchy continuum.

\medskip{}
 Once that the amplitudes (and hence the solution) have been determined,
the incident, reflected and transmitted flux can be computed according
to equations (\ref{Flux-Cauchy-incident-2})-(\ref{Flux-relaxed-transmitted-1}),
in which the amplitudes calculated for the constraint considered in
this Subsection are used.

\begin{figure}[H]
\begin{centering}
\includegraphics[scale=0.7]{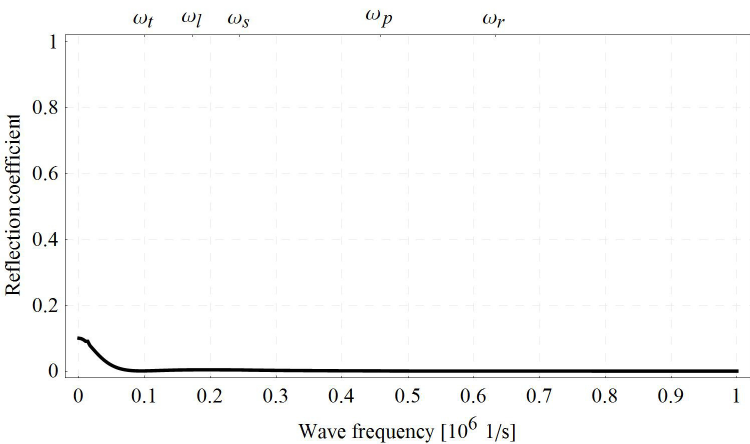} 
\par\end{centering}

\protect\caption{\label{fig:RPMind}\textcolor{black}{Cauchy/Mindlin interface: macro
clamp with fixed microstructure. Reflection coefficient as function
of frequency for incident P waves ($\bar{\alpha}_{1}=1,\ \bar{\alpha}_{2}=\bar{\alpha}_{3}=0$).
Due to the absence of band-gaps in Mindlin continua, a very low amount
of energy is reflected for any value of the frequency in the observed
interval. }}
\end{figure}

\begin{figure}[H]
\begin{centering}
\includegraphics[scale=0.7]{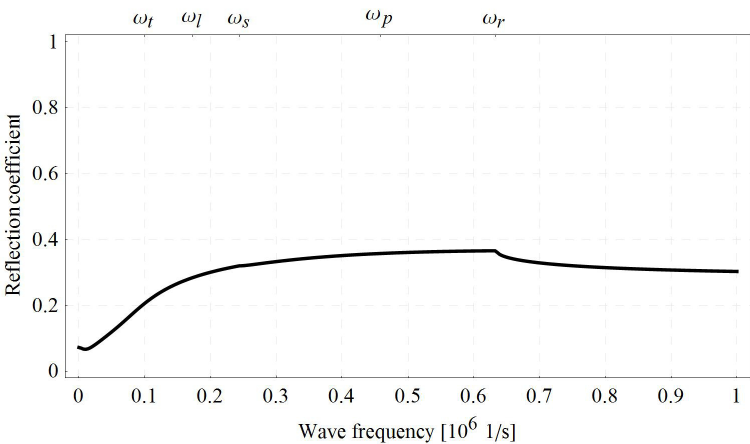} 
\par\end{centering}

\protect\caption{\label{fig:RSMind}\textcolor{black}{Cauchy/Mindlin interface: macro
clamp with fixed microstructure. Reflection coefficient as function
of frequency for incident S waves ($\bar{\alpha}_{1}=0,\ \bar{\alpha}_{2}=1,\ \bar{\alpha}_{3}=0$,
or equivalently $\bar{\alpha}_{1}=0,\ \bar{\alpha}_{2}=0,\ \bar{\alpha}_{3}=1$).
Due to the absence of band-gaps in Mindlin continua, a very low amount
of energy is reflected for any value of the frequency in the observed
interval. }}
\end{figure}

\begin{figure}[H]
\begin{centering}
\includegraphics[scale=0.7]{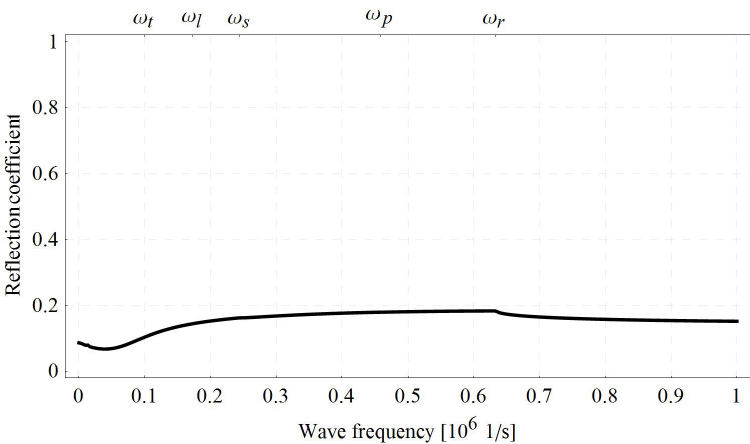} 
\par\end{centering}

\protect\caption{\label{fig:RPSMind}\textcolor{black}{Cauchy/Mindlin interface: macro
clamp with fixed microstructure. Reflection coefficient as function
of frequency for incident P+S wave ($\bar{\alpha}_{1}=1,\ \bar{\alpha}_{2}=1,\ \bar{\alpha}_{3}=1$).
Due to the absence of band-gaps in Mindlin continua, a very low amount
of energy is reflected for any value of the frequency in the observed
interval. }}
\end{figure}

Figures \ref{fig:RPMind}, \ref{fig:RSMind} and \ref{fig:RPSMind}
show the behavior of the reflection coefficient as a function of frequency
for longitudinal, transverse and generic (longitudinal + transverse)
incident waves, respectively. It can be noticed that no frequency
interval can be identified for which complete reflection occur, due
to the fact that no band-gaps are allowed in standard micromorphic
media (see Figure \ref{fig:DispersionTotal-Mind}). More than this,
it can be noticed that, very few energy is reflected back in the Cauchy
medium for all values of frequencies. This means that the fact of
fixing the microstructure at the interface, does not allow for micro
resonant modes that trigger energy reflection. As a result, the standard
micromorphic medium almost behaves as a Cauchy continuum with respect
to wave reflection and transmission since no microstructure-related
local resonances are activated.

\subsubsection{Macro clamp with free microstructure}

We recall from Subsection (\ref{sub:Macro-clamp-fixed_micro}) that
the 12 jump conditions to be imposed at $x_{1}=0$ for the considered
connection are 
\[
\left\llbracket u_{i}\right\rrbracket =0,\qquad t_{i}-f_{i}=0,\qquad\tau_{ij}^{+}=0,\qquad i=1,2,3,\ \ j=2,3,
\]
which in terms of the new variables (see Eqs. (\ref{eq:Cauchy_force}),
(\ref{eq:JumpRelaxed-1_1}), (\ref{eq:P_change_var}), (\ref{eq:Unknowns1})
and (\ref{eq:Unknowns2})) and introducing the tangent vectors $\gtau_{1}=(0,1,0)$
and $\gtau_{2}=(0,0,1)$, read\medskip{}
 
\[
\left\llbracket u_{1}\right\rrbracket =0,\quad\left\llbracket u_{2}\right\rrbracket =0,\quad\left\llbracket u_{3}\right\rrbracket =0,
\]
\medskip{}
 
\begin{gather}
\begin{pmatrix}\lambda_{e}+2\mu_{e}\vspace{1.2mm}\\
0\vspace{1.2mm}\\
0
\end{pmatrix}\cdot\left(\vv_{1}^{+}\right)'+\begin{pmatrix}0\vspace{1.2mm}\\
-2\mu_{e}\vspace{1.2mm}\\
-(3\lambda_{e}+2\mu_{e})
\end{pmatrix}\cdot\vv_{1}^{+}=(\lambda_{\rm macro}+2\mu_{\rm macro})\,\left(u_{1}^{-}\right)'\vspace{1.2mm}\nonumber \\
\vspace{1.2mm}\label{eq:JumpRelaxed-1_1-1-1-1}\\
\begin{pmatrix}\mu_{e}+\mu_{c}\vspace{1.2mm}\\
0\vspace{1.2mm}\\
0
\end{pmatrix}\cdot\left(\vv_{\alpha}^{+}\right)'+\begin{pmatrix}0\vspace{1.2mm}\\
-2\mu_{e}\vspace{1.2mm}\\
2\mu_{c}
\end{pmatrix}\cdot\vv_{\alpha}^{+}=\mu_{\rm macro}\,\left(u_{\alpha}^{-}\right)',\qquad\alpha=2,3,\nonumber 
\end{gather}
\medskip{}
 
\begin{flalign}
\tilde{\tau}_{11} & =\begin{pmatrix}0\vspace{1.2mm}\\
\mu_{e}L_{g}^{2}\vspace{1.2mm}\\
\mu_{e}L_{g}^{2}
\end{pmatrix}\cdot\left(\vv_{1}^{+}\right)'=0, & \tilde{\tau}_{12}= & \begin{pmatrix}0\vspace{1.2mm}\\
\mu_{e}L_{g}^{2}\vspace{1.2mm}\\
\mu_{e}L_{g}^{2}
\end{pmatrix}\cdot\left(\vv_{2}^{+}\right)'=0, & \tilde{\tau}_{13}= & \begin{pmatrix}0\vspace{1.2mm}\\
\mu_{e}L_{g}^{2}\vspace{1.2mm}\\
\mu_{e}L_{g}^{2}
\end{pmatrix}\cdot\left(\vv_{3}^{+}\right)'=0,\vspace{1.2mm}\nonumber \\
\vspace{1.2mm}\nonumber \\
\tilde{\tau}_{22}= & \begin{pmatrix}0\vspace{1.2mm}\\
-\frac{\mu_{e}L_{g}^{2}}{2}\vspace{1.2mm}\\
\mu_{e}L_{g}^{2}
\end{pmatrix}\cdot\left(\vv_{1}^{+}\right)'+\frac{\mu_{e}L_{g}^{2}}{2}\left(\vv_{6}^{+}\right)'=0, & \tilde{\tau}_{21}= & \begin{pmatrix}0\vspace{1.2mm}\\
\mu_{e}L_{g}^{2}\vspace{1.2mm}\\
-\mu_{e}L_{g}^{2}
\end{pmatrix}\cdot\left(\vv_{2}^{+}\right)'=0, & \tilde{\tau}_{23}=\  & \mu_{e}L_{g}^{2}\left(\left(\vv_{4}^{+}\right)'+\left(\vv_{5}^{+}\right)'\right)=0,\vspace{1.2mm}\label{eq:P_change_var-1-1-1}\\
\vspace{1.2mm}\nonumber \\
\tilde{\tau}_{33}= & \begin{pmatrix}0\vspace{1.2mm}\\
-\frac{\mu_{e}L_{g}^{2}}{2}\vspace{1.2mm}\\
\mu_{e}L_{g}^{2}
\end{pmatrix}\cdot\left(\vv_{1}^{+}\right)'-\frac{\mu_{e}L_{g}^{2}}{2}\left(\vv_{6}^{+}\right)'=0, & \tilde{\tau}_{32}= & \,\mu_{e}L_{g}^{2}\left(\left(\vv_{4}^{+}\right)'-\left(\vv_{5}^{+}\right)'\right)=0, & \tilde{\tau}_{31} & =\begin{pmatrix}0\vspace{1.2mm}\\
\mu_{e}L_{g}^{2}\vspace{1.2mm}\\
-\mu_{e}L_{g}^{2}
\end{pmatrix}\cdot\left(\vv_{3}^{+}\right)'=0.\nonumber 
\end{flalign}
Inserting the wave-form solutions (\ref{eq:WaveFormCauchy-1-1}) and
(\ref{eq:WaveForm-1-1}) with $x_{1}=0$ in such expressions for the
jump conditions the unknown amplitudes can be determined. As for the
previous case, we do not report here their explicit form for the sake
of conciseness. Nevertheless, we notice that from the boundary conditions
$\tau_{23}=\tau_{32}=0,$ it follows that $\beta_{4}=\beta_{5}=0$
and hence ${v_{4}=\vv_{5}=0}$ for all $x_{1}$. This means that,
as for the previous constraint, the modes $\vv_{4}$ and $\vv_{5}$
cannot be activated at the considered interface between a relaxed
micromorphic medium and a Cauchy continuum.\medskip{}
 Once that the amplitudes (and hence the solution) have been determined,
the incident, reflected and transmitted flux can be computed according
to Eqs. (\ref{Flux-Cauchy-incident-2})-(\ref{Flux-relaxed-transmitted-1}),
in which the amplitudes calculated for the constraint considered in
this Subsection are used.

\begin{figure}[H]
\begin{centering}
\includegraphics[scale=0.7]{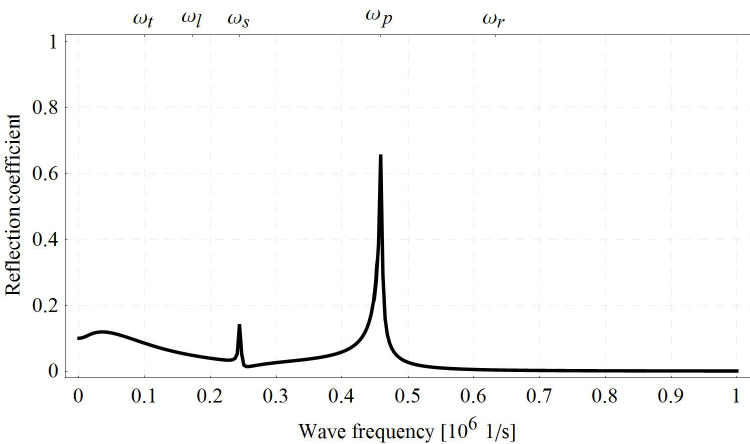} 
\par\end{centering}

\protect\caption{\label{fig:RPMind1}\textcolor{black}{Cauchy/Mindlin interface: macro
clamp with free microstructure. Reflection coefficient as function
of frequency for incident P waves ($\bar{\alpha}_{1}=1,\ \bar{\alpha}_{2}=\bar{\alpha}_{3}=0$).
A local resonance can be observed corresponding to the frequency $\omega_{p}$
due to the fact that the microstructure is free to vibrate.}}
\end{figure}

\begin{figure}[H]
\begin{centering}
\includegraphics[scale=0.7]{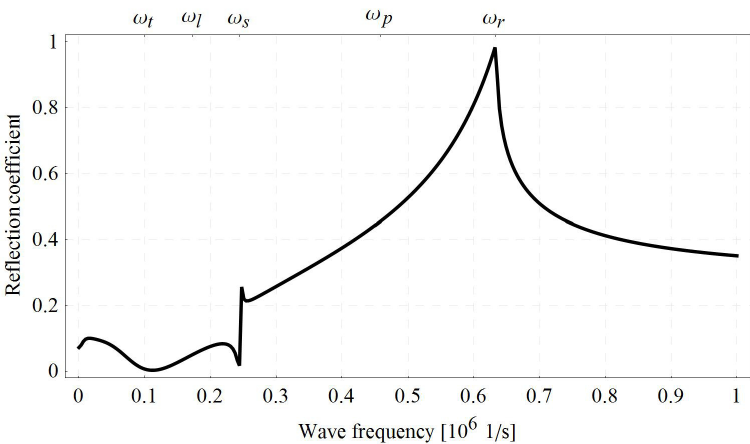} 
\par\end{centering}

\protect\caption{\label{fig:RSMind1}\textcolor{black}{Cauchy/Mindlin interface: macro
clamp with free microstructure. Reflection coefficient as function
of frequency for incident S waves ($\bar{\alpha}_{1}=0,\ \bar{\alpha}_{2}=1,\ \bar{\alpha}_{3}=0$,
or equivalently $\bar{\alpha}_{1}=0,\ \bar{\alpha}_{2}=0,\ \bar{\alpha}_{3}=1$).
A local resonance can be observed corresponding to the frequency $\omega_{r}$
due to the fact that the microstructure is free to vibrate.}}
\end{figure}

\begin{figure}[H]
\begin{centering}
\includegraphics[scale=0.7]{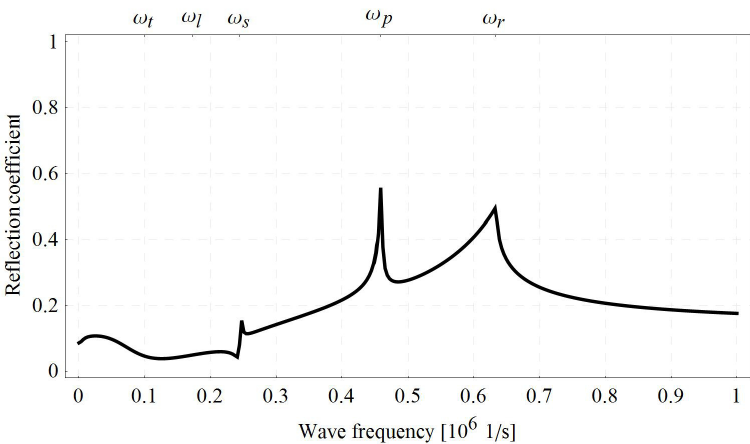} 
\par\end{centering}

\protect\caption{\label{fig:RPSMind1}\textcolor{black}{Cauchy/Mindlin interface: macro
clamp with free microstructure. Reflection coefficient as function
of frequency for incident P+S wave ($\bar{\alpha}_{1}=1,\ \bar{\alpha}_{2}=1,\ \bar{\alpha}_{3}=1$).
Two local resonances can be observed corresponding to the frequencies
$\omega_{p}$ and $\omega_{r}$ due to the fact that the microstructure
is free to vibrate.}}
\end{figure}

Figures \ref{fig:RPMind1}, \ref{fig:RSMind1} and \ref{fig:RPSMind1}
show the behavior of the reflection coefficient as function of the
frequency for longitudinal, transverse and generic (longitudinal +
transverse) waves, respectively. As for the previous case, no complete
reflection frequency intervals can be identified, due to the impossibility
of standard micromorphic models to predict frequency band-gaps. On
the other hand, due to the fact that the microstructure is left free
to vibrate at the considered interface some local resonances can be
identified at $\omega_{p}$ and $\omega_{r}$ for longitudinal and
transverse waves respectively. This is coherent with the results obtained
in Subsection \ref{sub:freeMicro} for the analogous constraint imposed
at a Cauchy/relaxed-micromorphic interface. Moreover, some local resonances
also occur in the vicinity of the cut-off frequency $\omega_{s}$
which, in the case of relaxed micromorphic continuum, was the upper
bound of the band-gap. This means that, in a standard Mindlin continuum,
the microstructural elements do not possess enough freedom to vibrate
independently of the matrix so that no complete frequency band-gap
can be triggered.

\section{Conclusions}

In this paper we present a comprehensive treatise on the setting-up
of jump conditions to be imposed at surfaces of discontinuity in relaxed
micromorphic and in standard Mindlin micromorphic continua, \textcolor{black}{also
recalling the analogous conditions for classical Cauchy media. This
general theoretical framework allows the correct setting up of:}
\begin{itemize}
\item \textcolor{black}{the jump conditions to be imposed at internal surfaces
embedded in relaxed micromorphic, Mindlin or Cauchy continua,}
\item \textcolor{black}{as a particular case of the previous point, the
possible connections between any combination of Mindlin, relaxed micromorphic
or Cauchy media.}
\end{itemize}
\textcolor{black}{We hence focus our attention on the particular case
of interfaces between a classical Cauchy continuum, on one side, and
a relaxed micromorphic one (or, alternatively, a standard Mindlin
one) on the other side. We study all the different types of possible
connections at such interfaces, with particular attention to what
we call ``}\textit{\textcolor{black}{macro internal clamp with fixed
microstructure}}\textcolor{black}{'' and ``}\textit{\textcolor{black}{macro
internal clamp with free microstructure}}\textcolor{black}{''. Both
these constraints guarantee continuity of the macroscopic displacement
at the considered interface and differ for the type of boundary conditions
which are imposed at the level of the micros}tructure. In particular,
the microstructure is kept fixed in the first constraint, while it
is free to vibrate in the second one.

We show that, independently of the chosen constraint, the relaxed
micromorphic medium always presents frequency intervals for which
complete reflection occurs, while the standard Mindlin's model does
not allow for this possibility. This is directly related to the fact
that frequency band-gaps are allowed in the relaxed micromorphic model,
while they are not possible in standard Mindlin continua.

On the other hand, the different boundary conditions that are imposed
at the level of the microstructure, may allow for extra resonant behaviors
producing reflection peaks at higher frequencies. In particular, such
additional microstructure-related resonance frequencies can be identified
when the microstructure is left free to vibrate at the considered
interface.

\textcolor{black}{Even if this is not the main aim of the present
paper, we explicitly remark that the suitable tuning of the constitutive
parameters of the relaxed micromorphic model may modify the local
resonant behaviors presented in Figures \ref{fig:RP1}-\ref{fig:RPS1},
eventually giving rise to a second band gap for the considered constraint
of }\textit{\textcolor{black}{macro internal clamp with free microstructure}}\textcolor{black}{.
The possibility of modeling an additional band-gap by suitably tuning
the parameters of the relaxed model is worth of note, since inverse
measurements may be envisaged for real band-gap metamaterials \cite{NonLoc}.}

\section*{Acknowledgements}

Angela Madeo thanks INSA-Lyon for the funding of the BQR 2016 \textquotedbl{}Caracth\'erisation
m\'ecanique inverse des m\'etamat\'eœriaux: mod\'elisation, identification
exp\'erimentale des param\'etres et \'evolutions possibles\textquotedbl{}.
The work of I.D. Ghiba was supported by a grant of the Romanian National
Authority for Scientific Research and Innovation, CNCS-UEFISCDI, project
number PN-II-RU-TE-2014-4-1109.


\begin{thebibliography}{10}
\bibitem{Achenbach}Achenbach J.D., (1973). ``Wave propagation in
elastic solids'', Elsevier Science Publishers, Amsterdam.

\bibitem{phononic1}Armenise M.N., Campanella C.E., Ciminelli C.,
Dell'Olio F., Passaro V.M.N., (2010). ``Phononic and photonic band
gap structures: modelling and applications\textquotedblright . Physics
Procedia Vol. 3, pp. 357-364.

\bibitem{berezow}Berezowsky A., Giorgio I., Della Corte A., (2015).
``Interfaces in micromorphic materials: Wave transmission and reflection
with numerical simulations''. Mathemetics and Mechanics of Solids,
DOI: 10.1177/1081286515572244.

\bibitem{Chen1}Chen Y., Lee J.D., Eskandarian A., (2004). ``Atomistic
viewpoint of the applicability of microcontinuum theories''. International
Journal of Solids and Structures, Vol. 41:8, pp. 2085-2097.

\bibitem{Chen2}Chen, Y., Lee J.D, (2003). ``Determining material
constants in micromorphic theory through phonon dispersion relations''.
International Journal of Engineering Science, Vol. 41:8, pp. 871-886.

\bibitem{FdISepp} dell'Isola F., Seppecher P., (1995). ``The relationship
between edge contact forces, double force and interstitial working
allowed by the principle of virtual power''. Comptes Rendus Academie
des Sciences II, Mec. Phys. Chim. Astron., Vol. 321, pp. 303-308.

\bibitem{EdgeIsolaSepp} dell'Isola F., Seppecher P., (1997). \newblock
``Edge contact forces and quasi-balanced power''. Meccanica, Vol.
32:1, pp. 33-52.

\bibitem{Seppecher} dell'Isola F., Madeo A., Seppecher P., (2009).
``Boundary conditions at fluid-permeable interfaces in porous media:
a variational approach\textquotedblright . International Journal of
Solids and Structures, Vol. 46, pp. 3150-3164.

\bibitem{FdIPlacidi}dell'Isola F., Madeo A., Placidi L., (2012).
``Linear plane wave propagation and normal transmission and reflection
at discontinuity surfaces in second gradient 3D continua\textquotedblright .
Zeitschrift f\"ur Angewandte Mathematik und Mechanik, Vol. 92:1,
pp. 52-71.

\bibitem{Dontsov1}Dontsov E.V. Tokmashev R.D., Guzina B.G., (2013)
A physical perspective of the length scales in gradient elasticity
through the prism of wave disperison, International Journal of Solids
and Structures, 50, 3674-3684. 

\bibitem{EringenBook}Eringen A. C., (2001). ``Microcontinuum field
theories''. Springer-Verlag, New York.

\bibitem{Ghiba} Ghiba I.D., Neff P., Madeo A., Placidi L., Rosi G.,
(2015). ``The relaxed linear micromorphic continuum: existence, uniqueness
and continuous dependence in dynamics\textquotedblright . Mathematics
and Mechanics of Solids, Vol. 20:10, pp. 1171-1197.

\bibitem{Gonella_Micro_inerzia}Gonella S., Greene M.S., Liu W.K.,
(2011). Characterization of heterogeneous solids via wave methods
in computational microelasticity, Journal of the Mechanics and Physics
of Solids, 59, 959-974.

\bibitem{Hui1}Hui T., Oskay C., (2013). A nonlocal homogenization
model for wave dispersion in dissipative composite materials. International
Journal of Solids and Structures, 50, 38-48. 

\bibitem{Hui2}Hui T., Oskay C. (2014), A high oder homogenization
model for transient dynamics of heterogeneous media including micro-inertia
effects, Computer Methods in Applied Mechanics and Engineering, 273,
181-203.

\bibitem{phononic}Liu Z., Zhang X., Mao Y., Zhu Y., Yang Z., Chan
C., Sheng P., (2000b). ``Locally resonant sonic materials\textquotedblright .
Science Vol. 289 (5485), pp. 1734-1736.

\bibitem{Lucklum}Lucklum R., Ke M., Zubtsov M., (2012). ``Two-dimensional
phononic crystal sensor based on a cavity mode\textquotedblright .
Sensors and Actuators B, Vol. 171-172, pp. 271-277.

\bibitem{Gavrilyuk} Madeo A., Gavrilyuk S., (2010). ``Propagation
of acoustic waves in porous media and their reflection and transmission
at a pure fluid/porous medium permeable interface\textquotedblright .
European Journal of Mechanics A/Solids, Vol. 29:5, pp. 897-910.

\bibitem{Sciarra2} Madeo A., dell'Isola F., Ianiro N. and Sciarra
G., (2008). ``A variational deduction of second gradient poroelasticity
II: An application to the consolidation problem\textquotedblright ,
Journal of Mechanics of Materials and Structures, Vol. 3:4, pp. 607-625.

\bibitem{damage} Madeo A., Placidi L., Rosi G., (2014). ``Towards
the design of meta-materials with enhanced damage sensitivity: second
gradient porous materials\textquotedblright . Research in Nondestructive
Evaluation, Vol. 25:2, pp. 99-124.

\bibitem{BandGaps1} Madeo A., Neff P., Ghiba I.D., Placidi L., Rosi
G., (2015). ``Wave propagation in relaxed micromorphic continua:
modeling metamaterials with frequency band gaps\textquotedblright .
Continuum Mechanics and Thermodynamics, Vol. 27, pp. 551-570.

\bibitem{BandGaps2}Madeo A., Neff P., Ghiba I.D., Placidi L., Rosi
G., (2014). ``Band gaps in the relaxed linear micromorphic continuum\textquotedblright .
Zeitschrift f\"ur Angewandte Mathematik und Mechanik, Vol. 27, pp.
551-570.

\bibitem{NonLoc}\textcolor{black}{Madeo A., Barbagallo
G., Placidi L., D\textquoteright Agostino M.V., Neff P., (2016). \textquotedblleft First
evidence of non-locality in real band-gap metamaterials: determining
constitutive parameters in the relaxed micromorphic model\textquotedblright .
http://arxiv.org/abs/1603.02258 (to appear in Proceedings of the Royal
Society of London A).}

\bibitem{Ghiba1}Neff P., Ghiba I.D., Madeo A., Placidi L., Rosi G.,
(2014). ``A unifying perspective: the relaxed linear micromorphic
continuum\textquotedblright . Continuum Mechanics and Thermodynamics,
Vol. 26, pp. 639-681.

\bibitem{Mindlin} Mindlin R.D., (1964). ``Micro-structure in linear
elasticity''. Archive for Rational Mechanics and Analysis, Vol. 16:1,
pp. 51-78.

\bibitem{Kutznetzova}Pham K., Kouznetsova V.G., Geers M.G.D., (2013).
``Transient computational homogenization for heterogeneous materials
under dynamic excitation\textquotedblright . Journal of the Mechanics
and Physics of Solids, Vol. 61, pp. 2125-2146.

\bibitem{Placidi} Placidi L., Rosi G., Giorgio I., Madeo A., (2013).
``Reflection and transmission of plane waves at surfaces carrying
material properties and embedded in second gradient materials\textquotedblright .
Mathematics and Mechanics of Solids, Vol. 19:5, pp. 555-578.

\bibitem{TheseSeppecher} Seppecher P., (1987). ``Etude d'une Modelisation
des Zones Capillaires Fluides: Interfaces et Lignes de Contact''.
Ph.D-Thesis, Ecole Nationale Superieure de Techniques Avancï¿œes, Universitï¿œ
Pierre et Marie Curie, Paris.

\bibitem{phononic3}Spadoni A., Ruzzene M., Gonella S., Scarpa F.,
(2009). ``Phononic properties of hexagonal chiral lattices\textquotedblright .
Wave Motion, Vol. 46, pp. 435-450.

\bibitem{phononic2}Steurer W., Sutter-Widmer D., (2007). Topical
Review: ``Photonic and phononic quasicrystals\textquotedblright .
Journal of Physics D: Applied Physics., Vol. 40, R229-R247.

\bibitem{phononicTorq}Man W., Florescu M., Matsuyama K., Yadak P.,
Nahal G., Hashemizad S., Williamson E., Steinhardt P., Torquato S.,
Chaikin P., (2013). ``Photonic band gap in isotropic hyperuniform
disordered solids with low dielectric contrast\textquotedblright .
Optics Express, Vol. 21:17, pp. 19972-19981.

\bibitem{Guyader} Rosi G., Madeo A., Guyader J.-L., (2013). ``Switch
between fast and slow Biot compression waves induced by ''second
gradient microstructure\textquotedbl{} at material discontinuity surfaces
in porous media\textquotedblright , International Journal of Solids
and Structures, Vol. 50:10, pp. 1721-1746.

\bibitem{Sciarra1}Sciarra G., dell'Isola F., Ianiro N., Madeo A.,
(2008). ``A variational deduction of second gradient poroelasticity
I: general theory\textquotedblright , Journal of Mechanics of Materials
and Structures, Vol. 3:3, pp. 507-526.

\bibitem{Geers1}Sridhar A., Kouznetsova V.G., Geers M.G.D, (2016).
``Homogenization of locally resonant acoustic metamaterials towards
an emergent enriched continuum''. Computational Mechanics, Accepted.

\bibitem{SteigmannOgden}Steigmann D.J., Ogden R.W., (1997). ``Plane
deformation of elastic solids with intrinsic boundary elasticity''.
Proceedings of the Royal Society of London Series A. Vol. 453, pp.
853-877.

\bibitem{guzina2}Wautier A., Guzina B.B., (2015). On the second order
homogenization of wave motion in periodic media and the sound of a
chessboard. Journal of the Mechanics and Physics of Solids, 78, 382-414.s\end{thebibliography}
\end{document}